\documentclass[twocolumn,twocolappendix,tighten]{aastex631}
\usepackage{graphicx}	
\usepackage{url}
\usepackage{epstopdf}
\usepackage{color}
\usepackage{textcomp}
\usepackage{gensymb}
\usepackage{multirow}
\usepackage{ragged2e}
\usepackage{hyperref}
\usepackage{url}
\usepackage{amsmath}
\usepackage{booktabs}
\usepackage{comment}
\usepackage{afterpage}
\usepackage{mathtools}
\usepackage{makecell}
\usepackage{tabularx}
\usepackage{bold-extra}
\usepackage{xspace}
\usepackage{relsize}
\usepackage[utf8]{inputenc}
\usepackage{newunicodechar}
\usepackage[encapsulated]{CJK}
\usepackage{ucs}
\usepackage{soul}
\newcommand{\cntext}[1]{\begin{CJK}{UTF8}{gbsn}#1\end{CJK}}
\usepackage[T1]{fontenc}
 
\DeclareGraphicsExtensions{.pdf,.png,.jpeg}
\def\lsim{\mathrel{\raise.3ex\hbox{$<$\kern-.75em\lower1ex\hbox{$\sim$}}}}
\def\gsim{\mathrel{\raise.3ex\hbox{$>$\kern-.75em\lower1ex\hbox{$\sim$}}}}
\def\gtwid{\mathrel{\raise.3ex\hbox{$>$\kern-.75em\lower1ex\hbox{$\sim$}}}}
\def\proptwid{\mathrel{\raise.3ex\hbox{$\propto$\kern-.75em\lower1ex\hbox{$\sim$}}}}

\newcommand{\dmc}{DMC\xspace}
\newcommand{\themis}{\textsc{Themis}\xspace}
\newcommand{\difmap}{\texttt{DIFMAP}\xspace}
\newcommand{\ehtim}{\texttt{eht-imaging}\xspace}
\newcommand{\polsolve}{\texttt{polsolve}\xspace}

\newcolumntype{P}[1]{>{\centering\arraybackslash}p{#1}}

\DeclareRobustCommand{\okina}{%
  \raisebox{\dimexpr\fontcharht\font`A-\height}{%
    \scalebox{0.8}{`}%
  }%
}
\newunicodechar{ʻ}{\okina}

\def\M87{M87\xspace}
\def\m87{M87$^*$\xspace}
\def\sgra{Sgr~A$^*$\xspace}
\def\ehtim{\texttt{eht-imaging}\xspace}

\def\3C279{3C\,279\xspace}

\usepackage{xcolor}

\defcitealias{PaperI}{Paper~I}
\defcitealias{PaperII}{Paper~II}
\defcitealias{PaperIII}{Paper~III}
\defcitealias{PaperIV}{Paper~IV}
\defcitealias{PaperV}{Paper~V}
\defcitealias{PaperVI}{Paper~VI}
\defcitealias{PaperVII}{Paper~VII}
\defcitealias{PaperVIII}{Paper~VIII}
\defcitealias{Roelofs_2023}{R23} % polarimetric m-ring modeling companion paper

\begin{document}
% \pagenumbering{roman}

\nocite{PaperI}
\nocite{PaperII}
\nocite{PaperIII}
\nocite{PaperIV}
\nocite{PaperV}
\nocite{PaperVI}
\nocite{PaperVII}
\nocite{PaperVIII}
\nocite{Roelofs_2023}

\title{
First M87 Event Horizon Telescope Results IX: Detection of Near-Horizon Circular Polarization}
\shorttitle{First M87 Event Horizon Telescope Results IX}

\author[0000-0002-9475-4254]{Kazunori Akiyama}
\affiliation{Massachusetts Institute of Technology Haystack Observatory, 99 Millstone Road, Westford, MA 01886, USA}
\affiliation{National Astronomical Observatory of Japan, 2-21-1 Osawa, Mitaka, Tokyo 181-8588, Japan}
\affiliation{Black Hole Initiative at Harvard University, 20 Garden Street, Cambridge, MA 02138, USA}

\author[0000-0002-9371-1033]{Antxon Alberdi}
\affiliation{Instituto de Astrofísica de Andalucía-CSIC, Glorieta de la Astronomía s/n, E-18008 Granada, Spain}

\author{Walter Alef}
\affiliation{Max-Planck-Institut für Radioastronomie, Auf dem Hügel 69, D-53121 Bonn, Germany}

\author[0000-0001-6993-1696]{Juan Carlos Algaba}
\affiliation{Department of Physics, Faculty of Science, Universiti Malaya, 50603 Kuala Lumpur, Malaysia}

\author[0000-0003-3457-7660]{Richard Anantua}
\affiliation{Black Hole Initiative at Harvard University, 20 Garden Street, Cambridge, MA 02138, USA}
\affiliation{Center for Astrophysics $|$ Harvard \& Smithsonian, 60 Garden Street, Cambridge, MA 02138, USA}
\affiliation{Department of Physics \& Astronomy, The University of Texas at San Antonio, One UTSA Circle, San Antonio, TX 78249, USA}
%\affiliation{Center for Computational Astrophysics, Flatiron Institute, 162 Fifth Avenue, New %York, NY 10010, USA}

\author[0000-0001-6988-8763]{Keiichi Asada}
\affiliation{Institute of Astronomy and Astrophysics, Academia Sinica, 11F of Astronomy-Mathematics Building, AS/NTU No. 1, Sec. 4, Roosevelt Rd., Taipei 10617, Taiwan, R.O.C.}

\author[0000-0002-2200-5393]{Rebecca Azulay}
\affiliation{Departament d'Astronomia i Astrofísica, Universitat de València, C. Dr. Moliner 50, E-46100 Burjassot, València, Spain}
\affiliation{Observatori Astronòmic, Universitat de València, C. Catedrático José Beltrán 2, E-46980 Paterna, València, Spain}
\affiliation{Max-Planck-Institut für Radioastronomie, Auf dem Hügel 69, D-53121 Bonn, Germany}

\author[0000-0002-7722-8412]{Uwe Bach}
\affiliation{Max-Planck-Institut für Radioastronomie, Auf dem Hügel 69, D-53121 Bonn, Germany}

\author[0000-0003-3090-3975]{Anne-Kathrin Baczko}
\affiliation{Department of Space, Earth and Environment, Chalmers University of Technology, Onsala Space Observatory, SE-43992 Onsala, Sweden}
\affiliation{Max-Planck-Institut für Radioastronomie, Auf dem Hügel 69, D-53121 Bonn, Germany}

\author{David Ball}
\affiliation{Steward Observatory and Department of Astronomy, University of Arizona, 933 N. Cherry Ave., Tucson, AZ 85721, USA}

% ...

\collaboration{0}{The Event Horizon Telescope Collaboration}

\begin{abstract}
% ApJL Length Limit: 250 Words
% To decide: whether we should abbreviate circular polarization as CP everywhere!
Event Horizon Telescope (EHT) observations have revealed a bright ring of emission around the supermassive black hole at the center of the M87 galaxy. EHT images in linear polarization have further identified a coherent spiral pattern around the black hole, produced from ordered
magnetic fields threading the emitting plasma. Here, we present the first analysis of circular polarization using EHT data, acquired in 2017, which can potentially provide additional insights into the magnetic fields and plasma composition near the black hole. Interferometric closure quantities provide convincing evidence for the presence of circularly polarized emission on event-horizon scales. We produce images of the circular polarization using both traditional and newly developed methods.
All methods find a moderate level of resolved circular polarization across the image ($\langle|v|\rangle < 3.7\%$), consistent with the low image-integrated circular polarization fraction measured by the ALMA array ($|v_{\rm int}| < 1\%$). Despite this broad agreement, the methods show substantial variation in the morphology of the circularly polarized emission, indicating that our conclusions are strongly dependent upon the imaging assumptions because of the limited baseline coverage, uncertain telescope gain calibration, and weakly polarized signal.
We include this upper limit in an updated comparison to general relativistic magnetohydrodynamic (GRMHD) simulation models. This analysis reinforces the previously reported preference for magnetically arrested accretion flow models. We find that most simulations naturally produce a low level of circular polarization consistent with our upper limit, and that Faraday conversion is likely the dominant production mechanism for circular polarization at 230 GHz in \m87.
\end{abstract}

\keywords{Galaxies: individual: M87; Radio interferometry; Very long baseline interferometry; Polarimetry; Supermassive black holes; Low-luminosity active galactic nuclei} 

\nocite{PaperI}
\nocite{PaperII}
\nocite{PaperIII}
\nocite{PaperIV}
\nocite{PaperV}
\nocite{PaperVI}
\nocite{PaperVIII}

%\NewPageAfterKeywords

%\tableofcontents
%\clearpage
%\newpage

%=================================================================================================================================================================
\section{Introduction}
\label{sec:intro}
%=================================================================================================================================================================

The Event Horizon Telescope (EHT) has produced
the first images of the event-horizon-scale millimeter emission around the supermassive black hole in the core of the massive elliptical galaxy M87 
at the center of the Virgo Cluster. 
Using very-long-baseline interferometry (VLBI) at 230 GHz, these initial EHT observations from 2017 recovered a ring-like structure with a diameter similar to the predicted ``black hole shadow'' of a $M_{\rm BH} \approx6.5\times10^9 M_\odot$ black hole at the distance of \m87 ($D\approx16.8$\,Mpc). The resolved total intensity images of the ring were consistent with models of synchrotron emission from ultra-hot magnetized plasma near the event horizon \citep[][hereafter Papers~I-VI]{PaperI,PaperII,PaperIII,PaperIV,PaperV,PaperVI}.

The EHT observes in full polarization, recording 
simultaneous data from orthogonally polarized
feeds at each antenna.
Images of the near-horizon linearly polarized radiation 
were published and analyzed in \citet[][hereafter Paper VII]{PaperVII}.
These linear polarimetric images provide essential new information about the magnetic field structure  near the event horizon of M87's supermassive black hole, indicating that the near-horizon magnetic fields are ordered and dynamically important  \citep[][hereafter Paper VIII]{PaperVIII}.

In this Letter we report on the search for resolved \emph{circularly} polarized radiation (CP)
on event-horizon scales in \m87 from EHT observations in 2017. 
The circularly polarized signal from synchrotron radiation near the black hole should contain unique information about the magnetic field and the nature of the radiating particles that cannot be inferred from total intensity or linear polarization alone. These include the possibility of directly measuring the strength of the magnetic field  
and determining whether the observed radiation is mainly from an electron-positron or an electron-ion plasma \citep[e.g.,][]{Wardle+1998}.
However, the circularly polarized signal is expected to be weaker than the linear polarization
\citep{Jones_ODell_1977,Jones_1988},
requiring high sensitivity and accurate calibration of each antenna to be detected.

Previous radio and millimeter-wavelength observations of CP in \m87 are quite limited, whether at low or high angular resolution. \citet{Homan2006}, using the Very Long Baseline Array (VLBA) at 15\,GHz, measured fractional circular polarization  in the core of $-0.49 \pm 0.10\,\%$. Interestingly, they found no linear polarization (LP $< 0.07\,\%$), reversing the Stokes parameter hierarchy described above. M87 is the only source in their list to show such behavior. 
On the other hand, using the Very Large Array (VLA) at 8.4\,GHz, \citet{Bower2002} detected weak fractional linear polarization (LP = $1.74 \pm 0.06\,\%$) but no circular polarization 
(|CP| $< 0.1\,\% $).
Single dish measurements of M87 with the Effelsberg 100m radio telescope at the same range of frequencies showed a similar trend with LP $\sim1.5\%$ and |CP| $< 0.2\%$ \citep{Myserlis2018}, while at 86 GHz the POLAMI monitoring program observed a fairly stable LP $\sim 5\%$ and CP$\sim-1.5\%$ over a period of 12 years with the IRAM 30m telescope \citep{Thum2018}.
Simultaneous observations during the 2017 EHT campaign with the Atacama Large Millimeter/submillimeter Array (ALMA) did not result in a significant detection of the unresolved fractional CP at 221\,GHz \citep[$\approx -0.3 \pm 0.6\,\% $;][]{Goddi2021}.

Beyond \m87, accurate measurements of 
circular polarization are generically difficult to obtain in VLBI. 
\citet{Homan2006} detected CP at the level of $3\sigma$ or better in 17 sources out of their sample of 133 Active Galactic Nuclei (AGN)
using a “gain transfer” technique \citep{homan_wardle_1999}, in which all sources are used for polarization calibration.
These detections at $15\,$GHz all had fractional polarizations between $0.25\,\%$ and $0.70\,\%$.
\citet{Gabuzda+2008} also detected circular polarization in eight AGN jets with $15\,$GHz VLBI measurements, and associated the observed CP signs with CP production by Faraday conversion in helical jet magnetic fields.  
At higher frequencies up to 43 GHz, \citet{Vitrishchak_2008} found circular polarization fractions up to $\sim1\%$ in a sample of AGN cores resolved with VLBI.

This paper presents the details of the 2017 EHT observations and data calibration for circular polarization, procedures and results for circular polarimetric imaging, and their theoretical interpretation in constraining parameters in a library of simulation models. 
In \autoref{sec:data}, we summarize the EHT\,2017 observations, describe evidence for the detection of
circular polarization in \m87, and describe our a priori calibration procedure.  
In \autoref{sec:methods},
we describe our methods of circular polarimetric image reconstruction and gain calibration from EHT data. 
In \autoref{sec:m87results}, we present image reconstructions from all methods across all EHT 2017 datasets. 
We derive an upper limit on the average, resolved circular polarization fraction in \m87 at 230 GHz.  

In \autoref{sec:theo}, we examine circular polarization in a library of General Relativistic Magnetohydrodynamic (GRMHD) simulations of \m87. We add upper limits to the 
circular polarization on event-horizon scales from our EHT observations
to the list of constraints applied to theoretical models in \citetalias{PaperVIII}. We discuss the effect of these new constraints on our preferred models for \m87's accretion flow, and we investigate the physical origin of circular polarization in passing GRMHD models. 
We summarize the work in \autoref{sec:conc}.

%=================================================================================================================================================================
\section{EHT Observations of \m87} % Data}
\label{sec:data}
%=================================================================================================================================================================

\subsection{Conventions}
\label{sec:conventions}

A radio interferometer, such as the EHT, samples the Fourier plane (visibility domain) of the brightness distribution (image) on the sky \citep[e.g.,][]{TMS}. 
Each image domain Stokes parameter ($\mathcal{I},\mathcal{Q},\mathcal{U},\mathcal{V}$) has a corresponding visibility, which we denote with a tilde (e.g., $\tilde{\mathcal{V}}$). At most stations, the EHT data are recorded in two orthogonal circular polarizations, right and left $(R, L)$. Interferometric visibilities are computed through the complex correlation between each pair of sites $(j,k)$ and polarization (i.e., $R_j R_k^*$, $R_j L_k^*$, $L_j R_k^*$, and $L_j L_k^*$). These measurements can be transformed to the Stokes representation through linear algebraic relations. In particular, the visibility domain circular polarization $\tilde{\mathcal{V}}$ on the $j-k$ baseline is given by
\begin{equation}
   \tilde{\mathcal{V}}_{jk} = 0.5 \left( R_j R_k^* - L_j L_k^*  \right) .
\label{eq:V_from_circular_basis}
\end{equation}
In practice, polarimetric measurements are corrupted through a multitude of systematic effects, which can be conveniently represented in a matrix formalism \citep[e.g.,][]{Hamaker_1996,Smirnov_2011}:
\begin{equation}
\boldsymbol{\rho}'_{jk} = {\bf J}_j  \boldsymbol{\rho}_{jk} {\bf J}_{k}^{ \dagger} \, . \label{eqn:JonesFormalism}
\end{equation}
In this expression, $\boldsymbol{\rho}_{jk}$ is a $2{\times}2$ matrix of the true visibilities on the $j-k$ baseline, $\boldsymbol{\rho}_{jk}'$ gives the measured visibilities, and $\mathbf{J}$ is a complex time- and site-dependent $2{\times}2$ matrix that describes the aggregate systematic effects. The latter can be further decomposed into a product of three terms,  $\mathbf{J}\!=\!\mathbf{G}\mathbf{D}\mathbf{\Phi}$, where $\mathbf{G}$ is a diagonal matrix that describes the time-dependent ``gains'' of the two feeds, $\mathbf{D}$ is a matrix with diagonal entries that are unity and off-diagonal entries that describe ``leakage'' between the feeds, and the matrix $\mathbf{\Phi}$ describes the overall rotation of the feeds. For studies of circular polarization with circularly polarized feeds, the gain matrix is the most important source of contamination, predominantly through the gain ratio $G_{R/L}$,
\begin{equation}
    \mathbf{G}\!\equiv\!\begin{pmatrix} 
G_R & 0 \\ 0 & G_L \end{pmatrix} = G_L \begin{pmatrix} 
G_{R/L} & 0 \\ 0 & 1 \end{pmatrix}   \, .
\label{eq:JonesPol}
\end{equation}
In particular, while the gains $G_L$ and $G_R$ have rapid variations, especially in their phase, the gain ratio $G_{R/L}$ can be stable over timescales of multiple days. 
For additional details on polarimetric relations, these representations, and our conventions, see \citetalias{PaperVII}.

\subsection{Observations and data reduction}
\label{sec:observations}

The EHT observed \m87 on 5, 6, 10, and 11 April 2017 with a VLBI array of 7 telescopes located at 5 geographical sites: 
ALMA and the Atacama Pathfinder Experiment (APEX) in the Atacama Desert in Chile; the Large Millimeter Telescope Alfonso Serrano (LMT) on the Volc\'{a}n Sierra Negra in Mexico; the IRAM 30\,m telescope (PV) on Pico Veleta in Spain; the Submillimeter Telescope (SMT) on Mt. Graham in Arizona (USA); and the Submillimeter Array (SMA) and the James Clerk Maxwell Telescope (JCMT) on Maunakea in Hawaiʻi. The South Pole Telescope (SPT) also participated in the EHT observations but cannot see \m87. 
  
Each telescope recorded two frequency bands, each 2\,GHz wide, centered at 229.1\,GHz (high band, HI) and at 227.1\,GHz (low band, LO). Most sites recorded right and left circular polarization simultaneously, except for the JCMT which recorded a single circular polarization each night, and ALMA, which recorded orthogonal linear polarizations that were subsequently converted to a circular basis \citep{Marti_2016,Matthews_2018,Goddi_2019}. 
\citetalias{PaperII} provides a detailed description of the EHT array and observations.
 
The EHT data sets were first correlated and then calibrated using two independent pipelines, \texttt{EHT-HOPS} \citep{Blackburn_2019} and CASA \texttt{rPICARD} \citep{Janssen_2019}, for the stabilization of the source signal in time and frequency \citep[e.g.,][]{2022Janssen}. The data presented in this paper correspond to the \texttt{EHT-HOPS} pipeline, following a verification of the inter-pipeline consistency \citepalias{PaperIII}, although in \autoref{app:polar_gains} we also give a brief summary of the gain calibration in the CASA \texttt{rPICARD} pipeline, and we compare imaging results from \texttt{EHT-HOPS} and \texttt{rPICARD} pipeline data in \autoref{sec:m87results}. The amplitude flux density calibration was performed with custom-built \texttt{EHT-HOPS} post-processing scripts, based on the metadata provided by the participating telescopes. A more extensive description of other aspects of the EHT \m87 data set calibration was presented in \citetalias{PaperIII}. The data sets analyzed in this paper are identical as the ones used in \citetalias{PaperVII} and \citetalias{PaperVIII}, following the same polarimetric calibration procedures.
The full-Stokes calibrated VLBI data from the 2017 EHT observations of M87 are publicly available through the EHT data portal\footnote{http://eventhorizontelescope.org/for-astronomers/data} under the code 2023-D01-01\citep{M87poldata}.

\subsection{$R/L$ gain calibration}
\label{sec:polarimetric_gains}

We calibrated the gain ratio $G_{R/L}$ for each site using multi-source and multi-day polynomial fits to the differences between $RR^*$ and $LL^*$ visibilities. Since ALMA observes in a linear-polarization basis, it provides the absolute electric vector position angle (EVPA) information and it is used as the reference station; its $G_{R/L}$ is fixed to unity \citep{Marti_2016,Goddi_2019}.\footnote{A clockwise 45 degree rotation was also applied to all observations across the whole array (i.e., 90 degrees added to the $G_{R/L}$ phases of all stations), in order to account for the orientation of the ALMA Band 6 cartridges with respect to the antenna mounts \citepalias{PaperVII}.}

For the other sites, the $R/L$ gains were modeled separately for each band using visibilities on baselines to ALMA, using custom-built Python scripts. The use of multiple sources (10 sources were observed with ALMA during the EHT~2017 campaign) in the fitting procedure helps to distinguish between instrumental effects, which are largely source-independent, and  contributions from circular polarization, which are source-dependent \citep{Steel2019}. 

For most sites, $G_{R/L}$ could be successfully modeled using a constant value across the full observing campaign, separately for each frequency band. The sites that required a time-dependent $G_{R/L}$ were APEX, which showed a linear phase drift between $RR^*$ and $LL^*$ visibilities, and the SMA, which showed irregular phase variation that we modeled using third order polynomials specific to each day and frequency band. We calibrated the amplitudes $|G_{R/L}|$ assuming a constant value for each site. 

In \autoref{app:polar_gains}, we provide more details on this strategy for relative gain calibration, as well as an example plot detailing how gains were estimated for the LMT
(\autoref{fig:RL_gain}). We also compare the above strategy for relative gain calibration used for the main results in this paper with the alternative strategy employed by the  CASA-based \texttt{rPICARD} pipeline for EHT data reduction, which calibrates relative gains assuming the intrinsic $\mathcal{V}=0$.

\begin{figure}[t]
\centering
\includegraphics[width=0.48\textwidth,angle=0]{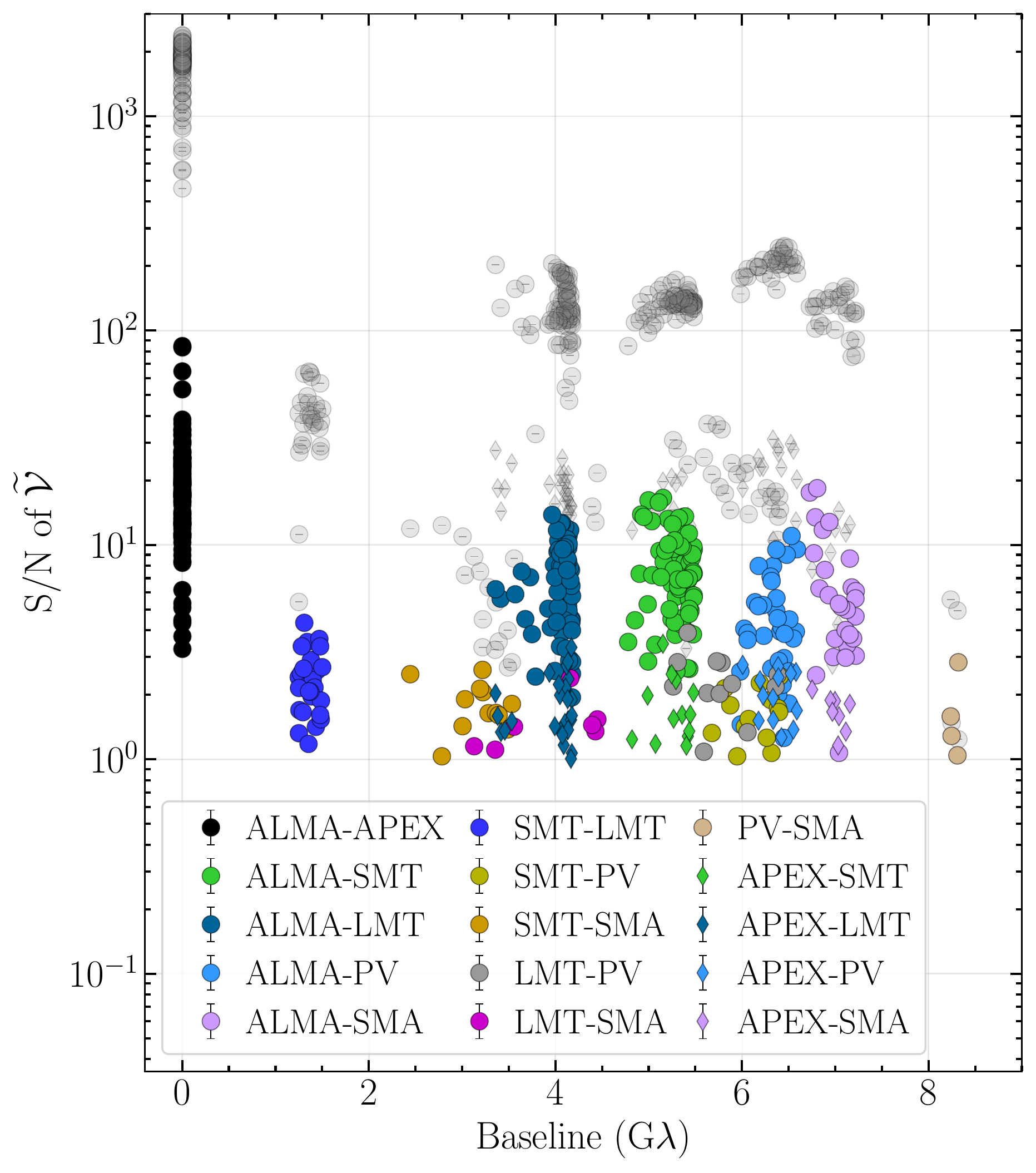}
\caption{Stokes $\tilde{\mathcal{V}}$ visibility S/N in scan-averaged EHT 2017 data, as a function of the projected baseline length. Data from both frequency bands, and all observing days are shown. No systematic uncertainties, like the imperfect calibration of the gain ratios $G_{R/L}$, were accounted for. Hence, the plotted S/N represents upper limits on the $\tilde{\mathcal{V}}$ detections. Gray points in the background indicate the S/N of Stokes $\tilde{\mathcal{I}}$ detections on the corresponding baselines.} \label{fig:m87_uv_coverage_snr}
\end{figure}

\begin{figure*}[t]
\centering
\includegraphics[width=0.95\textwidth]{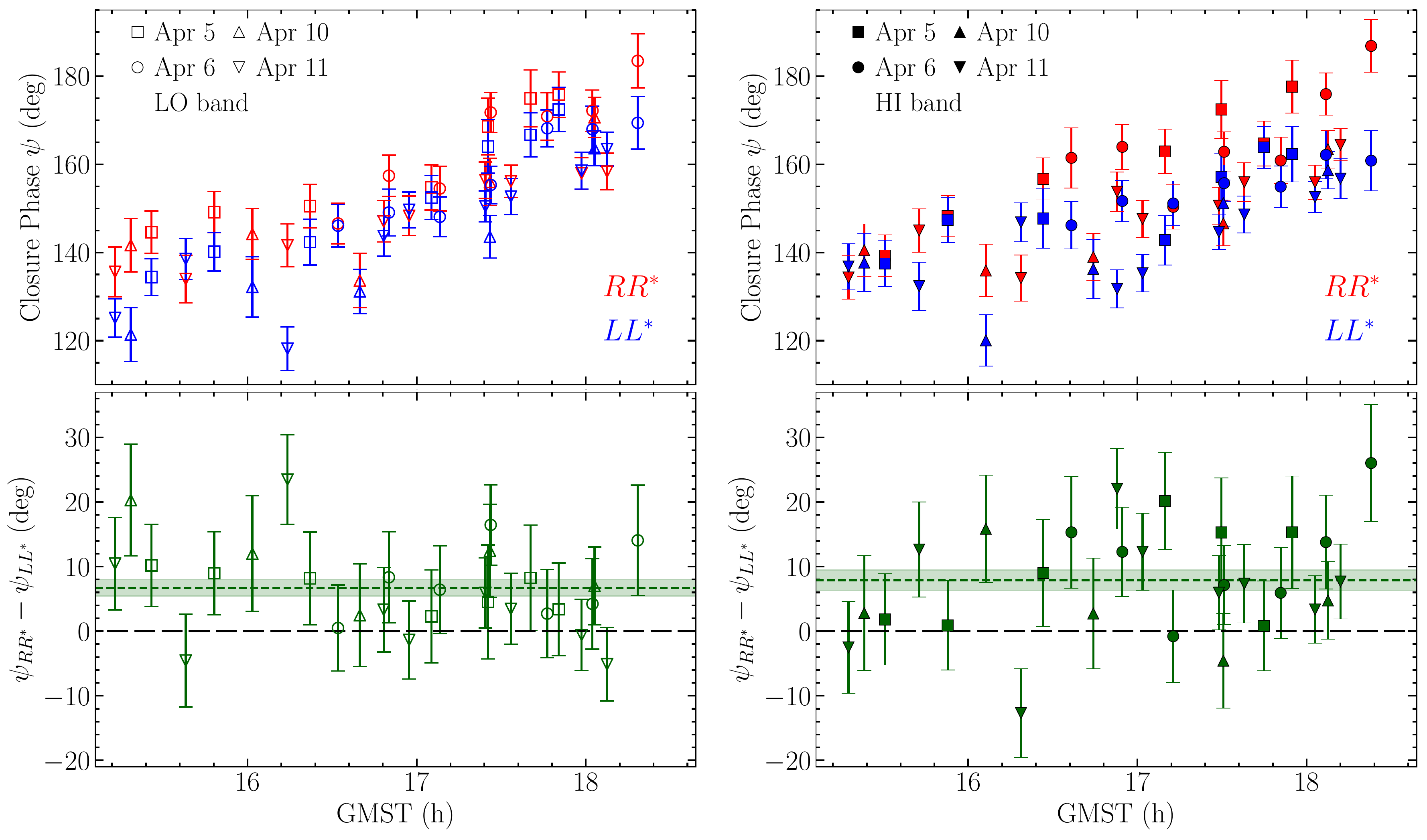}
 \caption{Closure phases observed on the triangle ALMA-SMT-PV triangle during M87$^*$ observations on 2017 Apr 5-11 in low band (left column) and in high band (right column).
Top: closure phases of scan-averaged visibilities for all epochs, $RR^*$ in red; $LL^*$ in blue. Bottom: difference of closure phases between $RR^*$ and $LL^*$. The zero level of the closure difference (i.e., no $\tilde{\mathcal{V}}$ detected) is marked with a black dashed line. A light green band shows the $RR^*-LL^*$ difference, inferred by-band, under the constant difference assumption.} 
\label{fig:ClosDiff}
\end{figure*}

\subsection{Evidence for circular polarization}
\label{sec:resolvedcp}

Under a perfect calibration, we could directly interpret the signal to noise (S/N) ratio of $\tilde{\mathcal{V}}$, calculated from observables with  \autoref{eq:V_from_circular_basis}, to identify robust detections of CP. In \autoref{fig:m87_uv_coverage_snr} we show the S/N of $\tilde{\mathcal{V}}$ as a function of a projected baseline length. Unfortunately, as circular polarization is encoded in the difference between $RR^*$ and $LL^*$ visibilities, residual errors in the calibration of $G_{R/L}$ will create spurious $\tilde{\mathcal{V}}$ signatures. As a consequence, the S/N will be inflated, and hence values shown in \autoref{fig:m87_uv_coverage_snr} can only be treated as upper limits. This is further illustrated in \autoref{app:polar_gains}, where we discuss a more aggressive calibration strategy for $G_{R/L}$, which reduces the S/N of $\tilde{\mathcal{V}}$ significantly.

There are, however, quantities that are robust against the effects of complex antenna gains. An example of such quantities are the closure phases \citep[e.g.,][]{TMS,Blackburn_Closure}, defined for a triangle of antennas ($A$, $B$, and $C$) as
\begin{equation}
\psi = \phi_{AB} + \phi_{BC} + \phi_{CA},
\end{equation}
\noindent where $\phi_{ij}$ is the visibility phase measured by baseline $i-j$. 
One way to minimize the effects from possible antenna mis-calibrations and 
infer the presence of resolved source-intrinsic circular polarization in our observations 
is to compute the difference of closure phases between the $RR^*$ and $LL^*$ visibilities; by construction, the difference of these closures will not be affected by $G_{R/L}$ terms at any antenna.\footnote{Note that care must be taken when time-averaging visibilities before computing closures, since the averaging of mis-calibrated visibilities can introduce systematics into the post-average closure distributions \citep[e.g.,][]{Marti_2008}.}

In \autoref{fig:ClosDiff} (top panels), we show the closure phases, $\psi$, of the antenna triangle formed by ALMA, SMT, and PV, computed for the whole campaign (left column, empty markers for low band; right column, filled markers for high band). Closure phases for the $RR^*$ and $LL^*$ visibilities are shown in red and blue, respectively. It can be seen that there is an offset in the closure phases between the two polarization channels, which generates a non-zero closure difference (indicated in green in the bottom panels). This difference in the $RR$ and $LL$ closure phases is present in all epochs and in both bands.

For this antenna triangle, the average offset in closure phases between $RR^*$ and $LL^*$ (combining all epochs and assuming a constant residual value) is $6.7\pm1.3$\,deg in the low band and $7.9\pm 1.6$\,deg in the high band, indicated with green bands in the bottom panels. The offset is consistent despite each band being calibrated independently. Moreover, the measured offset is difficult to explain with the conservative systematic non-closing error budget discussed in Section 8.4 of \citetalias{PaperIII}, and hence it implies a tentative detection of a weak
Stokes $\tilde{\mathcal{V}}$ at the level of S/N $\sim 5$. 

While this measurement implies the presence of a
fractional CP reaching $\sim 3$\% somewhere in the visibility domain, the measurement can not be directly translated into a quantitative image domain constraint. The ALMA-SMT-PV triangle shown is the one that produces closure-phase differences with the most clear deviation from zero. 
In \autoref{app:all_triplets}, we show the results for all other triangles including ALMA. None of these other triangles shows an unambiguous detection of a nonzero closure phase difference like that seen on the ALMA-SMT-PV triangle, suggesting that SMT-PV is the baseline most sensitive to the CP signatures in \m87.

Even though the $RR^*$ and $LL^*$ closure phases are robust against antenna gains, they may be affected by instrumental polarimetric leakage (antenna $D$-terms, the $\mathbf{D}$ matrix term in \autoref{eq:JonesPol}). However, the effect of $D$-term uncertainties in the parallel-hand visibilities is much smaller than in the cross-hands \citep[e.g.,][]{Smirnov_2011}, which implies that Stokes $\mathcal{V}$ is much less affected by instrumental polarization than Stokes $\mathcal{Q}$ and $\mathcal{U}$. 
To verify a negligible impact of the polarimetric leakage on the closure phase signal, we have
compared the closure-phase values between the data with and without $D$-term calibration. For all triangles related to ALMA, the effect of the $D$-terms on the $RR^*-LL^*$ closure differences is always less than the standard deviation of the thermal noise on the closure phase difference. 
For the triangle shown in \autoref{fig:ClosDiff}, the maximum effect of the $D$-terms is only 0.42\,$\sigma$. 

Hence, we can conclude that the closure phase differences indicating the presence of circular polarization on EHT baselines, presented in \autoref{fig:ClosDiff}, are robust against both antenna gains and polarimetric leakage. In \autoref{app:closuretraces}, we discuss evidence for polarization in ``closure trace'' products \citep{BroderickPesce_2020}, quantities that are insensitive to \emph{all} station-based systematic factors, including $D$-terms. 

%=================================================================================================================================================================
\section{Polarimetric Imaging Methods}
\label{sec:methods}
%=================================================================================================================================================================

\begin{deluxetable*}{lll}[th!]
    \tablecaption{Summary of the Imaging and Modeling Methods Used\label{tab:method_summary}}
    \tablehead{\colhead{Method} & \colhead{Leakage Assumptions} & \colhead{Gain Assumptions}}
    \startdata
    \centering
        CLEAN Imaging: & & \\
          \quad \polsolve     & Priors from \citetalias{PaperVII}   & Self-calibration (gains \& $D$-terms). ALMA $G_{R/L} \equiv 1$   \\
          \quad \difmap       & \citetalias{PaperVII} values fixed & Self-calibration, assuming $\mathcal{V}=0$. ALMA $G_{R/L} \equiv 1$ \\
        Bayesian Raster: & & \\
          \quad \dmc                & None (leakage fitted)     &  ALMA gain phases are 0 (all other gains fitted)\\
          \quad \themis       &  None (leakage fitted)   &    $G_{\rm R} \equiv G_{\rm L}$              \\
        Geometric Modeling: & & \\
          \quad \ehtim  & \citetalias{PaperVII} values fixed     & None (only analyzes closure products); \\        
          && $v_{\rm net}$ fixed to ALMA measurement.\\
\enddata
    \tablecomments{Each method produces images of \m87 in all four Stokes parameters. The circular polarization signal is weak and strongly depends on calibration assumptions, so we summarize the primary differences in calibration assumptions among the methods. For additional details (e.g., priors adopted for fitted values), see the more detailed descriptions in \autoref{app:methods}.}
\end{deluxetable*}

As discussed in \autoref{sec:data}, the circular polarization signal in \m87 is both weak and sensitive to calibration errors. Consequently, the inferred circular polarization structure can be sensitive to assumptions made about the residual calibration errors as well as choices made in producing images from the measured visibilities. To assess the potential impact of these effects, we produce fully polarized images of \m87 using five methods. We summarize these methods and their assumptions in \autoref{tab:method_summary}. In \autoref{app:methods}, we present more detail on each method's assumptions and procedures used for circular polarization image reconstructions. 

The five methods we use can be divided into three general categories. The first category uses adaptations of the standard CLEAN imaging algorithm together with iterative ``self-calibration'' to solve for calibration errors; we use the software \difmap (\autoref{sec:difmap}) and \polsolve (\autoref{sec:polsolve}) as two implementations of this approach. The second category uses a Bayesian forward modeling approach, jointly solving for both a polarized image raster and for residual calibration errors and providing estimates for the posterior distributions of each; we use the software \themis and \dmc (\autoref{sec:DMCThemis}). The third category also uses a Bayesian forward modeling approach, but it uses a simple geometric model for the sky image and only fits VLBI ``closure'' quantities to constrain the circular polarization structure; we use the software \ehtim (\autoref{sec:geometric}).

In addition to the differences inherent in each approach as a result of the underlying method (CLEAN vs raster fitting vs geometric modeling, Bayesian exploration vs fitting a single image), our methods face additional choices on how they deal with polarimetric leakage ($D$-terms) and residual gain ratios ($G_{R/L}$). Constraints on the $D$-terms were derived and discussed extensively in \citetalias{PaperVII}; these constraints have been confirmed with analysis of EHT observations of AGN in \citet{Issaoun2022} and \citet{Jorstad2023}. As a result, some methods (\difmap, and $m$-ring modeling) chose to directly apply the \citetalias{PaperVII} $D$-term results to the data and not treat polarimetric leakage further; in contrast, \dmc and \themis fully explored uncertainties in the $D$-terms under flat priors as part of their Bayesian posterior exploration. \polsolve performed $D$-term self-calibration as a part of its imaging procedure using priors motivated by the \citetalias{PaperVII} results.

Most significantly for circular polarization, each method had freedom to chose how to approach residual uncertainties in the gain ratios $G_{R/L}$, in both amplitude and phase. Three imaging methods -- \polsolve, \difmap and \dmc -- solved for separate $G_R$ and $G_L$ terms as part of their self-calibration (\polsolve and \difmap) or Bayesian forward modeling (\dmc) approach. In contrast, \themis did not solve for separate right- and left-circular gains, but only solved for an overall gain term $G=G_R=G_L$, absorbing any uncertainty in the residual gain ratios into the recovered image structure. The geometric $m$-ring modeling method did not solve for any gain terms, as it directly fit gain-insensitive closure quantities in right- and left- circular polarizations independently. These choices in the treatment of the gain ratios can have a large impact on the recovered circular polarization structure and its uncertainty; for instance, while the methods are otherwise similar in their Bayesian approach and treatment of the $D$-terms, the weak priors on the gain ratios in \dmc result in larger error bars in the recovered Stokes $\mathcal{V}$ structure as compared to \themis, which locks all gain ratios to unity. 

Before applying our imaging methods to EHT observations of \m87, we first tested each method on synthetic data taken from GRMHD simulation images of \m87 on EHT2017 baselines. We present these tests in \autoref{sec:synthdata}. Our results on synthetic data suggest that our imaging methods are generally not capable of unambiguously determining the horizon-scale structure of circular polarization in \m87, given the low fractional polarization in the source (and GRMHD models), as well as the poor $u-v$ coverage of the EHT in 2017. Indeed, when turning to real data in \autoref{sec:m87results}, we also find that our methods are unable to arrive at a single consistent image of the circular polarization in \m87. However, we are able to use the images presented in \autoref{sec:m87results} to derive an upper limit for the circular polarization fraction on horizon-scales.

%=================================================================================================================================================================
\section{\m87 Imaging Results}
\label{sec:m87results}
%=================================================================================================================================================================
\begin{figure*}[t]
\centering
\includegraphics[width=\textwidth]{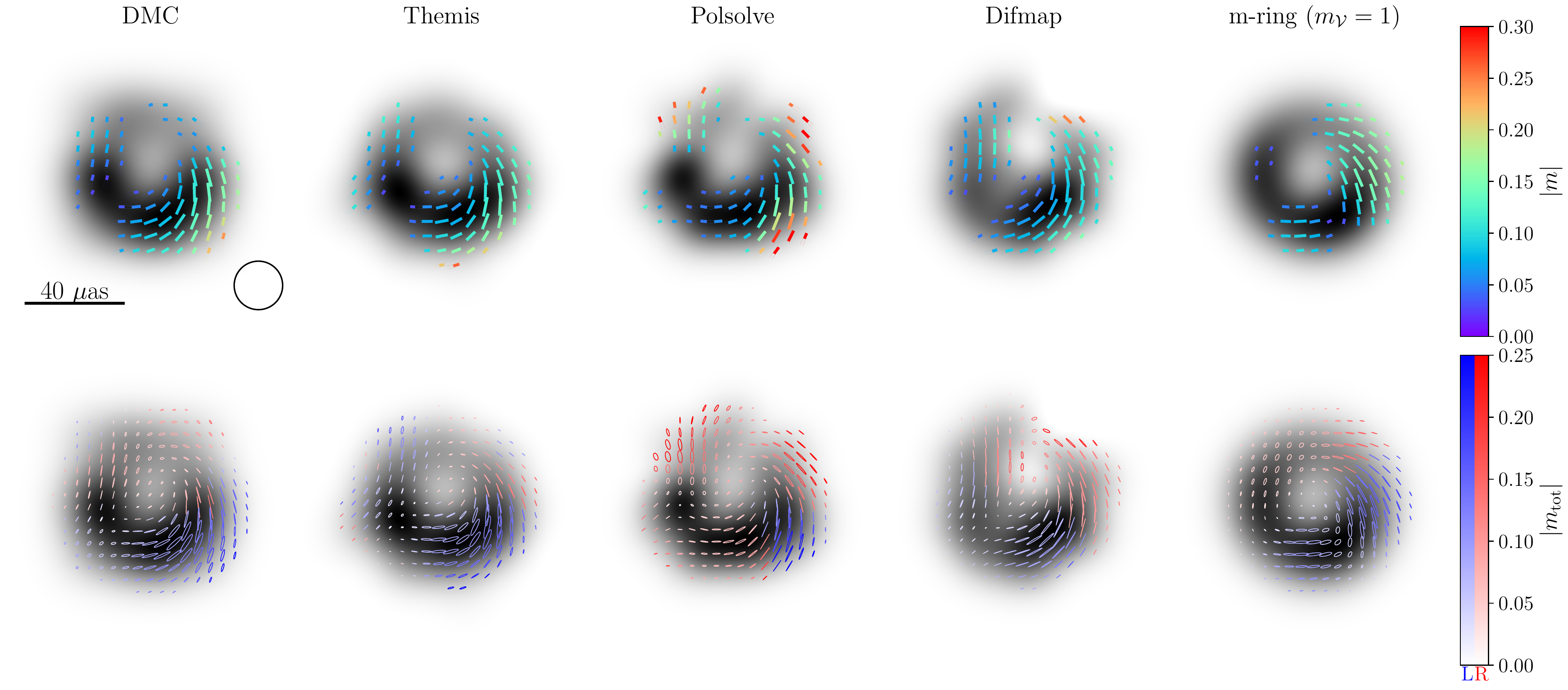} 
\caption{Reconstructions of 2017 EHT \m87 data from April 11, low band. The top row shows total intensity images from all reconstruction methods in grayscale and fractional linear polarization in colored ticks as in \citetalias{PaperVII}. The second row shows the same grayscale total intensity image overlaid with colored ellipses indicating the \emph{total} polarization fraction $|m_{\rm tot}| = \sqrt{\mathcal{Q}^2 + \mathcal{U}^2 + \mathcal{V}^2}/\mathcal{I}$. The size of each ellipse indicates the total polarized brightness; the orientation of each ellipse indicates the linear EVPA, and axis ratio indicates the relative fraction of circular polarization. The color of each ellipse indicates the sign of circular polarization.  
}
\label{fig:images_all_3601}
\end{figure*}

\begin{figure*}[t]
\centering
\includegraphics[width=\textwidth]{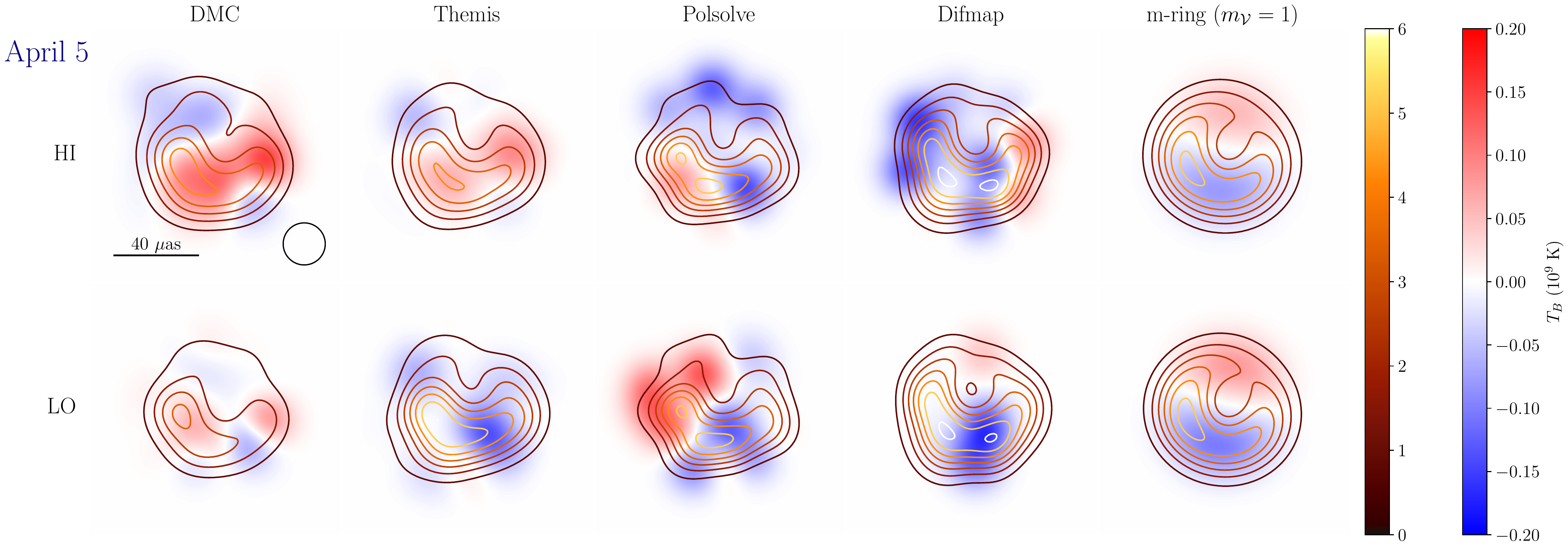} \\
{\color{gray}\rule{\textwidth}{2pt}} \\
\includegraphics[width=\textwidth]{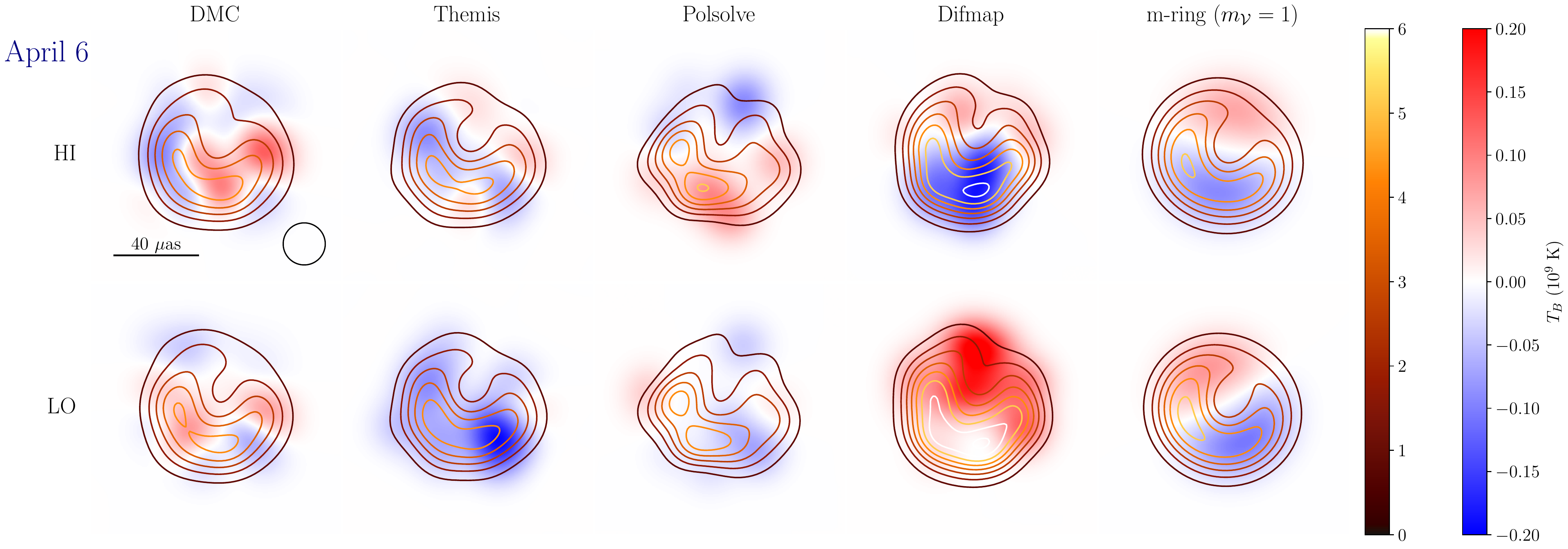} \\
\caption{Circular polarization imaging results from 2017 EHT observations of \m87 on April~5 (top two rows) and April~6 (bottom two rows). 
Images of circular polarization on these consecutive days are expected to be nearly identical, as is seen in total intensity and linear polarization.
The top/bottom row in each pair shows results from imaging the high/low-band data. 
In each panel, total intensity is indicated in the colored linear-scale contours, and the Stokes $\mathcal{V}$ brightness is indicated in the diverging colormap, 
with red/blue indicating a positive/negative sign. 
The colorbar ranges are fixed in both plots (and in \autoref{fig:images_1011}). For posterior exploration methods (\dmc, \themis, $m$-ring fitting), the posterior-average image is shown. All images are blurred with the same $20\,\mu$as FWHM Gaussian, shown with the black inset circle in the upper-left panels. 
}
\label{fig:images_56}
\end{figure*}

\begin{figure*}[t]
\centering
\includegraphics[width=\textwidth]{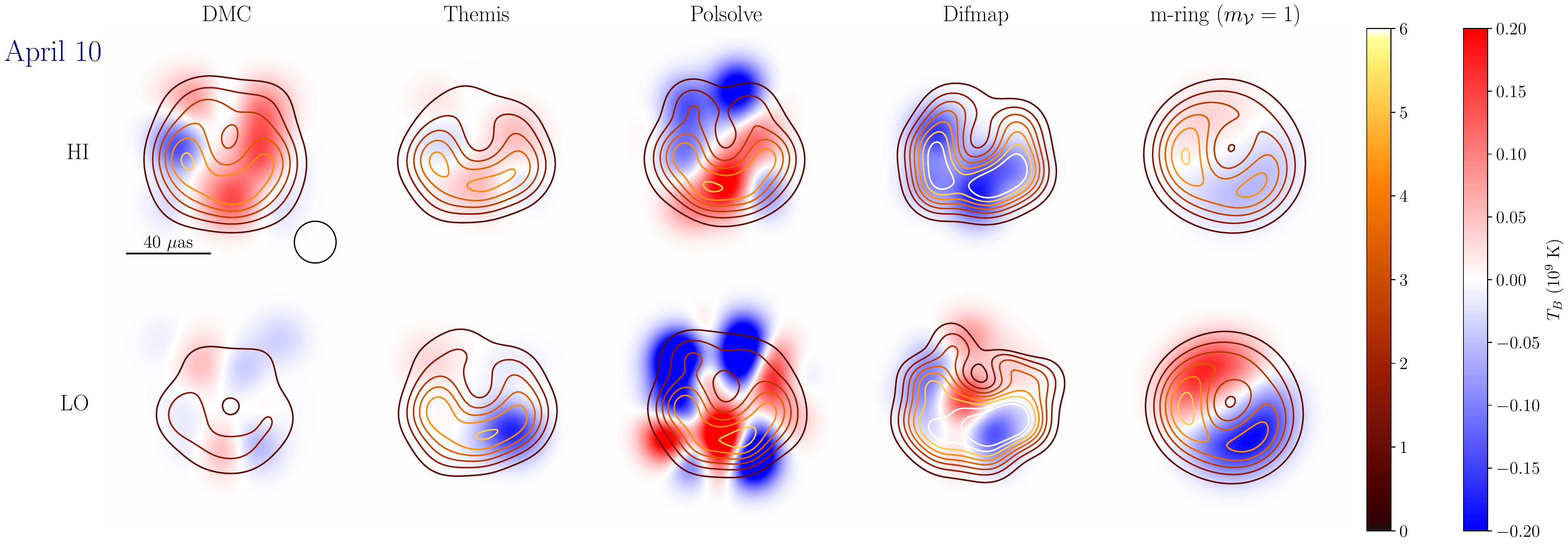} \\
{\color{gray}\rule{\textwidth}{2pt}} \\
\includegraphics[width=\textwidth]{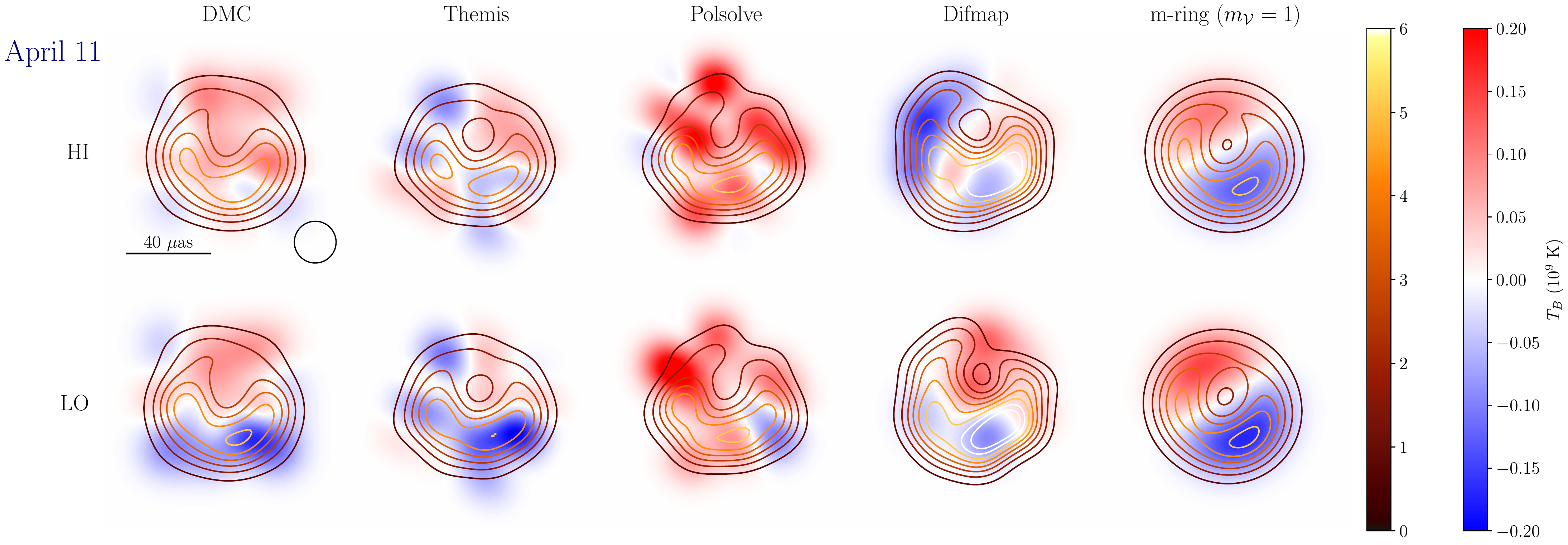} \\
\caption{The same as \autoref{fig:images_56}, but for 2017 EHT observations of \m87 on April~10 (top two rows) and April~11 (bottom two rows).}
\label{fig:images_1011}
\end{figure*}

\begin{figure*}[t]
\centering
\includegraphics[width=\textwidth]{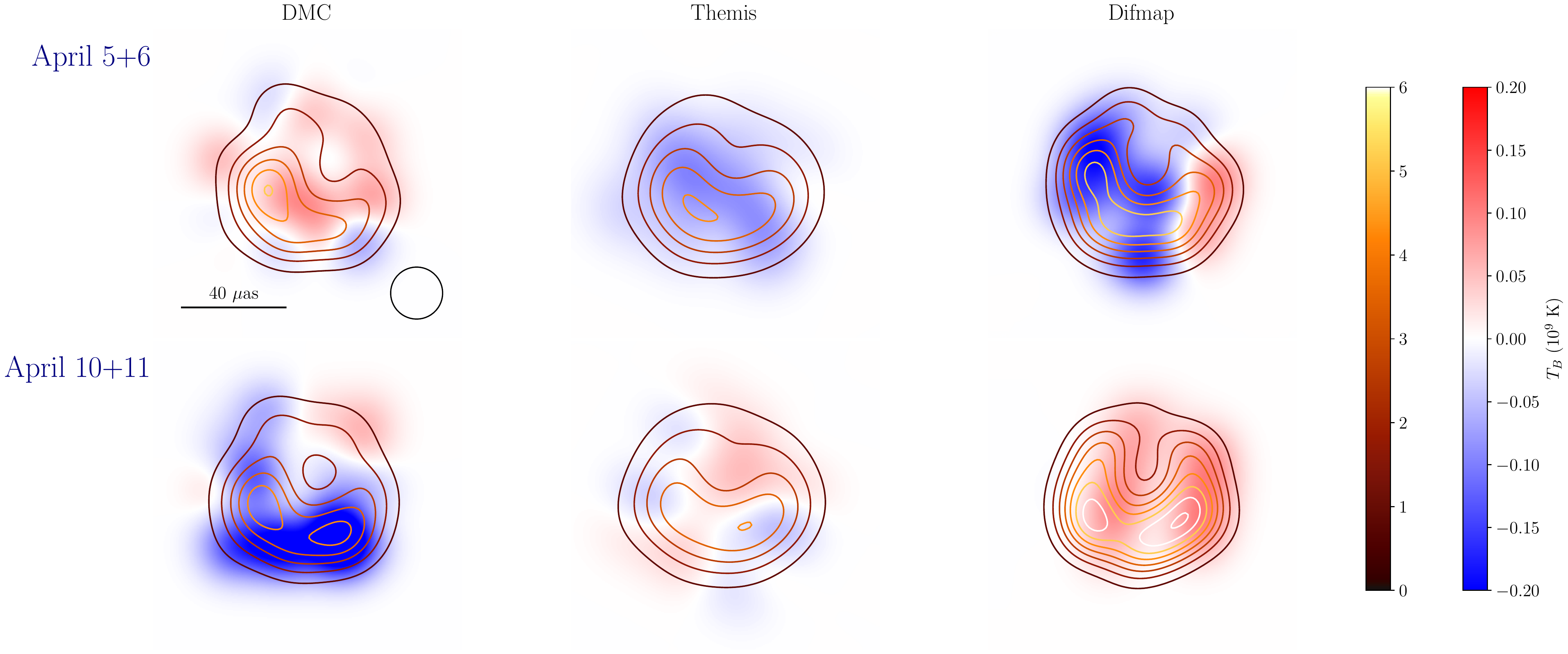} \\
\caption{\m87 imaging results combining days and bands. The top row shows results from three methods on a data set combining April 5 and 6 observations, both low and high band. The bottom row shows corresponding results from combining April 10 and 11 observations, low and high band. For \themis and \dmc we show the posterior-average image. Images are plotted in the same manner as in \autoref{fig:images_56}.}
\label{fig:images_multiday}
\end{figure*}

\subsection{Imaging results on individual days and bands}

Each imaging method introduced in \autoref{sec:methods} was used to produce Stokes $\mathcal{I}$,  $\mathcal{Q}$,  $\mathcal{U}$, and $\mathcal{V}$ images from the eight individual 2017 EHT data sets of \m87, corresponding to the four observation days in both low and high band. Imaging methods were free to use data that were pre-calibrated for the zero-baseline $D$-terms derived in \citetalias{PaperVII} or not, and to solve for residual $G_{R/L}$ gain errors or assume they are fixed to unity in self-calibration. The choices made by each imaging method are summarized in \autoref{tab:method_summary}.
Of the images presented here, the \dmc and \themis posteriors for the \m87 data are identical to those already presented for linear polarization in \citetalias{PaperVII}. The \polsolve results were generated with a similar procedure as the linear polarization results in \citetalias{PaperVII}, but with an additional series of imaging and self-calibration for recovering CP and the $G_{R/L}$ offsets. The \difmap and $m$-ring modeling results are new to this work.

\autoref{fig:images_all_3601} shows the April 11 low band reconstructions in both total intensity and linear polarization in the top row (in the style of \citetalias{PaperVII}) and in an ``ellipse-plot'' representation of the total linear plus circular polarization in the bottom row. The ``ellipse-plots'' illustrate the degree of linear polarization relative to circular polarization by the eccentricity of small ellipses plotted across the image.
As in \citetalias{PaperVII}, all imaging methods recover consistent images of total intensity and linear polarization. The $\approx40\,\mu$as diameter ring structure is recovered in all methods, as is the $\approx 15\,\%$ peak linear polarization fraction and predominantly azimuthal EVPA pattern in the south-west part of the image. Minor differences between images in the fractional linear polarization appear at the edges of the ring and in the $m$-ring pattern, which is limited by a small number of degrees of freedom in the $m=3$ mode fit in linear polarization. The ellipse plots in the bottom row show that in all cases the circularly polarized brightness recovered is a small fraction ($\lesssim20\,\%$) of the total polarized brightness, which is indicated by the large axis ratio/eccentricity of each ellipse. The colors of each ellipse show the sign of circular polarization recovered at each point. For April 11 low band there is a consistent negative sign of $\mathcal{V}$ at the total intensity maximum in the south-west, but across the rest of the image there is little consistency between methods in the sign of $\mathcal{V}$. 

In \autoref{fig:images_56} and \autoref{fig:images_1011} we focus on the results of our EHT 2017 \m87 circular polarization imaging and modeling results by showing circular polarization images within contours indicating the $\mathcal{I}$ brightness. In the color map chosen, red corresponds to a positive sign of CP while blue corresponds to negative CP. 
As in the synthetic data tests, different methods show consistent structure in total intensity, but significantly different structure in circular polarization. 
The imaging results are most consistent for low-band data, where on most days most methods recover negative circular polarization at and near the total intensity maximum.
In general, however, the Stokes $\mathcal{V}$ structures across the image are not consistent from method-to-method. Furthermore, the $\mathcal{V}$ images are not consistent between bands when imaged with the same method. An exception is $m$-ring modeling, which consistently indicates an approximately North-South asymmetry, with more negative circular polarization in the South. These results must be interpreted carefully, however, as strong assumptions on the source structure are imposed in the choice of a simple $m=1$ model for fitting the circular polarization. The $m$-ring results are discussed in more detail in \citetalias{Roelofs_2023}.

To test whether the observed inconsistency in the recovered Stokes $\mathcal{V}$ images is mitigated by a different calibration strategy, we produced images with a subset of methods using data on April 5 and April 11 reduced with the CASA-based \texttt{rPICARD} pipeline \citep{Janssen_2019}. We made use of a new \texttt{rPICARD} calibration mode, where instrumental phase and delay offsets are solved initially to align the the $RR^*$, $RL^*$, $LR^*$, and $LL^*$ phases of the high and low bands. Subsequently, all fringe-fitting and phase calibration solutions are obtained from the combined data of the four correlation products and the two frequency bands. Additionally, unlike the data reduced with the \texttt{EHT-HOPS} pipeline used for \autoref{fig:images_56} and \autoref{fig:images_1011}, where a multi-day and multi-source fit is employed, the CASA data utilized here is calibrated with $G_{R/L}$ assuming $\mathcal{V}=0$. The station-based amplitude gains are solved every few minutes and do not affect closures. We note that the R-L gain calibration strategies employed for the HOPS/CASA data will likely underfit/overfit instrumental gain offsets. The images produced form the CASA data still show inconsistency across imaging methods and observing days, and do not change our main conclusions based on the \texttt{EHT-HOPS} images, which we adopt as fiducial for the rest of the paper. 
We show full results of this test in \autoref{append:casa}.

The general inconsistency among methods and across bands is in sharp contrast to the total intensity images presented in \citetalias{PaperIV} and linear polarization images from \citetalias{PaperVII}. This inconsistency is a result of the severe difficulties in recovering resolved circular polarization from sparse 2017 EHT observations with low S/N and $G_{R/L}$ calibration uncertainties. 
\subsection{Combining days and bands}
\label{sec:combined}

Given the lack of consistency in the Stokes $\mathcal{V}$ reconstructions presented in \autoref{fig:images_56} and \autoref{fig:images_1011} for individual EHT data sets, it is natural to wonder if by averaging data across frequency bands and in time we may increase S/N enough to more confidently recover circular polarization structure.
A subset of our imaging methods tested this hypothesis.
\autoref{fig:images_multiday} presents imaging results from \dmc, \themis, and \difmap on datasets combining high and low band EHT observations of \m87 on two pairs of days: April 5 and 6 2017 (top row) and April 10 and 11 (bottom row). The source structure in \m87 evolved slightly over the week of observations in 2017, but was stable on each pair of days combined here (\citetalias{PaperIV},\citetalias{PaperVII}). 
As in \autoref{fig:images_56} and \autoref{fig:images_1011}, we find no consistency in the reconstructions from the band- and day-averaged data. While the S/N is improved by a factor of two by averaging four data sets in each combined reconstruction, the low intrinsic circular polarization and residual $G_{R/L}$ uncertainty are still too large and result in inconsistent image reconstructions from the combined data. We focus on results derived from the individual day/band images in the rest of this work.

\subsection{Upper limit on the resolved circular polarization fraction}
\label{sec:upperlimit}

We quantify the average circular polarization fraction in resolved EHT images by the image-averaged fractional circular polarization magnitude $|\mathcal{V}/\mathcal{I}|$, weighted by the Stokes $\mathcal{I}$ intensity:
\begin{equation}
    \langle|v|\rangle = \frac{\int |\mathcal{V}/\mathcal{I}|\, \mathcal{I} \, dA}{\int \mathcal{I}\,dA},
    \label{eq:vavg}
\end{equation}
where the integral is over the whole area of the image $\mathcal{A}$. The definition of $\langle|v|\rangle$ in \autoref{eq:vavg} is in close analogy with the average  resolved linear polarization fraction $\langle|m|\rangle$ used in \citetalias{PaperVII} and \citetalias{PaperVIII}. 

We contrast this average polarization fraction on EHT scales with the unresolved circular polarization fraction $v_{\rm net}$
\begin{equation}
    v_{\rm net} = \frac{\int \mathcal{V} \, dA}{\int \mathcal{I}\,dA} .
    \label{eq:vnet}
\end{equation}
By definition, $\langle|v|\rangle$ depends on the image resolution, while $v_{\rm net}$ is invariant to convolution with a blurring kernel. The definition of $\langle|v|\rangle$ naturally down-weights contributions from regions where the total intensity image $\mathcal{I}$ is dim and noisy. However, $\langle|v|\rangle$ is by definition constrained to be positive and thus cannot indicate the predominant sign of circular polarization in an image. For weakly polarized/noisy images, $\langle|v|\rangle$ will also be biased to non-zero values  (see \autoref{app:CirPolFracBiases} for a discussion).

We chose to use $\langle|v|\rangle$ instead of alternative metrics like $\mathcal{V}/\mathcal{I}$ at the peak brightness location because our image reconstructions and GRMHD synthetic images often do not show a single peak in circular polarized intensity. Furthermore, image-integrated metrics are preferable for computing sensible posterior means and error bars from the results of Bayesian methods like \dmc, \themis, and $m$-ring model fitting.

\begin{figure*}[t!]
\centering
\includegraphics[width=\textwidth]{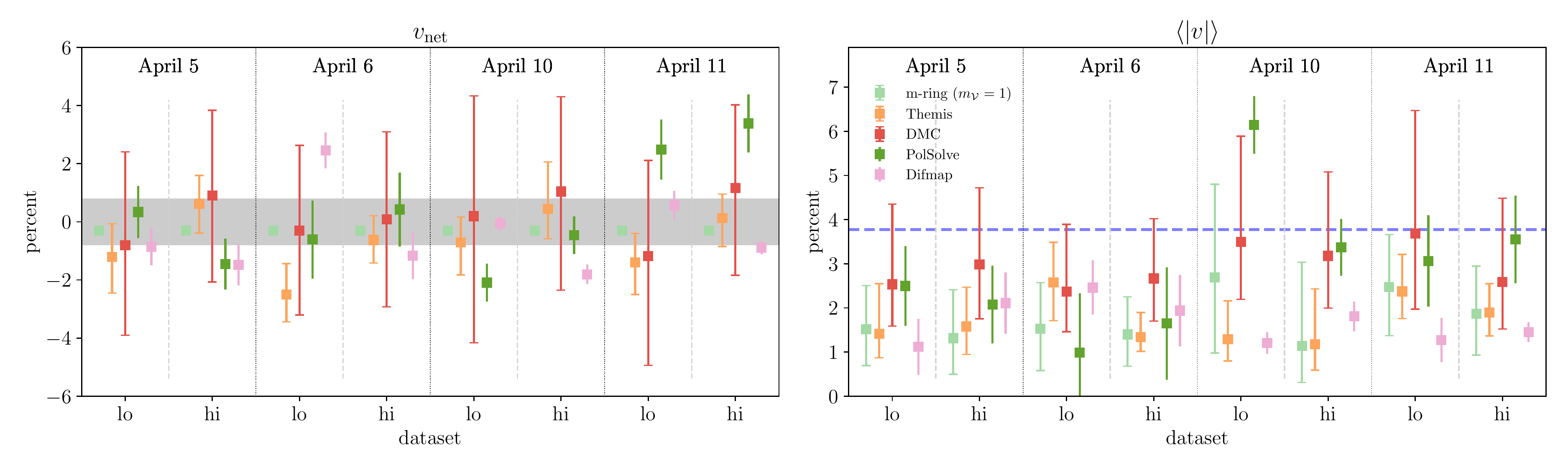} 
\caption{Image-integrated statistics from \m87 images. The left panel shows the net circular polarization fraction $v_{\rm net}$ computed from each method for the eight EHT \m87 datasets and the right panel shows the average resolved circular polarization fraction $\langle|v|\rangle$ computed after blurring each image with a $20\,\mu$as kernel. The results from the posterior exploration methods are presented with the median value and $2\sigma$ error bars (Note that $m$-ring modeling strictly enforces $|v_{\rm net}|=0$). The \difmap and \polsolve results are derived from the single fitted image and standard error propagation from measurements of the off-source image RMS in $\mathcal{V}$ and $\mathcal{I}$. In the left panel, the limits on $v_{\rm net}$ from ALMA observations reported in \citet{Goddi2021} (used to constrain GRMHD models in \citetalias{PaperVIII}) are indicated in the gray shaded region. In the right panel, the upper limit on $\langle|v|\rangle$ derived in this work ($\langle|v|\rangle < 3.7\,\%$, \autoref{tab:method_upperlimits}) is indicated by the dashed horizontal line.}
\label{fig:stats_realdata}
\end{figure*}

\autoref{fig:stats_realdata} shows the image-integrated and resolved circular polarization fractions for all reconstruction methods, days, and bands of EHT 2017 \m87 data. For the posterior exploration methods (\dmc, \themis, and $m$-ring modeling), we plot the posterior median of each quantity and error bars corresponding to $2.27$th and $97.7$th percentile of the posterior distribution (corresponding to $2\sigma$ error bars if the posterior were Gaussian). For \difmap and \polsolve, we plot the single value corresponding to the image results in \autoref{fig:images_56} and \autoref{fig:images_1011}; we derive approximate $2\sigma$ error bars for these methods based on the measured off-source residuals in $\mathcal{V}$ and $\mathcal{I}$ using standard Gaussian error propagation.
\dmc has particularly large error bars on fractional circular polarization because of its permissive priors on the relative  gain ratios $G_{R/L}$.

The values of the integrated circular polarization fraction $v_{\rm net}$ in \autoref{fig:stats_realdata} recovered by each imaging method 
are typically within the upper limits reported by \citet{Goddi2021} from ALMA observations, though some methods produce anomalously larger integrated polarization fractions on certain data sets (e.g. \polsolve on both April~11 data sets, \difmap and \themis on April~6 low band data). 
Our methods do not recover a consistent sign of the integrated circular polarization fraction $v_{\rm net}$.

The values of the resolved circular polarization fraction at $20\,\mu$as resolution $\langle|v|\rangle$ for most reconstructed images in the right panel of \autoref{fig:stats_realdata} are typically less than $4\,\%$. The synthetic data results in \autoref{fig:stats_synthetic} indicate that for 2017 EHT coverage and sensitivity most methods will always produce $\langle|v|\rangle\gtrsim 1\,\%$, even when the actual value is lower, as a consequence of uncertainties in the reconstruction in the low S/N regime, uncertainty in the residual gain ratios $G_{R/L}$, and an upward bias on the quantity $\langle|v|\rangle$ (\autoref{app:CirPolFracBiases}). As a result, and because of the lack of agreement among methods in recovering a consistent source structure in $\mathcal{V}$ in \autoref{fig:images_56} and \autoref{fig:images_1011}, we use these results only to obtain a conservative upper limit for the resolved circular polarization fraction.

We estimate a combined upper limit on $\langle|v|\rangle$ in \m87 from each method's eight measurements across the four days and two observing bands. We average the measurements of $\langle|v|\rangle$ within each method across these eight data sets using standard inverse-variance weighting, then compute the 99th percentile of the resulting distribution (assumed to be Gaussian). These $99\,\%$ upper limits for each method are reported in \autoref{tab:method_upperlimits}. 

We adopt the \dmc results as the most conservative upper limit on $\langle|v|\rangle$. The upper limit computed from \polsolve is slightly higher than that from \dmc ($3.8\,\%$ vs $3.7\,\%$). 
Nonetheless, we adopt the \dmc  value as our fiducial upper limit in this work. \dmc performs full Bayesian posterior exploration over the image intensity values and station gains. In contrast, \polsolve computes only a single image with error bars computed using idealized Gaussian error propagation.  
None of our interpretation in \autoref{sec:theo} is affected by choosing $3.8\,\%$ vs $3.7\,\%$ as our fiducial upper limit. For these reasons, 
we adopt the \dmc value $\langle|v|\rangle<3.7\,\%$ going forward as our conservative upper limit on the average resolved horizon-scale circular polarization fraction in \m87 at 230\,GHz. 

\begin{deluxetable}{l|c}[h]
    \tablecaption{99th Percentile $\langle|v|\rangle$ Upper Limits from Each Method\label{tab:method_upperlimits}}
    \tablehead{\colhead{Method} & \colhead{Combined upper limit on $\langle|v|\rangle$}}
    \startdata
    %\centering
        \dmc & $\mathbf{3.7 \%}$\\
        \themis & 2.0 \%\\
        \polsolve & 3.8 \% \\
        \difmap & 1.9 \%\\
        $m$-ring & 2.2 \%\\
   \enddata
    \tablecomments{We derive each upper limit using inverse-variance weighted averaging of the results in the right panel of \autoref{fig:stats_realdata}. We adopt the  conservative upper limit, $\langle|v|\rangle<3.7\,\%$ from \dmc, as our fiducial value. Note that in computing the upper limit from \polsolve, we exclude the anomalously highly polarized April 10 low band result, as it suffers from an anomalously large overall offset in the recovered $G_{R/L}$.}
\end{deluxetable}

% \begin{table}[h]
%     \centering
%     \begin{tabular}{@{}l|c}
%         \toprule
%         Method & Combined upper limit on $\langle|v|\rangle$ \\
%         \midrule
%         \dmc & $\mathbf{3.7 \%}$\\
%         \themis & 2.0 \%\\
%         \polsolve & 3.8 \% \\
%         \difmap & 1.9 \%\\
%         $m$-ring & 2.2 \%\\
%         \bottomrule
%     \end{tabular}
%     \caption{99th percentile $\langle|v|\rangle$ upper limits from each method, after combining results from 8 data sets covering 4 days and bands using inverse-variance weighted averaging of the results in the right panel of \autoref{fig:stats_realdata}. We adopt the  conservative upper limit, $\langle|v|\rangle<3.7\,\%$ from \dmc, as our fiducial value. Note that in computing the upper limit from \polsolve, we exclude the anomalously highly polarized April 10 low band result, as it suffers from an anomalously large overall offset in the recovered $G_{R/L}$. }
%     \label{tab:method_upperlimits}
% \end{table}

%=================================================================================================================================================================
\section{Theoretical Interpretation}
\label{sec:theo}
%=================================================================================================================================================================

On event horizon scales, circularly polarized images of hot accretion flows can encode valuable information about the plasma, including its magnetic field geometry and composition.  While our imaging methods are unable to unambiguously determine the horizon-scale structure of circular polarization, in \autoref{sec:m87results}
we establish an upper limit on the magnitude of the fractional circular polarization magnitude on scales of the EHT beam: $\langle|v|\rangle<3.7\%$. This upper limit, combined with the existing limit on the unresolved, source-integrated circular polarization fraction from ALMA ($|v_{\rm net}|<0.8\%$), can be used to constrain models of the emitting plasma and accretion flow around \m87, \emph{even without} additional information on the structure of the circularly polarized emission.

Recent work has revealed that intrinsically circularly polarized synchrotron emission, Faraday conversion and rotation, and twisted field geometries are all important in generating the CP image that we observe in millimeter wavelengths \citep{Tsunetoe+2020,Moscibrodzka+2021,Ricarte+2021,Tsunetoe+2021,Tsunetoe+2022}. 
For the plasma parameters of interest in \m87, millimeter synchrotron emission is intrinsically circularly polarized only at the $\sim$1\,\% level.
However, Faraday conversion also plays a role in exchanging linear and circular polarization states, and in fact dominates the Stokes~$\mathcal{V}$ production in many models.  Stokes~$\mathcal{V}$ generated by Faraday conversion is understood to be sensitive to the magnetic field geometry, which has been utilized to infer helical field structure in jets  \citep[e.g.,][]{Wardle&Homan2003,Gabuzda+2008}.  On event horizon scales, the connection between Stokes $\mathcal{V}$ and the magnetic field geometry is complicated by the geometries probed by geodesics that take non-trivial paths through the space-time, leading in particular to Stokes $\mathcal{V}$ sign flips of successive sub-images in some models \citep{Moscibrodzka+2021,Ricarte+2021}.

As with the previous papers in this series, our theoretical interpretation centers on ray-traced GRMHD simulations, which self-consistently generate the plasma that performs emission, absorption, and Faraday effects.  Since images of total intensity and linear polarization were studied in detail in \citetalias{PaperV} and \citetalias{PaperVIII} respectively, here we focus mainly on the astrophysics governing the generation of circular polarization.

\begin{figure*}[t!]
    \centering
    \includegraphics[width=\textwidth]{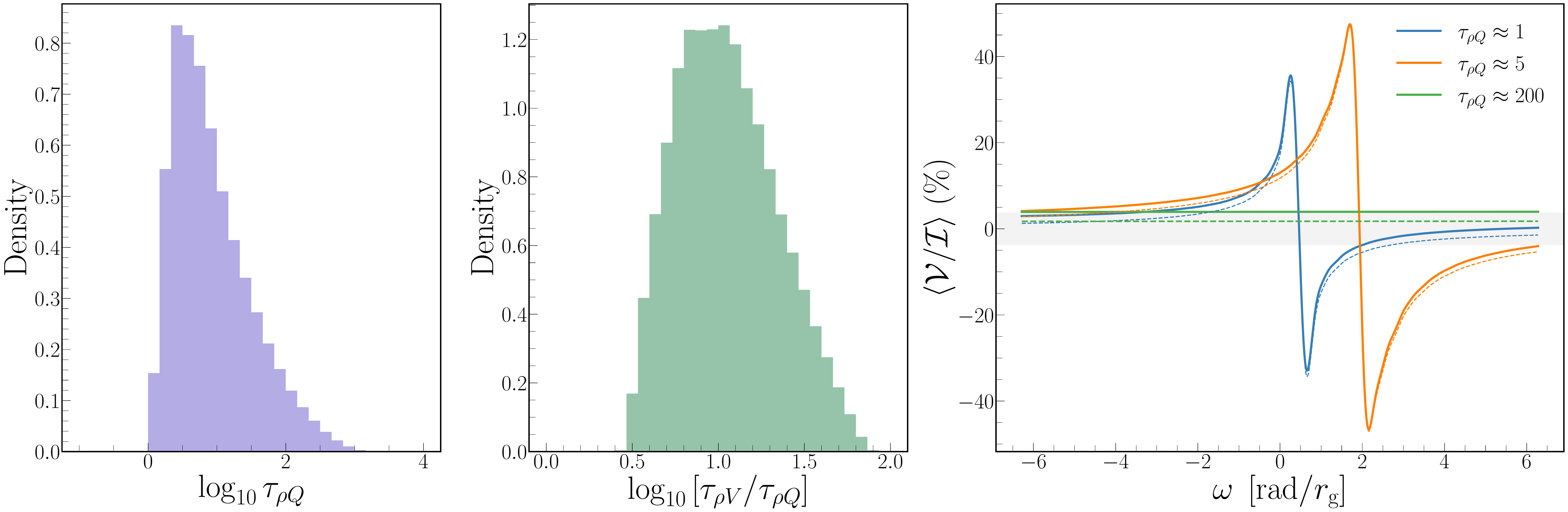}
    \caption{Circular polarization properties of passing \citetalias{PaperVIII} one-zone models. (Left) Distribution of the Faraday conversion optical depth $\tau_{\rho Q}$ in passing models. In all passing models $\tau_{\rho Q}>1$, indicating that most circular polarization is likely produced by Faraday conversion. (Center) Distribution of the ratio of the Faraday rotation to Faraday conversion optical depths. In all cases,  $\tau_{\rho V} > \tau_{\rho Q}$, indicating that with a constant field orientation in the emission region, circular polarization will dominate over linear polarization in these models. (Right) The average fractional circular polarization between $5\,r_{\rm g}$ and $10\,r_{\rm g}$ in one-zone models with a rotating magnetic field direction along the line of sight, as a function of the angular rotation frequency $\omega$. We show three different models: a model with low Faraday conversion depth (blue), a model with median conversion depth (orange), and a model with high conversion depth (green). Dashed lines show corresponding results for one-zone models with no intrinsic emission of circular polarization, $j_\mathcal{V}=0$.}
    \label{fig:onezone}
\end{figure*}

\begin{figure*}[ht!]
    \centering
    \includegraphics[width=\textwidth]{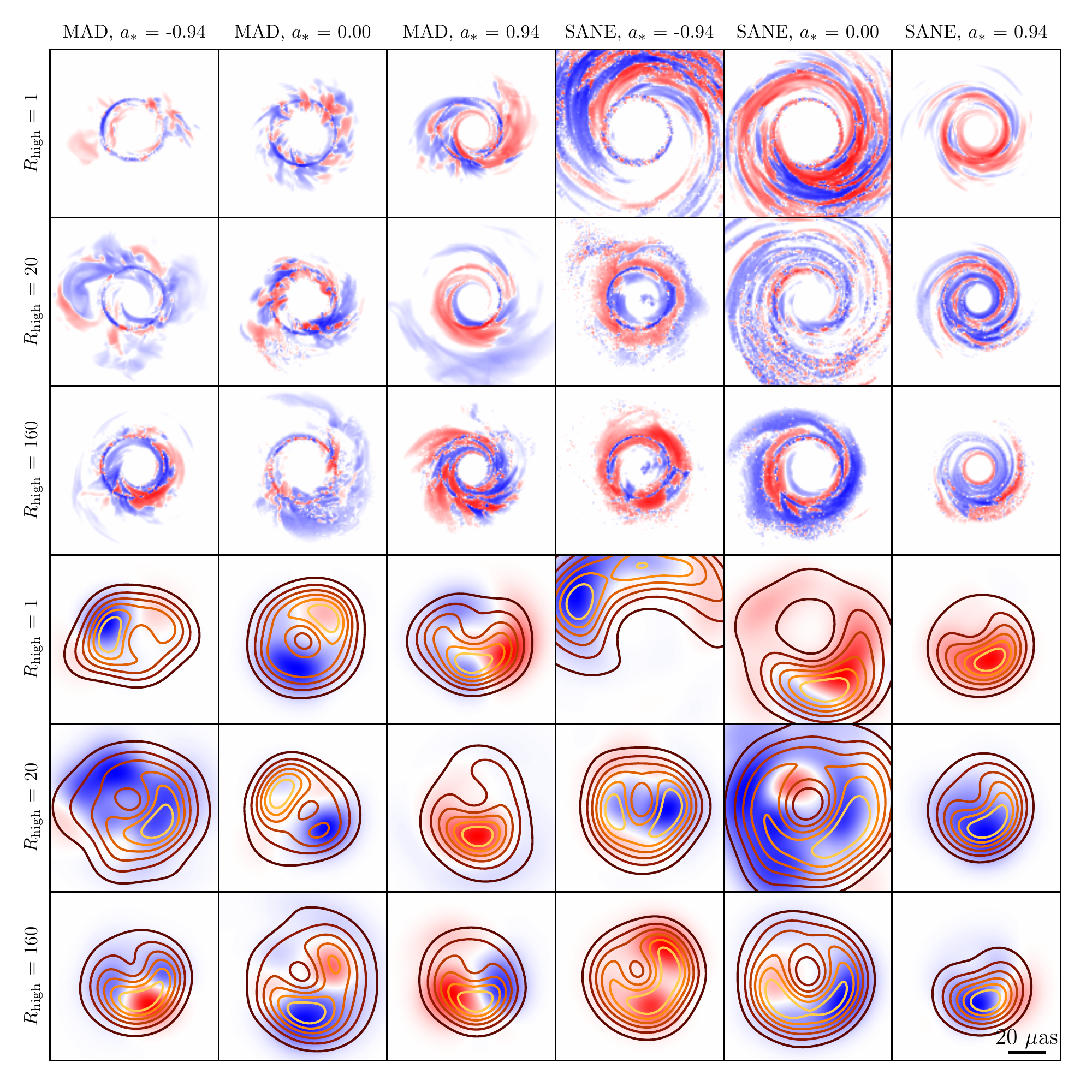}
    \caption{A random selection of representative snapshots from our GRMHD image library.  The color scales for each snapshot are normalized individually.  The first three rows are presented at native resolution in symmetric logarithmic scale with three decades in dynamic range shown to better visualize faint features.  The bottom three rows plot Stokes $\mathcal{I}$ in contours and Stokes $\mathcal{V}$ in color after blurring with a 20\,$\mu$as FWHM Gaussian, both in linear scale.  Models exhibit a wide variety of morphologies and almost always show sign reversals, both at perfect resolution and EHT resolution.}
    \label{fig:grmhd_gallery}
\end{figure*}

\subsection{One-zone models}
\label{sec:onezone}
In \citetalias{PaperV} and \citetalias{PaperVIII}, we used one-zone isothermal sphere models to derive order-of-magnitude estimates on the parameters of the synchrotron emitting plasma in M87*. In particular, applying constraints on the linear polarization, \citetalias{PaperVIII} 
found that one-zone models for \m87 imply
the dimensionless electron temperature $\Theta_e = k_B T_e / m_e c^2$ lies in a mildly relativistic regime, $2<\Theta_e<20$,
the magnetic field is in the range $1\lesssim |B| \lesssim 20$\,G, and the number density lies in the range $10^{4}\lesssim n_e \lesssim 10^{7} $\,cm$^{-3}$. 
While one-zone models neglect the critical effects of fluid velocity, gravitational redshift, and a non-uniform emitting region, they favor plasma parameters that are in general agreement with those found in favored GRMHD simulations \citepalias{PaperVIII}. 

Here, we explore some implications for circular polarization in the one-zone model space selected in \citetalias{PaperVIII}. We use the same model of a uniform $R=5r_{\rm g}$ isothermal sphere\footnote{For \m87, adopting $M_{\rm BH}=6.5\times10^9\,M_\odot$, the gravitational radius $r_{\rm g}=GM_{\rm BH}/c^2=9.6\times10^{14}$\,cm $=64$\,AU.} with a magnetic field at a pitch angle $\theta_B=\pi/3$ to the line-of-sight. We determine polarized emissivities $j_\mathcal{I},j_\mathcal{Q},j_\mathcal{V}$, absorption coefficients $\alpha_\mathcal{I},\alpha_\mathcal{Q},\alpha_\mathcal{U}$ and Faraday rotation/conversion coefficients $\rho_\mathcal{V},\rho_\mathcal{Q}$ using the fitting functions for relativistic thermal electron distributions in \citet{Dexter_2016}. 

First, we consider the importance of Faraday conversion in producing the observed circular polarization in these models. The left panel of \autoref{fig:onezone} shows the distribution of the Faraday conversion optical depth $\tau_{\rho_\mathcal{Q}}=2R\,\rho_\mathcal{Q}$ in one-zone models that pass the \citetalias{PaperVIII} constraints. In all cases $\tau_{\rho_\mathcal{Q}} > 1$, indicating that Faraday conversion dominates intrinsic emission in producing circular polarization. 

In the center panel of \autoref{fig:onezone}, we consider the ratio of the Faraday rotation to conversion optical depth, $\tau_{\rho_\mathcal{V}}/\tau_{\rho_\mathcal{Q}}$. In all passing models, this ratio is greater than unity, as we expect for a mildly relativistic plasma. As a consequence, these one-zone models typically produce \emph{more} circular polarization than linear polarization. This is because in one-zone models where Faraday effects are significant ($\tau_{\rho_\mathcal{Q}} > 1$, $\tau_{\rho_\mathcal{V}}>1$) but absorption is insignificant ($\tau_\mathcal{I}<1$), the ratio of circular to linear polarized intensity in the limit of large Faraday depth approaches the ratio of the rotativities: $|\mathcal{V}|/|\mathcal{P}|\rightarrow |\rho_\mathcal{V}| / |\rho_\mathcal{Q}|$ \citep[see Appendix C of][for exact one-zone solutions in this limit]{Dexter_2016}.
As a result, our 
one-zone models with uniform magnetic field orientations over-produce circular polarization relative to linear polarization, and we must invoke effects not considered in the one-zone models to explain the observed $|\mathcal{V}|<|\mathcal{P}|$.

We consider the effects of one such complication -- a spatially twisted magnetic field -- in the right panel of \autoref{fig:onezone}. Faraday conversion only converts linear polarization in the $\mathcal{U}$ Stokes parameter into $\mathcal{V}$, so a spatial rotation of the projected $B$-field can work with or against Faraday rotation to enhance or suppress the produced circular polarization. 
We add a constant rotation of the magnetic field direction at a spatial frequency $\omega$  to our one-zone models, and we solve for the final circular polarization fraction  $\mathcal{V}/\mathcal{I}$ as a function of $\omega$
in three models: one with a low conversion depth $(\tau_{\rho_\mathcal{Q}}\approx 1: |B|=11 \,\mathrm{G}, \Theta_e=7, n_e=10^4\,\mathrm{cm}^{-3})$; one with a median conversion depth $(\tau_{\rho_\mathcal{Q}}\approx 5: |B|=5 \,\mathrm{G}, \Theta_e=8, n_e=10^5\,\mathrm{cm}^{-3})$; and one with a large conversion depth $(\tau_{\rho_\mathcal{Q}}\approx 200: |B|=9 \,\mathrm{G}, \Theta_e=3, n_e=4\times10^6\,\mathrm{cm}^{-3})$.

For moderate rates of field rotation in the plane of the sky, only the model with the highest Faraday depth produces a constant circular polarization fraction at the values of $\omega$ considered. The other models produce $\mathcal{V}/\mathcal{I}$ that varies rapidly and changes sign as the rotation rate $\omega$ passes through the critical frequency $\omega_{\rm crit} = \rho_\mathcal{V}/2$. Thus, field twist through an inhomogenous emission region can have a significant impact on both the magnitude and sign of the observed circular polarization \citep[e.g.][]{Ricarte+2021, Moscibrodzka+2021}. We also show results for the same models artificially setting the intrinsic circular polarization emission to zero ($j_\mathcal{V}=0$, dashed lines). In all three cases, this change has a minor effect on the produced circular polarization fraction, again confirming that Faraday conversion dominates the production of circular polarization for plasma parameters appropriate for \m87. 

The fact that one-zone models overproduce $\mathcal{V}$ relative to linear polarization and the fact that a changing field geometry in the emission region has significant impact on $\mathcal{V}/\mathcal{I}$ motivates consideration of more complex models \citep[see, e.g.,][for a two-zone model of \m87's Faraday rotation]{Goddi2021}. We proceed next to consider circular polarization in GRMHD simulation images, which self-consistently include the effects of an inhomogenous emission region, field twist, and special and general relativistic redshift and parallel transport.

\subsection{GRMHD image libraries}
\label{sec:libraries}

\label{sec:lib}
The 3D GRMHD simulations used in this work were first considered in \citetalias{PaperV} and \citetalias{PaperVIII}.  We use the code {\sc ipole} \citep{Noble2007,Moscibrodzka&Gammie2018} to perform general relativistic radiative transfer (GRRT) following the methodology outlined in \citetalias{PaperVIII} and \citet{Wong_2022}.  We briefly summarize our library generation here, and refer to \autoref{app:lib_details} and these previous works for more details.

All images assume a fixed observing frequency of $230\,\mathrm{GHz}$, BH mass of $M_\mathrm{BH}=6.2\times10^9 \,M_\odot$, and distance of $16.9\,\mathrm{Mpc}$.\footnote{
Note that the values of the \m87 BH mass and distance used in the GRMHD library ($M_\mathrm{BH}=6.2\times10^9 \,M_\odot$,$D=16.9\,\mathrm{Mpc}$) following \citetalias{PaperV,PaperVIII}) are slightly different than the EHT's measured value from the ring diameter ($M_\mathrm{BH}=6.5\times10^9 \,M_\odot$, $D=16.8\,\mathrm{Mpc}$, \citetalias{PaperVI}) which we adopt in other Sections of this paper. Because we do not rely on the image size as a constraint on our models, we do not expect the 5\% difference in mass and 0.5\% difference in distance between the values adopted in our GRMHD simulations and \citetalias{PaperVI} measurements to affect our interpretation.}
The fluid density in the simulations is scaled to reproduce an average flux density of $F_\nu=0.5\,\mathrm{Jy}$ for each model, following the observed compact flux density in April 2017 \citepalias{PaperIV}. The observing inclination is tilted such that the approaching jet (parallel to the BH spin axis) is inclined at 17$^\circ$ with respect to our line of sight.  

Our simulation library probes 5 free parameters:  (i) the magnetic field state, either a strong field ``magnetically arrested disk'' (MAD) \citep{Bisnovatyi-Kogan&Ruzmaikin1974,Igumenshchev+2003,Narayan+2003} or weak field ``standard and normal evolution'' (SANE) \citep{Narayan+2012,Sadowski+2013}, (ii) the BH spin $a_* \in \{-0.94,-0.5,0,0.5,0.94\}$, where a negative sign denotes retrograde accretion, (iii-iv) $R_\mathrm{high} \in \{1,10,20,40,80,160\}$ and $R_\mathrm{low}\in \{1,10\}$ which modulate the ion-to-electron temperature ratio in different regions, 
and (v) the magnetic field polarity, which is either aligned or anti-aligned with respect to the disk angular momentum on large scales.  Each model is imaged at a cadence of 5\,$GM/c^3$ for a total duration ranging from 2500 to 5000\,$GM/c^3$ depending on the model.

In total, we compute 184796 image snapshots from ten GRMHD simulations. For each snapshot we compute a set of polarimetric observables to compare with the data and score our models.  \autoref{fig:grmhd_gallery} displays a random selection of snapshots visualized in circular polarization as a function of spin, magnetic field state, and $R_\mathrm{high}$.  All images have been rotated such that the oncoming jet is projected at a position angle of 288$^\circ$ east of north.  In the first three rows, we visualize the images in Stokes $\mathcal{V}$ at high resolution in symmetric logarithmic scale with three decades in dynamic range.  We find a wide variety of morphologies in near-horizon Stokes~$\mathcal{V}$ images, including sign flips in almost every snapshot.  Some images, such as the SANE $a_*=0$ models, exhibit ``noise-like'' regions with rapid sign flips among adjacent pixels.  This is equivalent to randomly oriented ticks in linear polarization: circular polarization in these regions is scrambled due to either large Faraday rotation\footnote{Although Faraday rotation does not affect circular polarization directly, it can scramble the EVPA of the linear polarization that is then converted into circular.} or Faraday conversion depths.  In the bottom three rows, we visualize these snapshots in linear scale with Stokes $\mathcal{I}$ contours, blurred with a Gaussian with a 20\,$\mu$as FWHM.  At EHT resolution, GRMHD simulations again predict a wide variety of morphologies including ubiquitous sign flips.

\subsection{The importance of magnetic field polarity}
\label{sec:polarity}

This is the first paper in this series to consider the magnetic field polarity as a free parameter.  We flip the magnetic field polarity in post-processing using the same GRMHD snapshots, since the equations of ideal GRMHD are invariant to a sign flip in the magnetic field vector.  The equations of polarized GRRT are \emph{not} invariant to sign of the magnetic field vector, however.  Reversing the polarity of the magnetic field reverses the sign of $\rho_\mathcal{V}$ (Faraday rotation) and $j_\mathcal{V}$ (circularly polarized emission), but it does not affect $\rho_\mathcal{Q}$ (Faraday conversion).  Consequently, flipping the magnetic field direction does not necessarily simply reverse the sign of circular polarization across the image.  

Throughout, we describe the magnetic field as ``aligned'' if its polarity is parallel to the angular momentum of the disk on large scales, or ``reversed'' if this polarity is anti-parallel.  Note that the magnetic field structure can be complicated and turbulent in the near-horizon emission region, especially in retrograde disks, so a magnetic field aligned with the angular momentum on large scales is not necessarily trivially aligned on event horizon scales.

\begin{figure}[h!]
    \centering
    \includegraphics[width=\linewidth]{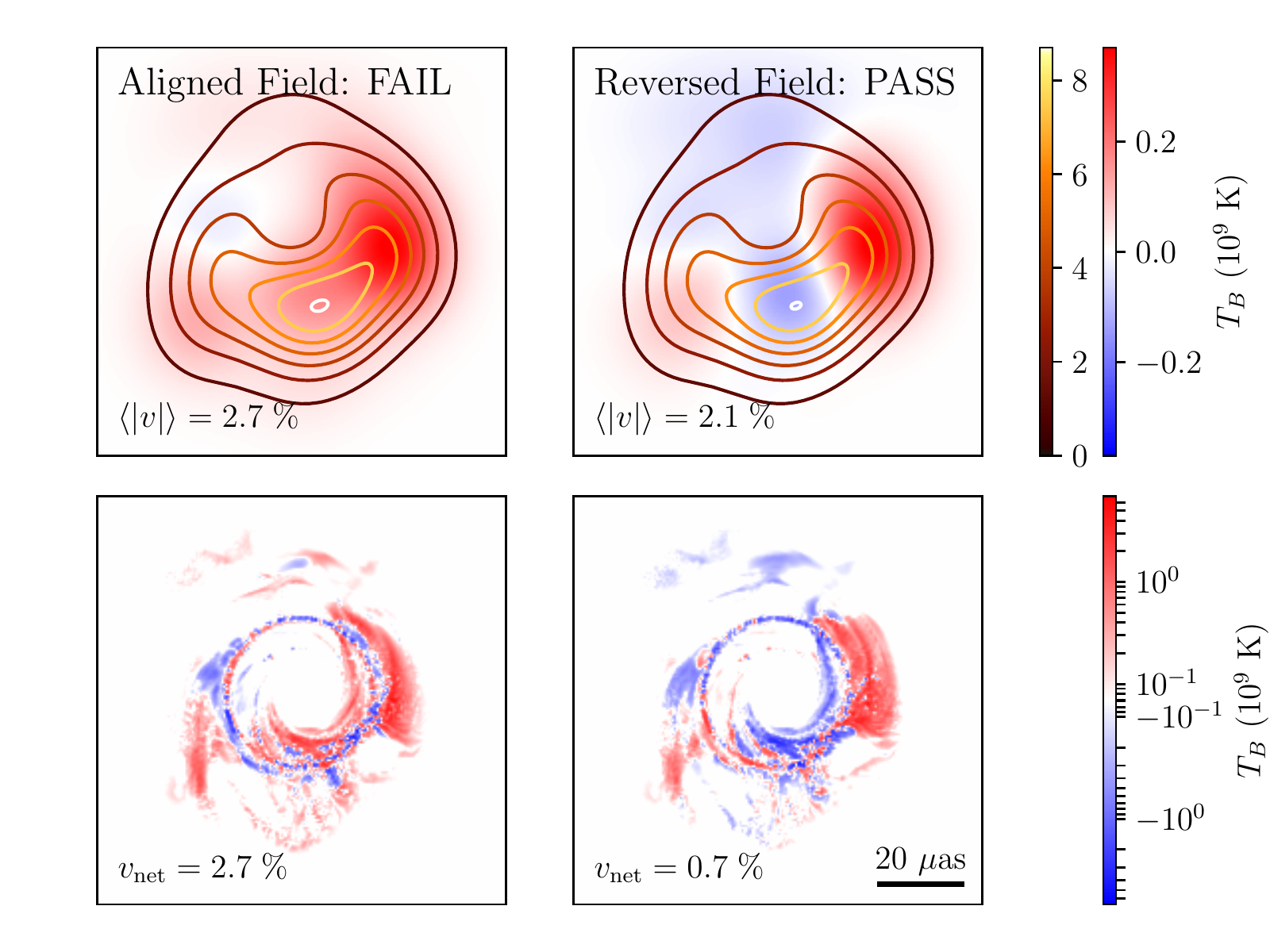}
    \caption{Example GRMHD snapshot (MAD $a_*=-0.94$ $R_\mathrm{low}=10$ $R_\mathrm{high}=160$) plotted with both magnetic field configurations, aligned (left) and reversed (right). The top panels show the images blurred to EHT resolution, and the bottom panels show the images at their native resolution. 
    As shown in the left panel, this snapshot fails simultaneous polarimetric constraints with the aligned field configuration, overproducing $v_\mathrm{net}$.  However, as shown in the right panel, flipping the magnetic field polarity produces some oppositely-signed regions that reduce $|v_\mathrm{net}|$.  Flipping the field has no effect on the total intensity image.}
    \label{fig:field_flip_example}
\end{figure}

In \autoref{fig:field_flip_example}, we visualize a snapshot of the MAD $a_*=-0.94$ $R_\mathrm{low}=10$ $R_\mathrm{high}=160$ model with both magnetic field polarities.  The top row is presented in linear scale blurred with a 20\,$\mu$as FWHM Gaussian kernel, while the bottom row is presented in symmetric logarithmic scale with 2 decades of dynamic range in intensity.  With a polarity aligned with the angular momentum of the disk as in previous work, the unresolved circular polarization fraction $v_\mathrm{net}=2.7\,$\% vastly exceeds the upper limit of $|v_\mathrm{net}| < 0.8$\,\% from ALMA observations \citep{Goddi2021}.  However, flipping the magnetic field polarity reverses the sign of circular polarization in a significant portion of the image, reducing $v_{\rm net}$ to $0.7\%$ and allowing the model to pass.  In fact, this snapshot simultaneously passes {\it all} polarimetric constraints considered in this work with the reversed magnetic field configuration. 

\autoref{fig:field_flip_example} illustrates that it is not easy to predict which regions of a given image change upon a reversal of the magnetic field direction.  We expect that regions dominated by intrinsic synchrotron emission should trivially flip sign, while regions dominated by Faraday conversion may remain unchanged unless Faraday rotation is significant along those geodesics.  We further explore the effect of flipping the magnetic field polarity on \emph{linear} polarization in \autoref{app:field_reversal}.  While there are noticeable differences in the distribution of the $\angle \beta_2$ parameter \citep{PWP} across all GRMHD models, the effect is less dramatic for linear polarization metrics than it is for circular. 

\begin{figure*}[th]
    \centering
    \includegraphics[width=0.45\textwidth]{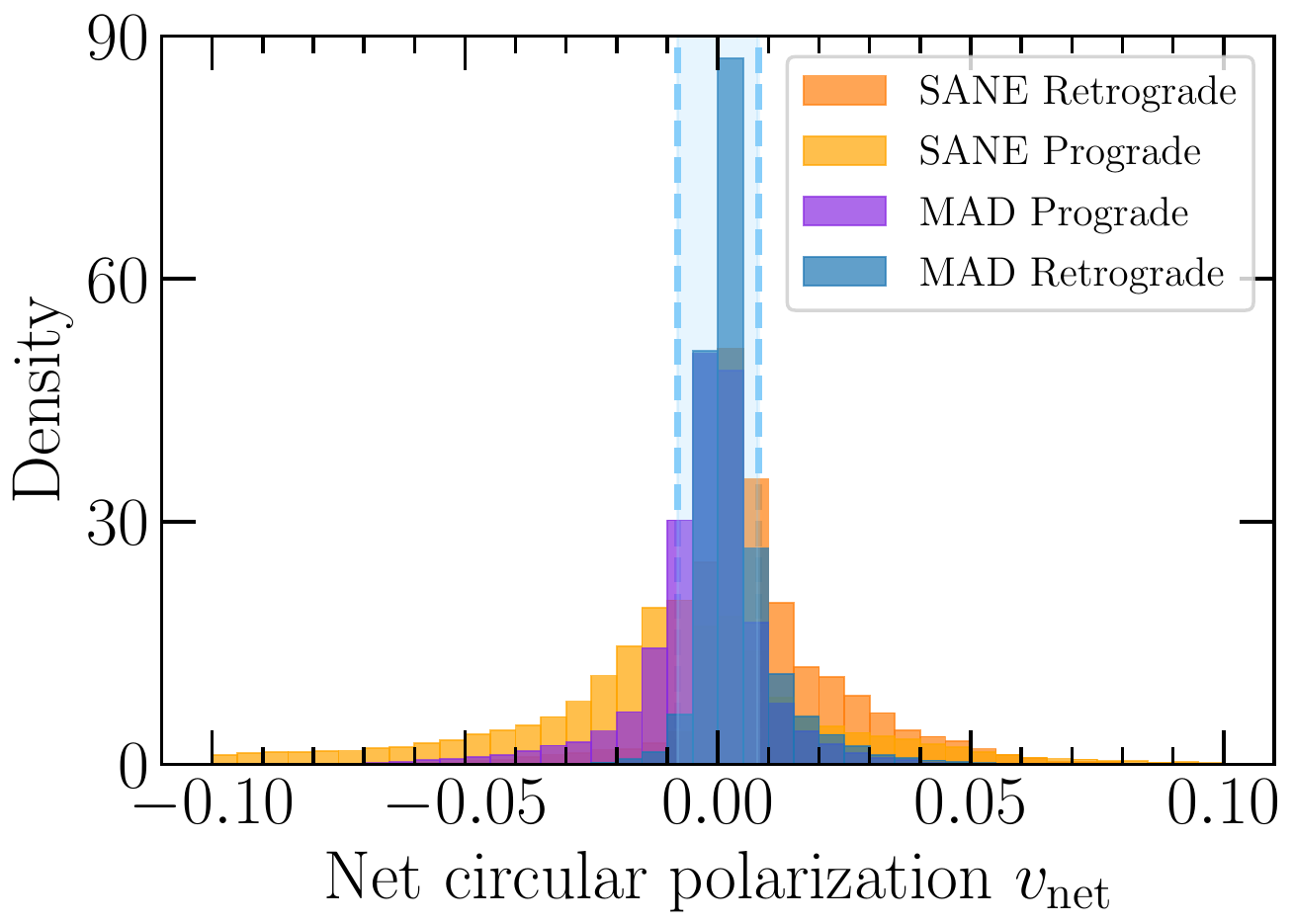} 
    \includegraphics[width=0.445\textwidth]{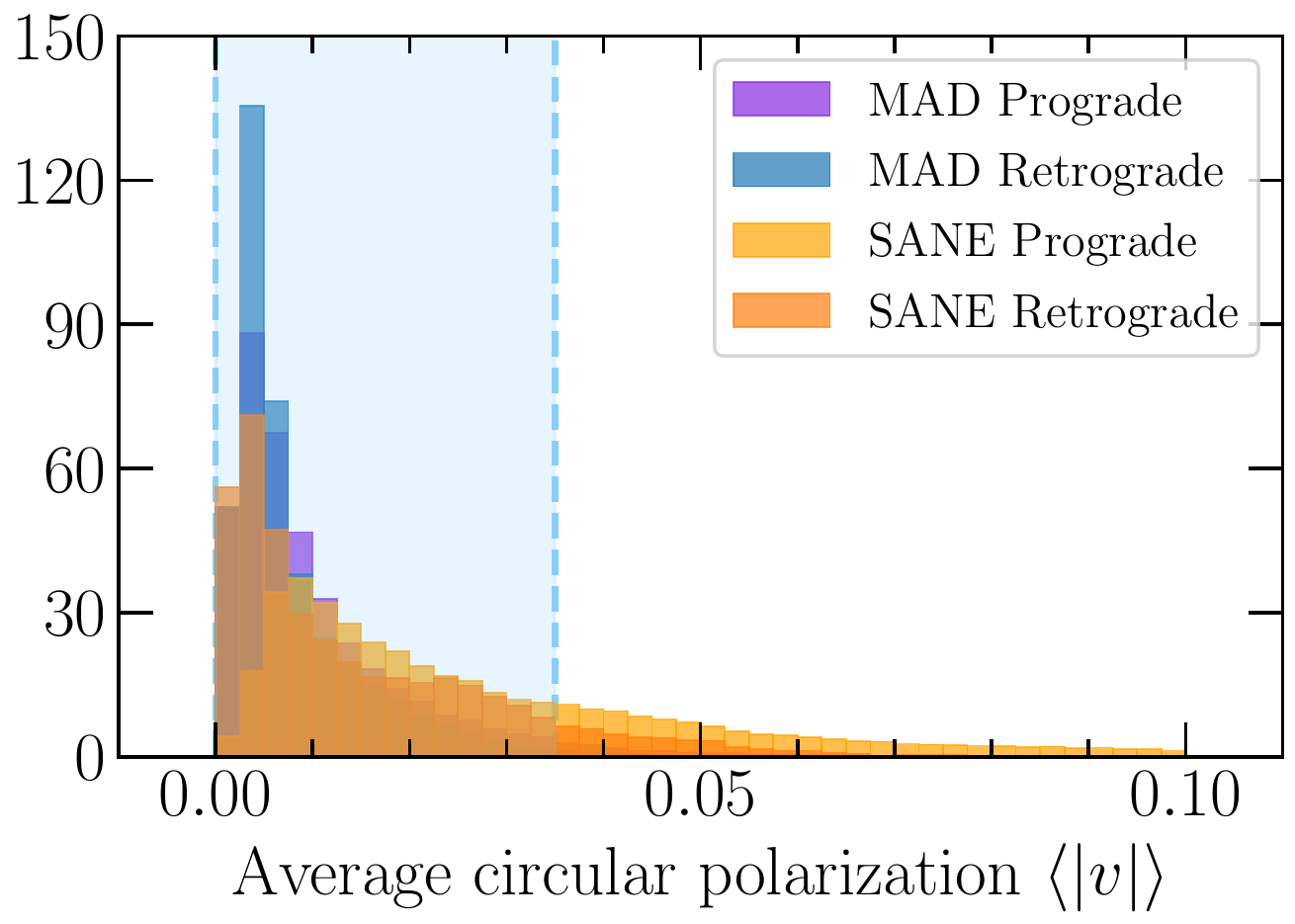} 
    \caption{Number density distribution of net circular polarization $v_\text{net}$ fraction (left panel) and image-averaged fractional circular polarization $\langle |v| \rangle$ (right panel) with an aligned and anti-aligned magnetic field respecting all images in the \m87 library. Allowed inferred ranges for ALMA-only data ($v_\text{net}$) and measured values of reconstructed polarimetric images of \m87 reconstructions ($\langle |v| \rangle$) are limited by the dashed lines.}
    \label{fig:cpnet}
\end{figure*}

\subsection{GRMHD simulation scoring}
\label{sec:constraints}

% \begin{table}[]
%     \centering
%     \begin{tabular*}{0.45\textwidth}{p{2.7cm}p{2.7cm}p{2.7cm}}\\
%     \\
%     \\
%     \hline
%     \hline
%     Parameter & Minimum & Maximum\\
%     \hline
%     $m_\text{net}$ & 1.0\,\% & 3.7\,\% \\
%     $v_\text{net}$ & -0.8\,\% & 0.8\,\% \\
%     $\langle |m| \rangle$ & 5.7\,\% & 10.7\,\% \\ %?
%     $|\beta_2|$ & 0.04 & 0.07 \\
%     $\angle \beta_2$ & -163 deg & -129 deg \\
%     \hline
%     $\langle |v| \rangle$ (This Work) & 0 & 3.7\,\% \\
%
%     \end{tabular*}
%     \caption{Observational constraints applied to our GRMHD image library.  Most of these constraints are inherited from \citetalias{PaperVII} and were previously used to constrain models in \citetalias{PaperVIII}.  This work adds the new upper limit on $\langle |v| \rangle$.}
%     \label{tab:constraints}
% \end{table}

\begin{deluxetable}{p{2.7cm}p{2.7cm}p{2.7cm}}[h]
    \tablecaption{Observational Constraints Applied to our GRMHD Image Library\label{tab:constraints}}
    \tablehead{\colhead{Parameter} & \colhead{Minimum} & \colhead{Maximum}}
    \tablewidth{0.45\textwidth}
    \startdata
    $m_\text{net}$ & 1.0\,\% & 3.7\,\% \\
    $v_\text{net}$ & -0.8\,\% & 0.8\,\% \\
    $\langle |m| \rangle$ & 5.7\,\% & 10.7\,\% \\ %?
    $|\beta_2|$ & 0.04 & 0.07 \\
    $\angle \beta_2$ & -163 deg & -129 deg \\
    \hline
    $\langle |v| \rangle$ (This Work) & 0 & 3.7\,\% \\
   \enddata
    \tablecomments{ Most of these constraints are inherited from \citetalias{PaperVII} and were previously used to constrain models in \citetalias{PaperVIII}.  This work adds the new upper limit on $\langle |v| \rangle$.}
\end{deluxetable}

In \citetalias{PaperVIII}, five observational metrics were used to score GRMHD models of \m87's accretion flow \textemdash (1) the unresolved linear polarization fraction $|m|_\mathrm{net}$; (2) the unresolved circular polarization fraction $|v|_\mathrm{net}$; (3) the image-average linear polarization fraction $\langle |m| \rangle$; (4) the amplitude $|\beta_2|$ and (5) the phase $\angle \beta_2$ of the second azimuthal Fourier coefficient of the linear polarization pattern
(see \citealt{PWP}, \citetalias{PaperVIII}). In this paper, we add one additional constraint to this set from our observational results in \autoref{sec:m87results}: an upper limit on the circular polarization fraction on EHT scales: $\langle |v| \rangle < 3.7\,\%$. We summarize the observational constraints used to score GRMHD models in \autoref{tab:constraints}.\footnote{
Note that in computing these image metrics from GRMHD simulation snapshots, we first blur the model images with a 20$\,\mu$as FWHM circular Gaussian beam.}

In \citetalias{PaperVIII}, we found that many GRMHD models which could satisfy both resolved and unresolved linear polarization constraints could also self-consistently satisfy the upper limit on the unresolved circular polarization, 
\mbox{$|v_\mathrm{net}|<0.8\,\%$}. 
In the left panel of \autoref{fig:cpnet}, we plot histograms of $v_\mathrm{net}$ for our models, now including flipped magnetic field configurations, which were not included in the original analysis of \citetalias{PaperVIII} (their Figure 8).  Both MAD and SANE models are capable of passing the upper limit on $v_{\rm net}$, but SANE models are more likely to fail. In the right panel of \autoref{fig:cpnet}, we plot histograms of the spatially resolved circular polarization $\langle |v| \rangle$, and we over-plot our allowed region from the imaging results in \autoref{sec:m87results}: $\langle|v|\rangle< 3.7\,\%$.  SANE models produce the largest values of $\langle |v| \rangle$, which extend to nearly 10\,\% in some cases.  We find that 87\,\% of the images that fail our new upper limit on $\langle |v| \rangle$ are SANE.

We use two different methods for applying observational constraints to our GRMHD images, as in \citetalias{PaperVIII}:
\begin{enumerate}
    \item \emph{Simultaneous} scoring:  If a single model snapshot simultaneously passes every polarimetric constraint, then it passes.  Otherwise, it is rejected. This method is relatively strict, as for a model to pass at least one image must satisfy all constraints.  See Section 5.2 of \citetalias{PaperVIII} for more detail on the simultaneous scoring procedure.
    \item \emph{Joint} scoring: We compute $\chi_{j,k}^2$ statistics for each of the six metrics $j$ for each GRMHD snapshot $k$ (\citetalias{PaperVIII}, equation 17). We then compute a likelihood $\mathcal{L}_j$ for each metric by calculating the fraction of snapshots where $\chi^2_{j,k}>\chi^2_{j,\mathrm{data}}$. The final model likelihood is the product of the individual likelihoods from each metric:$\mathcal{L}=\prod_j \mathcal{L}_j$. This method is relatively lenient, as 
    no single image is required to satisfy all constraints simultaneously. See Section 5.3 of \citetalias{PaperVIII} for more detail on the joint scoring procedure.
\end{enumerate}

\noindent In \citetalias{PaperVIII}, these methods produced slightly different results, but both methods favored MAD models.  

\begin{figure*}
    \centering
    \includegraphics[width=0.45\textwidth]{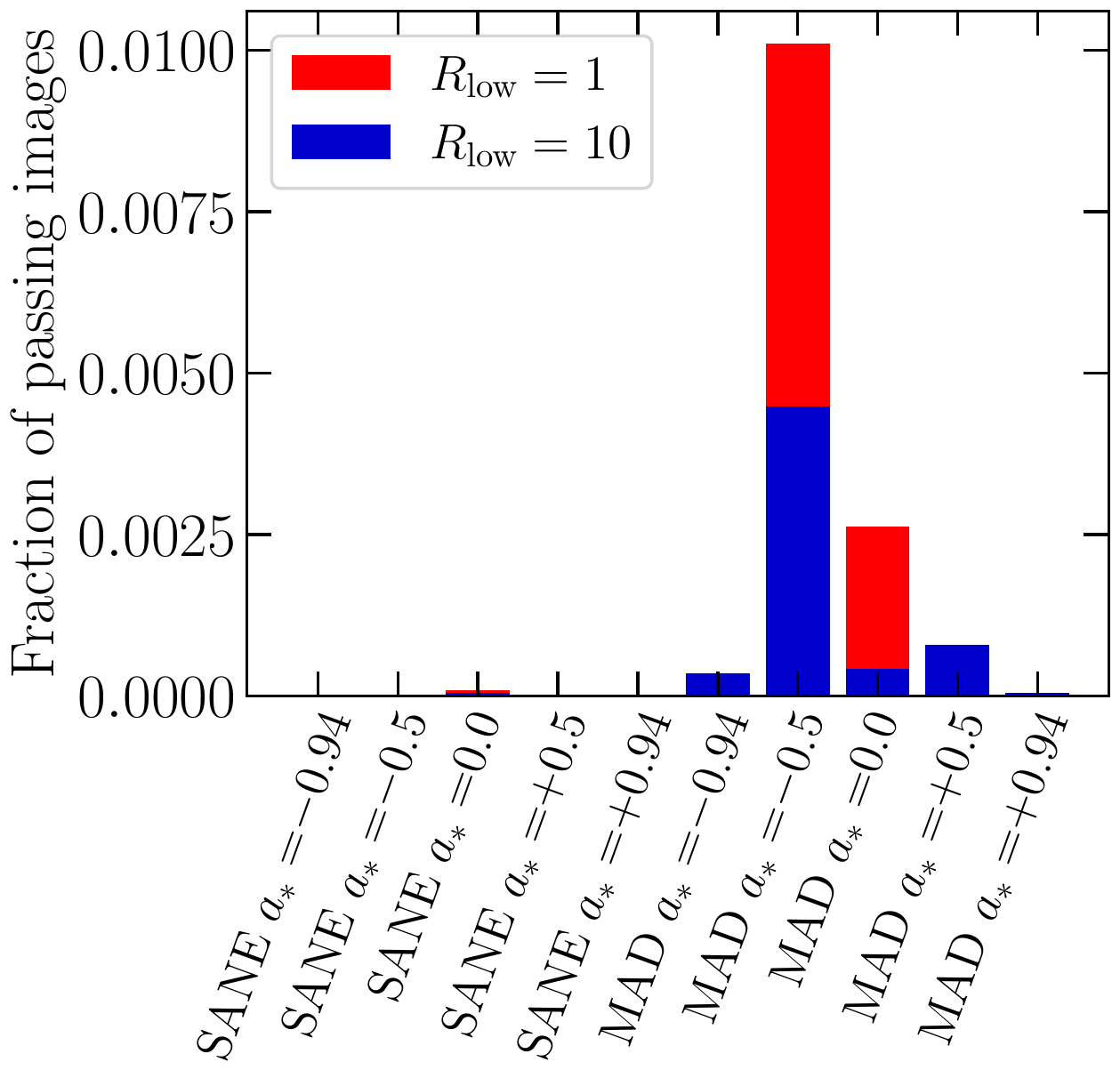}
    \includegraphics[width=0.43\textwidth]{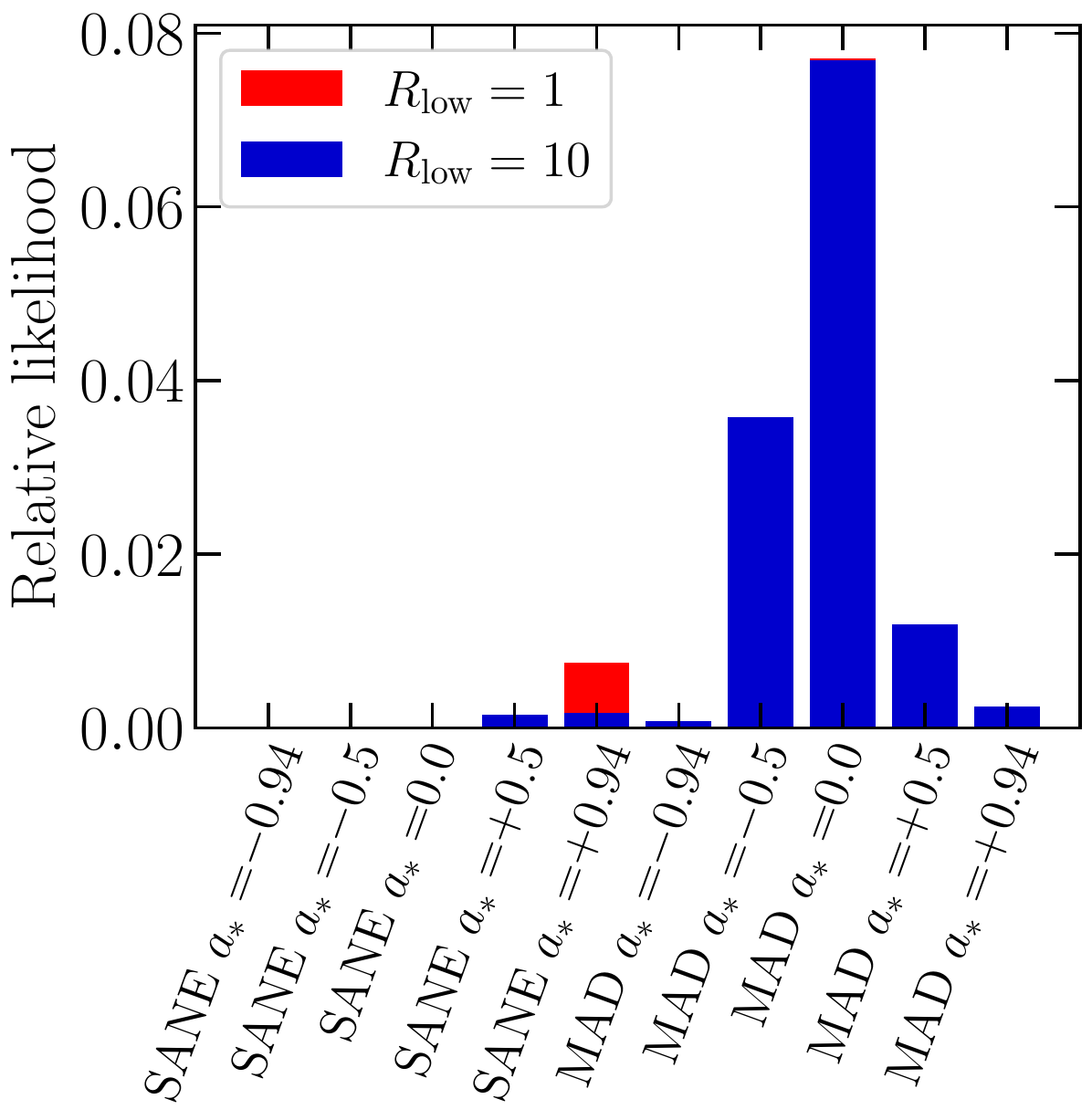}
    \caption{Scoring results for all the models using the simultaneous scoring method (left panel) and the joint distribution method (right panel). Passing fraction or relative likelihoods are summed over $R_{\mathrm{high}}$ and $\Vec{B}$ field alignment. MADs remain favored across both methods.}
    \label{fig:theory_model_scoring}
\end{figure*}

Our new model scoring results from both methods are shown in \autoref{fig:theory_model_scoring}.  This Figure is an update to Figure 12 of \citetalias{PaperVIII}, updated for both new observational results (the addition of the upper limit on $\langle|v|\rangle$) and new theoretical models (the set of reversed-polarity GRMHD snapshots).  As in \citetalias{PaperVIII}, we find that MAD models are strongly favored over their SANE counterparts.  

In the simultaneous scoring results (left panel of \autoref{fig:theory_model_scoring}), we find that a slight majority of the passing snapshots have aligned magnetic polarities with respect to the disk angular momentum vector.  Furthermore, we find that 83\,\% of passing snapshots have magnetic field polarities aligned with the spin of \m87, pointed away from us.  This is due not to Stokes $\mathcal{V}$ constraints, but rather from $\angle \beta_2$, for which the distributions shift slightly upon flipping the magnetic field polarity (see \autoref{app:field_reversal}).  While potentially interesting, especially for constraining the origins of the magnetic field \citep[e.g.,][]{Contopoulos+1998, Contopoulos+2022},
upon examining \autoref{fig:beta2ang} in \autoref{app:field_reversal}, it appears that this preference for an aligned field arises from the fact that $\angle \beta_2$ happens to be Faraday rotated out of the observed range more often in reversed field models than in aligned field models for the few spins that we sampled. It is possible that this effect may disappear if spin is sampled more densely, as $\angle \beta_2$ depends strongly on spin in GRMHD models \citep{PWP,Emami+2022,Qiu+2023}.  

Only a small improvement on the upper limit on $\langle |v| \rangle$ would have been necessary to rule out some currently passing snapshots.  Among the passing snapshots, we find $0.30\,\% \leq \langle |v| \rangle \leq 2.8\,\%$, with a median value of 0.48\,\%.  If the upper limit had instead been $\langle |v| \rangle < 1\,\%$, then 204 snapshots would have passed, down from 288.  If the upper limit had been $\langle |v| \rangle < 0.5\,\%$, then only 35 snapshots would have passed, all of which would have been MAD.

\begin{figure*}[ht!]
    \centering
    \includegraphics[width=\textwidth]{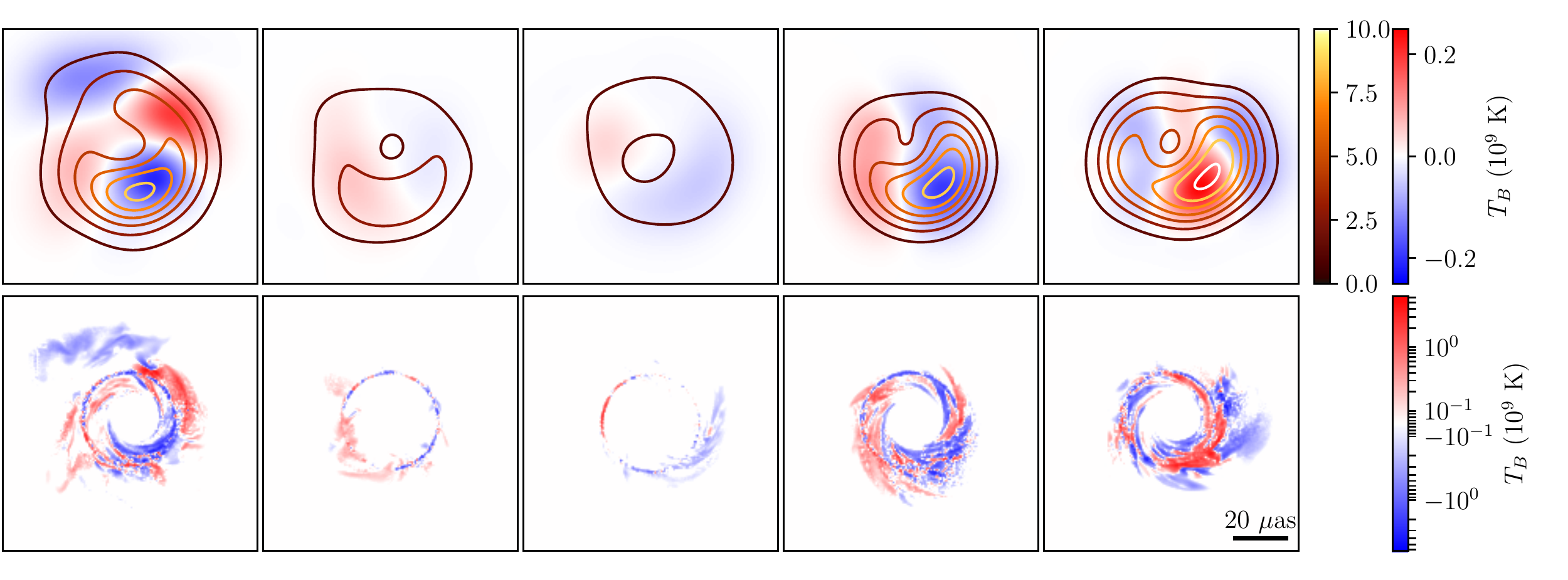} 
    \caption{Five passing snapshots visualized in circular polarization at EHT resolution in the top row, and at native simulation resolution in the bottom row in a symmetric logarithmic scale.  From left to right, these models correspond to (MAD $a_*=-0.94$ $R_\mathrm{high}=160$ $R_\mathrm{low}=10$ reversed field), (MAD $a_*=-0.5$ $R_\mathrm{high}=160$ $R_\mathrm{low}=1$ reversed field), (MAD $a_*=0$ $R_\mathrm{high}=80$ $R_\mathrm{low}=1$ aligned field), (MAD $a_*=+0.5$ $R_\mathrm{high}=160$ $R_\mathrm{low}=10$ aligned field), and (MAD $a_*=+0.94$ $R_\mathrm{high}=160$ $R_\mathrm{low}=10$ aligned field).  Sign reversals are ubiquitous in these models, even at spatial scales smaller than the EHT beam.  Some, but not all, Stokes $\mathcal{V}$ images exhibit a clear dipolar structure.}
    \label{fig:passing_examples}
\end{figure*}

In \autoref{fig:passing_examples}, we visualize in circular polarization five of the snapshots that pass simultaneous constraints.  Each of them are MAD models, of increasing spin, described in the figure caption.  Some, but not all, of the Stokes $\mathcal{V}$ morphologies are dominated by a dipole at EHT resolution. 
Perfect resolution images reveal ubiquitous sign reversals within the EHT beam and a rich morphology.  Some snapshots feature an inverted photon ring, discussed in detail in 
\citet{Moscibrodzka+2021,Ricarte+2021}.

In \autoref{app:stokesvscore}, we discuss the scoring results for both scoring strategies in more detail. We find that the new constraint on $\langle |v| \rangle$ has no effect on the results of the simultaneous scoring method and only a slight effect on the results from the joint scoring method.  The largest driver of quantitative differences in the scoring results from \citetalias{PaperVIII} is the inclusion of reversed magnetic field models, while updated radiative transfer coefficients and slight differences in the snapshots included in ray-tracing also play a minor role.

\subsection{Origins of Stokes $\mathcal{V}$ in passing models}
\label{sec:vorigins}

To better understand the mechanisms by which Stokes $\mathcal{V}$ is generated in passing GRMHD models, we conducted a series of tests using our radiative transfer code {\sc ipole}.  By artificially switching off certain radiative transfer coefficients, we can isolate their effects on the results on the resulting images. 

We present these tests in detail in \autoref{app:stokesvorigin}. 
In summary, we find that Faraday conversion is the dominant production mechanism for Stokes $\mathcal{V}$ in passing GRMHD models. Faraday rotation also plays a critical role, as the linear polarization that is later converted into circular polarization is scrambled on scales smaller than the EHT beam, leading to beam depolarization both in the linear and circular polarization images. Intrinsic or direct emission of circular polarization is sub-dominant to Faraday conversion, though it is not negligible in all models.

\subsection{Extensions to GRMHD models}
\label{sec:modelextensions}
We caution that as with the previous papers in this series, constraints on astrophysical models from EHT results come with systematic uncertainties due to
limits on our parameter space. As with previous theoretical interpretation papers in this series, we have limited our scoring to ideal GRMHD simulations with either perfectly aligned or anti-aligned disks, populated with ion-electron plasma with purely thermal distribution functions. 

While our $R$-$\beta$ prescription for setting the electron-to-ion temperature ratio broadly describes general trends as a function of plasma $\beta$, simulations with electron heating exhibit substantial zone-to-zone scatter and additional features that our prescription cannot reproduce \citep{Mizuno+2021,Dihingia+2023,Jiang+2023}. On the other hand, simulations including radiative cooling produce denser and cooler accretion disk mid-planes, which may help produce Faraday rotation \citep{Yoon+2020}. 

Tilted disks are known to imprint signatures on the total intensity image \citep{Chatterjee+2020}, and additional signatures may occur in polarization.
Circularly polarized emission is also sensitive to the content of the plasma and its distribution function, which we briefly explore in \autoref{app:alternative_plasmas}. While beyond the scope of our analysis, we expect that the polarized data published in this series will continue to constrain models probing additional aspects of BH and plasma astrophysics in future work.

%=================================================================================================================================================================
\section{Summary and Conclusion}
\label{sec:conc}
%=================================================================================================================================================================

We have presented an analysis of the circular polarization in \m87 at 1.3\,mm wavelength from 2017 EHT observations, the first analysis of circular polarization in any source using EHT data. By examining the difference between EHT closure phases in right and left circular polarization measurements, we find firm evidence for a weak, astrophysical (non-instrumental) circular polarization signal in \m87 on event-horizon scales.

To further analyze these data, we developed five different approaches to image analysis, testing each method on a suite of synthetic data from GRMHD simulations. When applied to the observations of \m87, these methods all find a moderate degree of image circular polarization ($\lsim 4\,\%$), which is consistent with expectations from the low degree of circular polarization ($\lsim 1\,\%$) seen in unresolved observations of \m87 with other facilities such as ALMA.
The details of our reconstructed images vary considerably among the five methods, indicating that the circular polarization structure is sensitive to choices in the imaging and calibration. Overall, we find that the structure cannot be reliably inferred without additional data or stronger assumptions. From our set of image reconstructions, we establish an upper limit on the (unsigned) resolved circular polarization fraction at the EHT's resolution of 20$\mu$as: $\langle|v|\rangle<3.7\%$.

Our results provide new constraints for models of the central supermassive black hole in M87 and its environment. We apply these constraints to a large library of images produced from ray-traced GRMHD simulations, which span a broad range of values for the black hole spin, weak (SANE) and strong (MAD) magnetic fields, 
updating our constraints relative to \citetalias{PaperVIII}. We consider both prograde and retrograde accretion flows, and both aligned and anti-aligned poloidal magnetic fields relative to the angular momentum of the accretion disk. 

We again find that strongly magnetized MAD models remain favored over weakly magnetized SANEs. As in \citetalias{PaperVIII} and \citet{Qiu+2023}, our scoring prefers spin values roughly around -0.5 and 0.0, driven largely by our constraints on $\angle \beta_2$.  That is, the toroidal morphology of the linear polarization ticks favor models with substantial poloidal fields.  Our scoring also favors models with larger ion-to-electron temperature ratio $R_\mathrm{high}$.
That is, the relatively low polarization fraction favors models that contain relatively dense and cool electrons to perform Faraday depolarization.  
For the model snapshots which simultaneously pass observational constraints, we find that Faraday conversion is typically more important than the circular polarization inherent to synchrotron emission in generating the circular polarization that we observe.
This is consistent with the conclusions of parsec-scale studies of other active galactic nuclei at lower frequencies \citep[e.g.][]{Jones_1988,Wardle+1998,Bower+1999,Gabuzda+2008}.
Faraday rotation also serves an indirect role in limiting the circular polarization fraction in many models, scrambling much of the linear polarization that could otherwise be converted into circular.

Theoretical models of \m87 exhibit significant variability.
In \autoref{sec:vvar}, we explore this variability in circular polarization metrics in passing GRMHD models. Future \m87 observations may present more favorable conditions for Stokes $\mathcal{V}$ imaging (\autoref{fig:variability_vnet}).  Thus, continued polarimetric monitoring of \m87 on horizon scales will allow us to place better constraints on physical models of the 230~GHz emitting region.
Sagittarius~A*, the black hole in the galactic center recently imaged by the EHT \citep{SgrAEHTCI}, exhibits a much larger $v_\mathrm{net} \approx -1$\,\% that may make prospects of Stokes $\mathcal{V}$ imaging more promising for this source \citep{Bower+2018,Goddi2021,Wielgus2022}, although its much more rapid time variability presents separate challenges \citep{SgrAEHTCIII,SgrAEHTCIV,Wielgus2022_light_curves}. 

Our results complete the analysis of EHT observations of \m87 in 2017, revealing the 230 GHz emission encircling the apparent shadow of the black hole in all four Stokes parameters. More recent EHT observations, with additional telescopes, significantly improved sensitivity, and higher observing frequencies, will provide crucial improvements to reveal unambiguous structure in the circular polarization and to characterize its variability. 

\facilities{EHT, ALMA, APEX, IRAM:30m, JCMT, LMT, SMA, ARO:SMT, SPT}
\software{ AIPS \citep{Greisen_2003},  ParselTongue \citep{Kettenis_2006},  GNU Parallel \citep{GNU}, \ehtim \citep{Chael_2016}, \difmap \citep{Shepherd_2011}, Numpy \citep{numpy_2020},  Scipy \citep{scipy}, Pandas \citep{pandas},  Astropy \citep{astropy_2013,astropy_2018}, Jupyter \citep{jupyter}, Matplotlib \citep{Hunter_2007}, \themis \citep{Broderick_2019}, \dmc \citep{DMC}, \polsolve \citep{polsolve}, 
{\sc HARM} \citep{2003ApJ...589..444G,Noble2006},
{\sc ipole} \citep{Noble2007,Moscibrodzka&Gammie2018}
}

%=================================================================================================================================================================
\section{Acknowledgements}
%\begin{acknowledgments}
The Event Horizon Telescope Collaboration thanks the following
organizations and programs: the Academia Sinica; the Academy
of Finland (projects 274477, 284495, 312496, 315721); the Agencia Nacional de Investigaci\'{o}n 
y Desarrollo (ANID), Chile via NCN$19\_058$ (TITANs), Fondecyt 1221421  and BASAL FB210003; the Alexander
von Humboldt Stiftung; an Alfred P. Sloan Research Fellowship;
Allegro, the European ALMA Regional Centre node in the Netherlands, the NL astronomy
research network NOVA and the astronomy institutes of the University of Amsterdam, Leiden University, and Radboud University;
the ALMA North America Development Fund; the Astrophysics and High Energy Physics programme by MCIN (with funding from European Union NextGenerationEU, PRTR-C17I1); the Black Hole Initiative, which is funded by grants from the John Templeton Foundation and the Gordon and Betty Moore Foundation (although the opinions expressed in this work are those of the author(s) 
and do not necessarily reflect the views of these Foundations); the Brinson Foundation; ``la Caixa'' Foundation (ID 100010434) through fellowship codes LCF/BQ/DI22/11940027 and LCF/BQ/DI22/11940030; the Fondo CAS-ANID folio CAS220010; 
Chandra DD7-18089X and TM6-17006X; the China Scholarship
Council; the China Postdoctoral Science Foundation fellowships (2020M671266, 2022M712084); Consejo Nacional de Humanidades, Ciencia y Tecnología (CONAHCYT, Mexico, projects U0004-246083, U0004-259839, F0003-272050, M0037-279006, F0003-281692, 104497, 275201, 263356); the Colfuturo Scholarship; 
the Consejer\'{i}a de Econom\'{i}a, Conocimiento, 
Empresas y Universidad 
of the Junta de Andaluc\'{i}a (grant P18-FR-1769), the Consejo Superior de Investigaciones 
Cient\'{i}ficas (grant 2019AEP112);
the Delaney Family via the Delaney Family John A.
Wheeler Chair at Perimeter Institute; Dirección General de Asuntos del Personal Académico-Universidad Nacional Autónoma de México (DGAPA-UNAM, projects IN112820 and IN108324); the Dutch Organization for Scientific Research (NWO) for the VICI award (grant 639.043.513), the grant OCENW.KLEIN.113, and the Dutch Black Hole Consortium (with project No. NWA 1292.19.202) of the research programme the National Science Agenda; the Dutch National Supercomputers, Cartesius and Snellius  
(NWO grant 2021.013); 
the EACOA Fellowship awarded by the East Asia Core
Observatories Association, which consists of the Academia Sinica Institute of Astronomy and
Astrophysics, the National Astronomical Observatory of Japan, Center for Astronomical Mega-Science,
Chinese Academy of Sciences, and the Korea Astronomy and Space Science Institute; 
the European Research Council (ERC) Synergy
Grant ``BlackHoleCam: Imaging the Event Horizon
of Black Holes" (grant 610058); 
the European Union Horizon 2020
research and innovation programme under grant agreements
RadioNet (No. 730562) and 
M2FINDERS (No. 101018682); the Horizon ERC Grants 2021 programme under grant agreement No. 101040021; 
the Generalitat
Valenciana (grants APOSTD/2018/177 and  ASFAE/2022/018) and
GenT Program (project CIDEGENT/2018/021); MICINN Research Project PID2019-108995GB-C22;
the European Research Council for advanced grant `JETSET: Launching, propagation and 
emission of relativistic jets from binary mergers and across mass scales' (grant No. 884631); the FAPESP (Funda\c{c}\~ao de Amparo \'a Pesquisa do Estado de S\~ao Paulo) under grant 2021/01183-8; 
the Institute for Advanced Study; the Istituto Nazionale di Fisica
Nucleare (INFN) sezione di Napoli, iniziative specifiche
TEONGRAV; 
the International Max Planck Research
School for Astronomy and Astrophysics at the
Universities of Bonn and Cologne; 
DFG research grant ``Jet physics on horizon scales and beyond'' (grant No. FR 4069/2-1);
Joint Columbia/Flatiron Postdoctoral Fellowship (research at the Flatiron Institute is supported by the Simons Foundation); 
the Japan Ministry of Education, Culture, Sports, Science and Technology (MEXT; grant JPMXP1020200109); %the Japanese Government (Monbukagakusho:MEXT) Scholarship; 
the Japan Society for the Promotion of Science (JSPS) Grant-in-Aid for JSPS
Research Fellowship (JP17J08829); the Joint Institute for Computational Fundamental Science, Japan; the Key Research
Program of Frontier Sciences, Chinese Academy of
Sciences (CAS, grants QYZDJ-SSW-SLH057, QYZDJSSW-SYS008, ZDBS-LY-SLH011); 
the Leverhulme Trust Early Career Research
Fellowship; the Max-Planck-Gesellschaft (MPG);
the Max Planck Partner Group of the MPG and the
CAS; the MEXT/JSPS KAKENHI (grants 18KK0090, JP21H01137,
JP18H03721, JP18K13594, 18K03709, JP19K14761, 18H01245, 25120007, 23K03453); the Malaysian Fundamental Research Grant Scheme (FRGS) FRGS/1/2019/STG02/UM/02/6; the MIT International Science
and Technology Initiatives (MISTI) Funds; 
the Ministry of Science and Technology (MOST) of Taiwan (103-2119-M-001-010-MY2, 105-2112-M-001-025-MY3, 105-2119-M-001-042, 106-2112-M-001-011, 106-2119-M-001-013, 106-2119-M-001-027, 106-2923-M-001-005, 107-2119-M-001-017, 107-2119-M-001-020, 107-2119-M-001-041, 107-2119-M-110-005, 107-2923-M-001-009, 108-2112-M-001-048, 108-2112-M-001-051, 108-2923-M-001-002, 109-2112-M-001-025, 109-2124-M-001-005, 109-2923-M-001-001, 110-2112-M-003-007-MY2, 110-2112-M-001-033, 110-2124-M-001-007, and 110-2923-M-001-001);
the Ministry of Education (MoE) of Taiwan Yushan Young Scholar Program;
the Physics Division, National Center for Theoretical Sciences of Taiwan;
the National Aeronautics and
Space Administration (NASA, Fermi Guest Investigator
grant 80NSSC20K1567, NASA Astrophysics Theory Program grant 80NSSC20K0527, NASA NuSTAR award 
80NSSC20K0645); 
NASA Hubble Fellowship 
grants HST-HF2-51431.001-A, HST-HF2-51482.001-A awarded 
by the Space Telescope Science Institute, which is operated by the Association of Universities for 
Research in Astronomy, Inc., for NASA, under contract NAS5-26555; 
the National Institute of Natural Sciences (NINS) of Japan; the National
Key Research and Development Program of China
(grant 2016YFA0400704, 2017YFA0402703, 2016YFA0400702); the National
Science Foundation (NSF, grants AST-0096454,
AST-0352953, AST-0521233, AST-0705062, AST-0905844, AST-0922984, AST-1126433, AST-1140030,
DGE-1144085, AST-1207704, AST-1207730, AST-1207752, MRI-1228509, OPP-1248097, AST-1310896, AST-1440254, 
AST-1555365, AST-1614868, AST-1615796, AST-1715061, AST-1716327,  %AST-1716536, 
OISE-1743747, AST-1816420, AST-1935980, AST-2034306, AST-2307887); 
NSF Astronomy and Astrophysics Postdoctoral Fellowship (AST-1903847); 
the Natural Science Foundation of China (grants 11650110427, 10625314, 11721303, 11725312, 11873028, 11933007, 11991052, 11991053, 12192220, 12192223, 12273022); 
the Natural Sciences and Engineering Research Council of
Canada (NSERC, including a Discovery Grant and
the NSERC Alexander Graham Bell Canada Graduate
Scholarships-Doctoral Program); the National Youth
Thousand Talents Program of China; the National Research
Foundation of Korea (the Global PhD Fellowship
Grant: grants NRF-2015H1A2A1033752, the Korea Research Fellowship Program:
NRF-2015H1D3A1066561, Brain Pool Program: 2019H1D3A1A01102564, 
Basic Research Support Grant 2019R1F1A1059721, 2021R1A6A3A01086420, 2022R1C1C1005255); 
Netherlands Research School for Astronomy (NOVA) Virtual Institute of Accretion (VIA) postdoctoral fellowships; 
Onsala Space Observatory (OSO) national infrastructure, for the provisioning
of its facilities/observational support (OSO receives
funding through the Swedish Research Council under
grant 2017-00648);  the Perimeter Institute for Theoretical
Physics (research at Perimeter Institute is supported
by the Government of Canada through the Department
of Innovation, Science and Economic Development
and by the Province of Ontario through the
Ministry of Research, Innovation and Science); the Princeton Gravity Initiative; the Spanish Ministerio de Ciencia e Innovaci\'{o}n (grants PGC2018-098915-B-C21, AYA2016-80889-P,
PID2019-108995GB-C21, PID2020-117404GB-C21); 
the University of Pretoria for financial aid in the provision of the new 
Cluster Server nodes and SuperMicro (USA) for a SEEDING GRANT approved toward these 
nodes in 2020; the Shanghai Municipality orientation program of basic research for international scientists (grant no. 22JC1410600);
the Shanghai Pilot Program for Basic Research, Chinese Academy of Science, 
Shanghai Branch (JCYJ-SHFY-2021-013);
the State Agency for Research of the Spanish MCIU through
the ``Center of Excellence Severo Ochoa'' award for
the Instituto de Astrof\'{i}sica de Andaluc\'{i}a (SEV-2017-
0709); the Spanish Ministry for Science and Innovation grant CEX2021-001131-S funded by MCIN/AEI/10.13039/501100011033; the Spinoza Prize SPI 78-409; the South African Research Chairs Initiative, through the 
South African Radio Astronomy Observatory (SARAO, grant ID 77948),  which is a facility of the National 
Research Foundation (NRF), an agency of the Department of Science and Innovation (DSI) of South Africa; 
the Toray Science Foundation; the Swedish Research Council (VR); the UK Science and Technology Facilities Council (grant no. ST/X508329/1); 
the US Department
of Energy (USDOE) through the Los Alamos National
Laboratory (operated by Triad National Security,
LLC, for the National Nuclear Security Administration
of the USDOE, contract 89233218CNA000001); and the YCAA Prize Postdoctoral Fellowship.

We thank
the staff at the participating observatories, correlation
centers, and institutions for their enthusiastic support.
This paper makes use of the following ALMA data:
ADS/JAO.ALMA\#2016.1.01154.V. ALMA is a partnership
of the European Southern Observatory (ESO;
Europe, representing its member states), NSF, and
National Institutes of Natural Sciences of Japan, together
with National Research Council (Canada), Ministry
of Science and Technology (MOST; Taiwan),
Academia Sinica Institute of Astronomy and Astrophysics
(ASIAA; Taiwan), and Korea Astronomy and
Space Science Institute (KASI; Republic of Korea), in
cooperation with the Republic of Chile. The Joint
ALMA Observatory is operated by ESO, Associated
Universities, Inc. (AUI)/NRAO, and the National Astronomical
Observatory of Japan (NAOJ). The NRAO
is a facility of the NSF operated under cooperative agreement
by AUI.
This research used resources of the Oak Ridge Leadership Computing Facility at the Oak Ridge National
Laboratory, which is supported by the Office of Science of the U.S. Department of Energy under contract
No. DE-AC05-00OR22725; the ASTROVIVES FEDER infrastructure, with project code IDIFEDER-2021-086; the computing cluster of Shanghai VLBI correlator supported by the Special Fund 
for Astronomy from the Ministry of Finance in China;  
We also thank the Center for Computational Astrophysics, National Astronomical Observatory of Japan. This work was supported by FAPESP (Fundacao de Amparo a Pesquisa do Estado de Sao Paulo) under grant 2021/01183-8.

APEX is a collaboration between the
Max-Planck-Institut f{\"u}r Radioastronomie (Germany),
ESO, and the Onsala Space Observatory (Sweden). The
SMA is a joint project between the SAO and ASIAA
and is funded by the Smithsonian Institution and the
Academia Sinica. The JCMT is operated by the East
Asian Observatory on behalf of the NAOJ, ASIAA, and
KASI, as well as the Ministry of Finance of China, Chinese
Academy of Sciences, and the National Key Research and Development
Program (No. 2017YFA0402700) of China
and Natural Science Foundation of China grant 11873028.
Additional funding support for the JCMT is provided by the Science
and Technologies Facility Council (UK) and participating
universities in the UK and Canada. 
The LMT is a project operated by the Instituto Nacional
de Astr\'{o}fisica, \'{O}ptica, y Electr\'{o}nica (Mexico) and the
University of Massachusetts at Amherst (USA). The
IRAM 30-m telescope on Pico Veleta, Spain is operated
by IRAM and supported by CNRS (Centre National de
la Recherche Scientifique, France), MPG (Max-Planck-Gesellschaft, Germany), 
and IGN (Instituto Geogr\'{a}fico
Nacional, Spain). The SMT is operated by the Arizona
Radio Observatory, a part of the Steward Observatory
of the University of Arizona, with financial support of
operations from the State of Arizona and financial support
for instrumentation development from the NSF.
Support for SPT participation in the EHT is provided by the National Science Foundation through award OPP-1852617 
to the University of Chicago. Partial support is also 
provided by the Kavli Institute of Cosmological Physics at the University of Chicago. The SPT hydrogen maser was 
provided on loan from the GLT, courtesy of ASIAA.

This work used the
Extreme Science and Engineering Discovery Environment
(XSEDE), supported by NSF grant ACI-1548562,
and CyVerse, supported by NSF grants DBI-0735191,
DBI-1265383, and DBI-1743442. XSEDE Stampede2 resource
at TACC was allocated through TG-AST170024
and TG-AST080026N. XSEDE JetStream resource at
PTI and TACC was allocated through AST170028.
This research is part of the Frontera computing project at the Texas Advanced 
Computing Center through the Frontera Large-Scale Community Partnerships allocation
AST20023. Frontera is made possible by National Science Foundation award OAC-1818253.
This research was done using services provided by the OSG Consortium~\citep{osg07,osg09} supported by the National Science Foundation award Nos. 2030508 and 1836650. 
%This research was carried out using resources provided by the Open Science Grid, which is supported by the National Science Foundation and the U.S. Department of Energy Office of Science. 
Additional work used ABACUS2.0, which is part of the eScience center at Southern Denmark University, and the Kultrun Astronomy Hybrid Cluster (projects Conicyt Programa de Astronomia Fondo Quimal QUIMAL170001, Conicyt PIA ACT172033, Fondecyt Iniciacion 11170268, Quimal 220002). 
Simulations were also performed on the SuperMUC cluster at the LRZ in Garching, 
on the LOEWE cluster in CSC in Frankfurt, on the HazelHen cluster at the HLRS in Stuttgart, 
and on the Pi2.0 and Siyuan Mark-I at Shanghai Jiao Tong University.
The computer resources of the Finnish IT Center for Science (CSC) and the Finnish Computing 
Competence Infrastructure (FCCI) project are acknowledged. This
research was enabled in part by support provided
by Compute Ontario (http://computeontario.ca), Calcul
Quebec (http://www.calculquebec.ca), and Compute
Canada (http://www.computecanada.ca). 
%CC acknowledges support from the Swedish Research Council (VR).

The EHTC has
received generous donations of FPGA chips from Xilinx
Inc., under the Xilinx University Program. The EHTC
has benefited from technology shared under open-source
license by the Collaboration for Astronomy Signal Processing
and Electronics Research (CASPER). The EHT
project is grateful to T4Science and Microsemi for their
assistance with hydrogen masers. This research has
made use of NASA's Astrophysics Data System. We
gratefully acknowledge the support provided by the extended
staff of the ALMA, from the inception of
the ALMA Phasing Project through the observational
campaigns of 2017 and 2018. We would like to thank
A. Deller and W. Brisken for EHT-specific support with
the use of DiFX. We thank Martin Shepherd for the addition of extra features in the Difmap software 
that were used for the CLEAN imaging results presented in this paper.
We acknowledge the significance that
Maunakea, where the SMA and JCMT EHT stations
are located, has for the indigenous Hawaiian people.

%\end{acknowledgments} 
%=================================================================================================================================================================

%=================================================================================================================================================================
%=================================================================================================================================================================
\clearpage
\appendix 

%=================================================================================================================================================================

\section{Strategies for Polarimetric Gain calibration}
\label{app:polar_gains}

The fiducial  multi-source procedure for polarimetric gains calibration, described in \autoref{sec:polarimetric_gains}, was employed for the whole series of EHT papers on the \m87 observations performed in 2017, as well as the multi-year \m87 variability study presented in \citet{Wielgus2020}, and the 3C\,279 quasar study \citep{Kim:2020}. 

\begin{figure}[t]
\centering
\includegraphics[width=0.48\textwidth,angle=0]{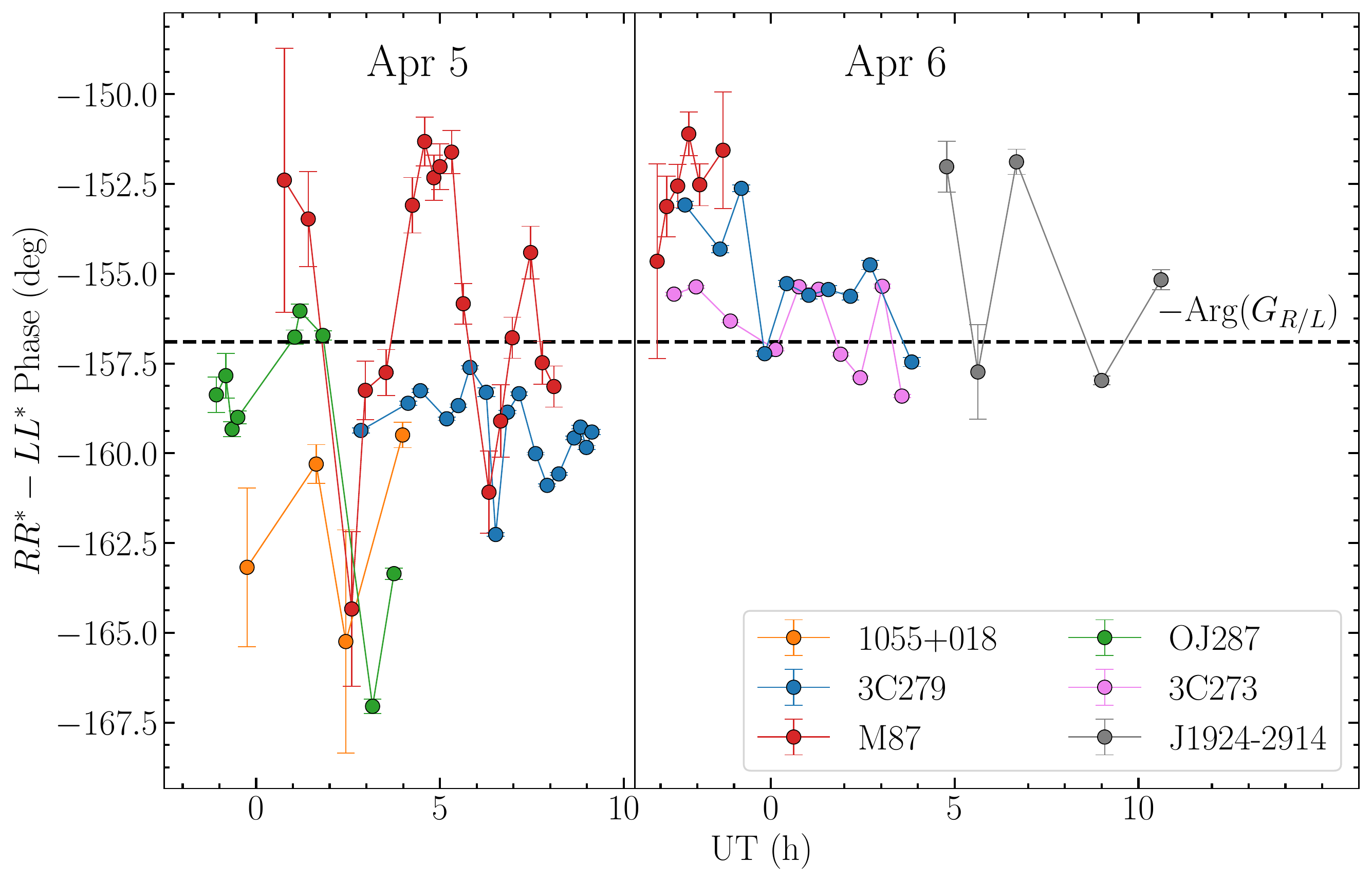}
\caption{Example of a multi-source, multi-day estimation of the $G_{R/L}$ phase for the LMT, using $RR^*$ and $LL^*$ high band visibilities measured on the ALMA-LMT baseline. For the actual estimate of a constant $G_{R/L}$ phase, 5 days and 10 sources were used. The origin of the small residual phases may be instrumental; however, they do not exhibit clear source-independent features. Alternatively, residuals may be caused by the presence of a small, source-specific} circular polarization.
\label{fig:RL_gain}
\end{figure}

\autoref{fig:RL_gain} shows an example of how we estimated the phase of $G_{R/L}$ for the LMT using this fiducial multi-source approach.  The figure shows visibility phase differences between $RR^*$ and $LL^*$ components recorded on the ALMA-LMT baseline, following the a priori correction for the field angle rotation. A near constant in time phase residual of about -157$^\circ$ is clearly present, consistent between observing days and for multiple observed sources. This residual is interpreted as the negative phase of the LMT instrumental gain ratio $G_{R/L}$ and subsequently corrected in the data. Similarly, $G_{R/L}$ gains corresponding to other sites were corrected  using the baseline from LMT to ALMA on multiple days and sources.

An alternative approach is based on calibrating complex $G_{R/L}$ to $RR^* = LL^*$, that is, to $\tilde{\mathcal{V}} = 0$. The latter strategy, which is simpler but more aggressive 
was used in the CASA-based \texttt{rPICARD} pipeline \citep{Janssen_2019}, where the phase difference between $RR^*$ and $LL^*$ visibilities is minimized through a polarimetric gains calibration.  
The $\tilde{\mathcal{V}} = 0$ self-calibration constitutes a very robust approach that remains valid for the recovery of the remaining Stokes parameters as long as $ \tilde{\mathcal{I}} \gg \tilde{\mathcal{V}}$. In addition to its use in the \texttt{rPICARD} pipeline, we have made a choice to employ this calibration variant for the \texttt{HOPS} data in the EHT papers on Sagittarius A* \citep{SgrAEHTCI}, Centaurus A \citep{Janssen2021}, J1924-2914 \citep{Issaoun2022}, and NRAO530 \citep{Jorstad2023}.

Note that self-calibration (multiplicative gain-calibration) to $\tilde{\mathcal{V}} = 0$ is not equivalent to removing the baseline signatures of $\tilde{\mathcal{V}} = 0$. In particular, this alternative calibration does not affect closure signatures, such as those presented in \autoref{fig:ClosDiff}. This is different than the EHT \m87 data set released with \citetalias{PaperIII} in 2019,\footnote{\url{https://datacommons.cyverse.org/browse/iplant/home/shared/commons_repo/curated/EHTC_FirstM87Results_Apr2019}} where all signatures of $\tilde{\mathcal{V}}$ were deliberately wiped out by replacing $RR^*$ and $LL^*$ on all baselines with their average, only retaining the Stokes $\tilde{\mathcal{I}}$ total intensity information intact. 

\begin{figure}[h!]
\centering
\includegraphics[width=0.9\linewidth,angle=0]{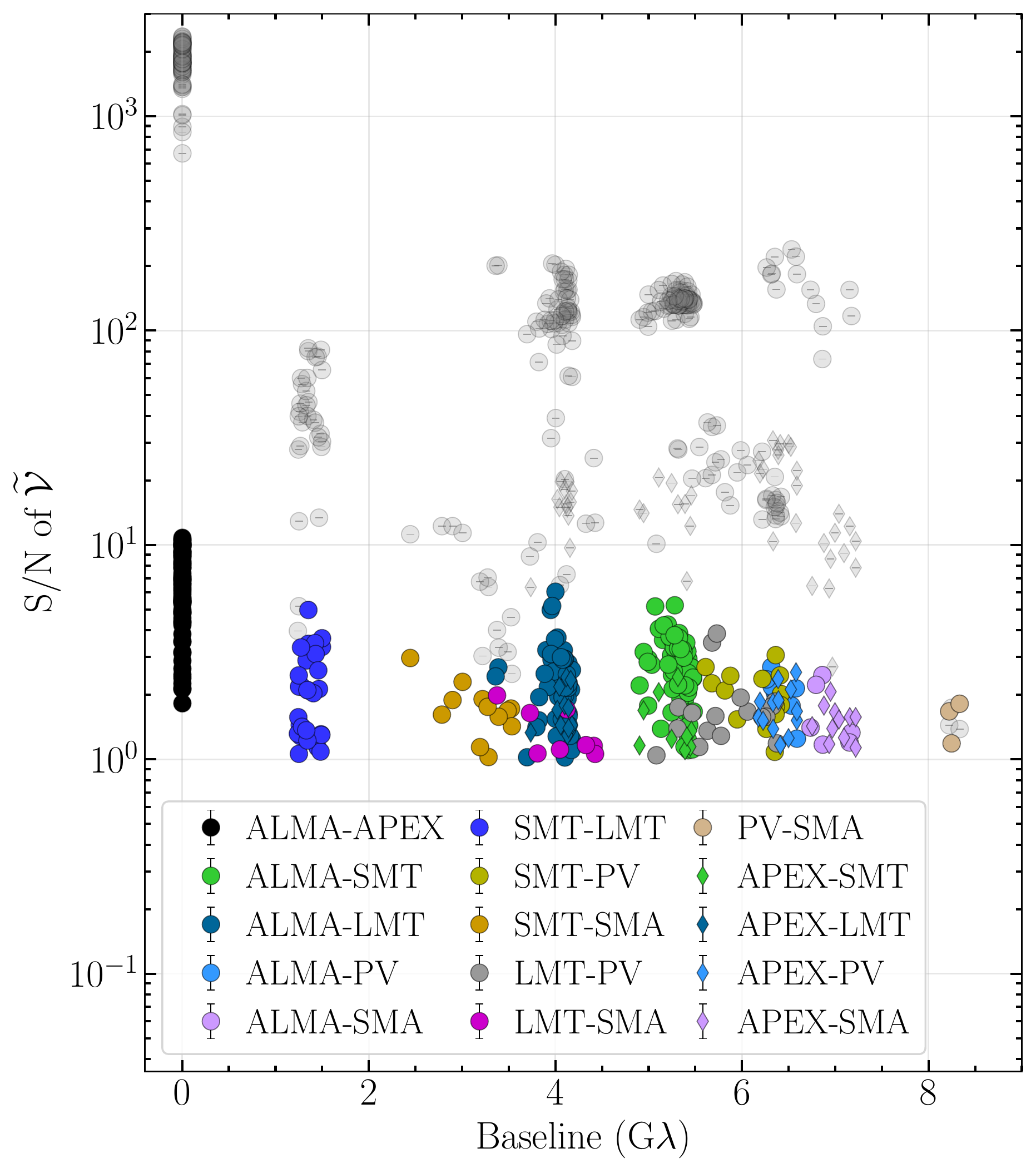}
\caption{S/N of the $\tilde{\mathcal{V}}$ observations as a function of the projected baseline length, for the data set self-calibrated to $\tilde{\mathcal{V}} = 0$. Gray points in the background correspond to the S/N of Stokes $\tilde{\mathcal{I}}$ detections.  } \label{fig:m87_uvdist_snr_ER6}
\end{figure}

\begin{figure*}[t!]
\centering
\includegraphics[width=0.98\textwidth,angle=0]{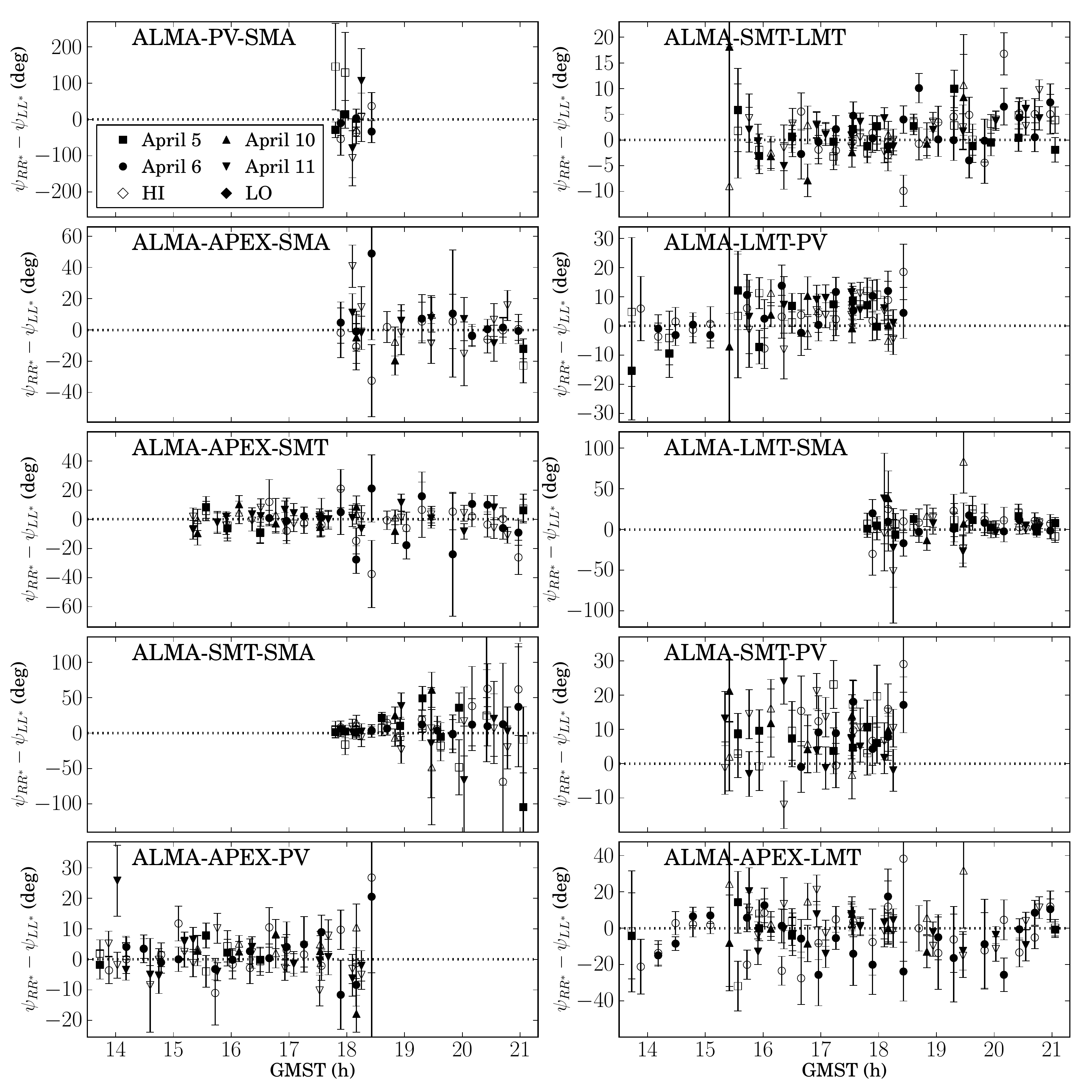}
\caption{Closure phase differences between scan-averaged $RR^*$ and $LL^*$ visibilities, for all the antenna triangles related to ALMA. All epochs and both bands (empty symbols for high band; filled symbols for low band) are shown. The zero line (i.e., no detection of spatially resolved Stokes $\mathcal{V}$) is shown as a dotted line.} \label{fig:all_triplets}
\end{figure*}

Despite certain advantages, self-calibration to $\tilde{\mathcal{V}} = 0$ may render circular polarization more difficult to recover and for that reason the multi-source polarimetric gains calibration procedure was selected as better suited for the conservative upper limits presented in this paper. In \autoref{fig:m87_uvdist_snr_ER6} we present the version of the \autoref{fig:m87_uv_coverage_snr}, but for the $\tilde{\mathcal{V}} = 0$ self-calibration data set. Significant decrease of $\tilde{\mathcal{V}}$ S/N is clearly visible, indicating, that high S/N detections seen in \autoref{fig:m87_uv_coverage_snr} could be related to systematic errors of the polarimetric gain calibration, rather than to robust signatures of circular polarization. S/N of the Stokes $\tilde{\mathcal{I}}$, on the other hand, is mostly insensitive to the $G_{R/L}$ calibration scheme.

%=================================================================================================================================================================

\section{Closure Phase Differences on ALMA Triangles}
\label{app:all_triplets}

\autoref{fig:ClosDiff} in the main text shows the closure-phase differences for the triangle ALMA-SMT-PV. For this triangle, an average deviation from zero at the level of $\sim 5\,\sigma$ is detected, which we interpret as evidence of the presence of a spatially-resolved (and asymmetric) Stokes $\mathcal{V}$ brightness distribution in M\,87*.

There is, though, a total of ten antenna triangles where ALMA is included, which are related to the closure phases with highest S/N. In \autoref{fig:all_triplets}, we show the closure-phase differences of scan-averaged visibilities for all these triangles. There are hints of departures from zero in some cases and for some time ranges (e.g., ALMA-SMT-LMT toward the end of the observations, as well as ALMA-APEX-LMT and ALMA-LMT-PV), but there is no clear detection above $3-4$\,$\sigma$. The detection of Stokes $\mathcal{V}$ from all the ALMA-related antenna triangles is, therefore, only tentative if we use the closure phase differences.

%=================================================================================================================================================================
\section{Closure Traces}
\label{app:closuretraces}

Closure traces are a set of complex closure quantities defined on quadrangles that are insensitive to %linear 
station-based corruptions \citep[see][]{BroderickPesce_2020}.  Of particular relevance to the current discussion of Stokes $\mathcal{V}$ is the fact that this includes the right- and left-hand complex station gains and leakage terms.  From these, a combination can be constructed on a single quadrangle, the conjugate closure trace product that differs from unity only in the presence of non-trivial polarization structure \citep[see][]{BroderickPesce_2020}.  For this reason, these conjugate closure trace products provide a robust direct test for source polarization.

\begin{figure}[t!]
    \centering
    \includegraphics[width=\columnwidth]{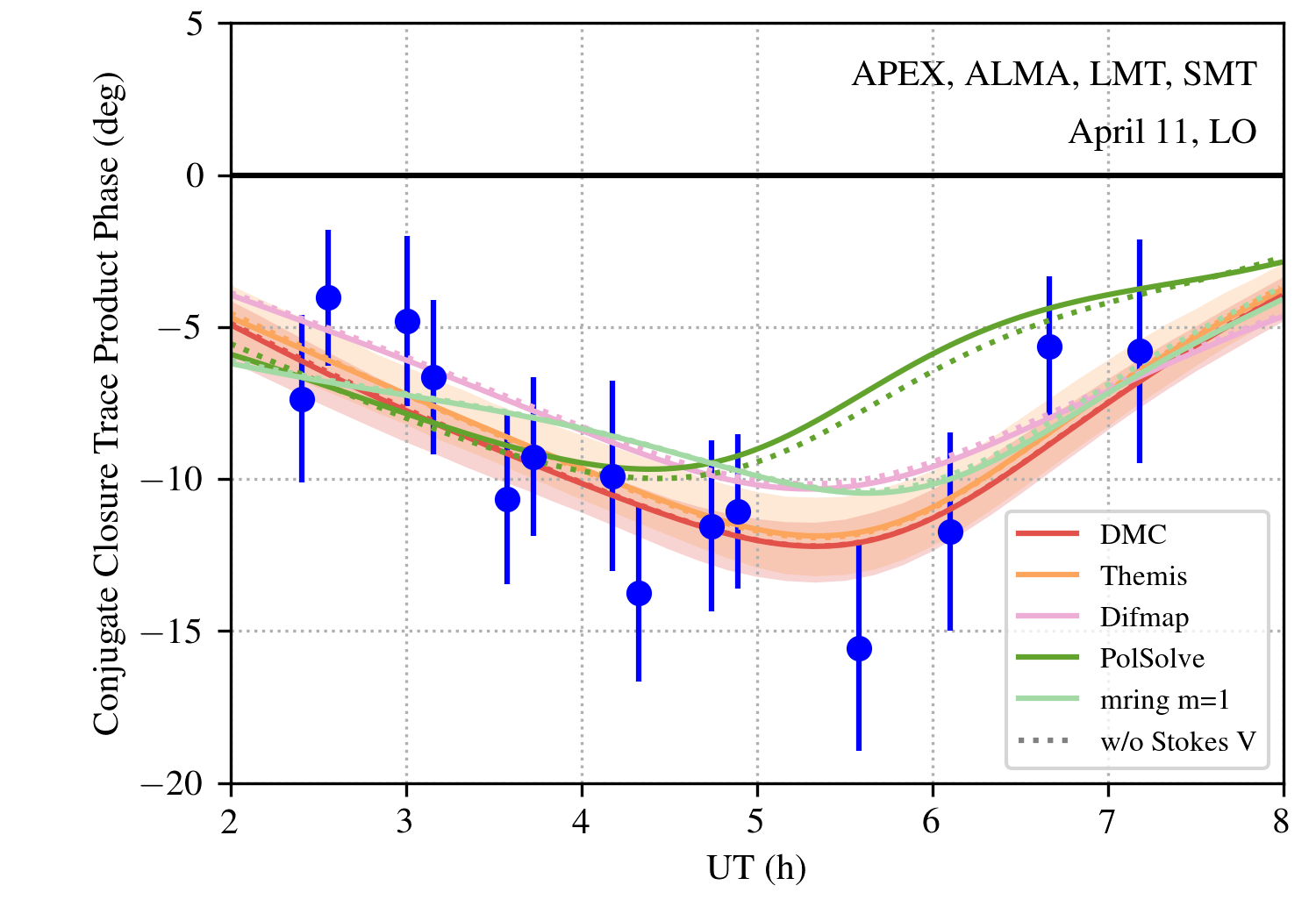}
    \caption{Phases of the conjugate closure trace product constructed on the APEX–ALMA–LMT–SMT and APEX–SMT–LMT–ALMA quadrangles from the 2017 April 11 low-band observations.  Colored lines show the same quantity computed from the mean images in \autoref{fig:images_all_3601}, with the $2\sigma$ regions for \themis and \dmc indicated by bands.  For comparison, the conjugate closure trace products when the Stokes $\mathcal{V}$ maps are ignored are shown by the corresponding dotted lines for each method.}
    \label{fig:cctpp}
\end{figure}

\autoref{fig:cctpp} compares the phases of the conjugate closure trace product generated on the APEX-ALMA-LMT-SMT and APEX-SMT-LMT-ALMA quadrangles from the image reconstructions in \autoref{fig:images_all_3601} directly to the 2017 April 11 low-band data.\footnote{The insensitivity of the closure traces to the gain ratios $|G_{R/L}|$ results in similar width $2\sigma$ bands in \autoref{fig:cctpp} for \themis and \dmc despite their different treatments of the station gains, and thus very different error bars in \autoref{fig:stats_realdata}.}  All reconstructions are broadly consistent with the conjugate closure trace products, similar to the results found in Appendix B of \citetalias{PaperVII}.

Because the closure traces are invariant to rotations on the Poincar\'e sphere, the conjugate closure trace products cannot isolate the contribution from the Stokes $\mathcal{V}$ maps. However, we also show by dotted lines the conjugate closure trace products for each reconstruction in \autoref{fig:images_all_3601} after setting the Stokes $\mathcal{V}$ to zero, indicating the gross magnitude of the impact of circular polarization.  For all reconstruction methods, the difference attributable to the Stokes $\mathcal{V}$ map is small.  For \themis and \dmc, the difference is small in comparison to the range of conjugate closure trace product phases spanned by the image posterior, implying that reconstruction of the Stoke $\mathcal{V}$ maps is strongly dependent on the assumed calibration priors, consistent with what is found in \autoref{sec:m87results}.

%=================================================================================================================================================================
\section{Method Summaries}
\label{app:methods}

In this Appendix, we describe each image reconstruction method introduced in \autoref{sec:methods} and summarized in \autoref{tab:method_summary}. 

\subsection{\difmap}
\label{sec:difmap}

The CLEAN algorithm \citep{Hogbom_1974} is a traditional VLBI imaging method which performs imaging via inverse modeling in all four Stokes parameters, $\mathcal{I, Q, U, \text{and}, V}$ for visibility data with dual polarization. The method is implemented
in the \difmap \citep{Shepherd_2011} and Astronomical Imaging Process System \citep[AIPS][]{Greisen_2003} software packages. Imaging in \difmap is an iterative process using a set of CLEAN windows until the residual $\chi^2$ between data and a model reach a minimum value.
For total intensity imaging, the process involves self-calibration of phase and amplitude of visibilities with the model at different steps of iterations, which results in calibrated visibility data
and a file containing a model of $\mathcal{I}$  delta functions characterized by the amplitude and position on the image. For linear and circular polarization imaging, no additional self-calibration procedure is performed, 
and \difmap generates models of $\mathcal{Q}$,$\mathcal{U}$, and $\mathcal{V}$ delta functions.

For EHT data, to obtain a Stokes $\mathcal{V}$ image 
we first perform an amplitude and phase calibration of R and L visibility data with the Stokes $\mathcal{I}$ image via the AIPS task CALIB, assuming that $(RR^*+LL^*)/2 = \tilde{\mathcal{I}}$. Then we apply the $D$-term correction to the visibility data and repeat the amplitude and phase calibration with the image assuming again $(RR^*+LL^*)/2 = \tilde{\mathcal{I}}$. For this study we use the \m87 $D$-terms, which are specified in Tables 3 \& 5 of Appendix~A in \citetalias{PaperVII}, applying them through an antenna table via AIPS tasks TBOUT and TBIN. At this stage, the calibrated data files are written out to \difmap for visual inspection and further editing, if needed.

After reading the edited visibility data back to AIPS, we proceed with the final R/L gain calibration, which is necessary to remove any residual instrumental circular polarization. To do that, we apply an additional amplitude and phase calibration of $R$ and $L$ visibility data {\it separately} with the $\mathcal{I}$ image using the CALIB task, assuming that  $RR^*=\tilde{\mathcal{I}}$ and $LL^*=\tilde{\mathcal{I}}$. This is necessary because \difmap does not calibrate R and L polarization hands separately. The resulting solution (SN) table ideally includes only a small residual correction for R and L that represents their gain offset for each antenna in the array. This SN table is processed outside AIPS to edit the $R$ and $L$ amplitude and phases with the assumption that the R/L gains of the ALMA station equals unity and hence any $R/L$ ratio seen by ALMA is ideally intrinsic to the source. Under this assumption, the $R$ and $L$ visibility data for all antennas are modified accordingly so that the average $R/L$ ratios better match the $R/L$ ratio seen by ALMA. We note that this approach corrects only for a $R/L$ offset for each antenna and is insensitive to possible time variations of the $R/L$ gains. The modified SN table is read back to AIPS with the task TBIN and used to produce the final data files (uvfits), using tasks CLCAL and SPLIT, which should be calibrated for both circular and linear polarization. Such calibrated uv-data are reloaded in \difmap and imaged in Stokes $\mathcal{V}$ using the set of windows employed for the Stokes $\mathcal{I}$ image. This is the procedure that we have used to produce Stokes $\mathcal{V}$ images for the 2017 \m87 data and synthetic data analyzed in this work. We have used the noise level (RMS) multiplied by a factor of 3 of Stokes $\mathcal{I}$ and $\mathcal{V}$ images as the uncertainties of the total and circular polarized intensities, respectively, to derive
uncertainties of parameters presented in \autoref{sec:synthdata} and \autoref{sec:m87results}.

\subsection{\polsolve}
\label{sec:polsolve}

The \polsolve algorithm is also based on CLEAN, but for polarimetric calibration uses the full Measurement Equation \citep[e.g.,][]{Smirnov_2011}, including (second order) non-linear corrections to the instrumental polarization \citep[see][for full details]{polsolve}. While the original \polsolve algorithm was not designed to include a Stokes $\mathcal{V}$ source model in the fitting of the instrumental polarization, the flexibility of CASA makes it possible to account for circular polarization by manually filling the {\em model} column of the measurement set and asking \polsolve to use it in the fitting. This strategy corresponds to the ``polarimetric self-calibration'' approach described in detail in \citet{polsolve}.

For the recovery of the Stokes $\mathcal{V}$ image with \polsolve, we have followed this iterative polarimetric self-calibration approach, starting from the data calibrated as described in \citetalias{PaperVII}. First, we have run the CASA-based CLEAN on the four Stokes parameters, by using a common mask based on the Stokes $\mathcal{I}$ image. The parameters used in this CLEAN are the same as those summarized for \polsolve in \citetalias{PaperVII}. Each run of (full-polarization) CASA-based CLEAN was followed by a (polarization-independent) self-cali\-bra\-tion (i.e., gain solutions were computed based on Stokes $\tilde{\mathcal{I}}$ only, using $RR^*$ and $LL^*$). Then, a $D$-term fitting with \polsolve was performed (this time, the full visibility matrix was used). This procedure was iterated until convergence (i.e., until the changes in the antenna gains and $D$-terms between consecutive iterations were negligible). After convergence, a final CLEAN image was generated.
This approach was the same for all the data presented in this publication (i.e., all the synthetic data tests and all the EHT observations).

\subsection{\dmc and \themis}
\label{sec:DMCThemis}

\themis \citep{Themaging:2020} and \dmc \citep{DMC} produced Stokes $\mathcal{V}$ maps concurrently with the linear polarization maps presented in \citetalias{PaperVII}, and details about the analyses are provided there.  Here, we briefly summarize both methods and emphasize the differences that are most relevant for imaging circular polarization.

\themis and \dmc formulate the imaging problem as one of Bayesian posterior exploration over the combined space of both image structures and station-based calibration quantities (i.e., complex gains and leakage terms).  The output of both codes is a set of MCMC samples from the joint posterior distribution over both the full-Stokes image structure and the calibration quantities.

For the purposes of Stokes $\mathcal{V}$ reconstruction, the most salient difference between \themis and \dmc is in their treatment of the relative right- and left-handed station gain quantities.  Whereas \themis holds that the right- and left-handed station gains are identical, \dmc explores potentially large differences between them.  With the exception of a reference station -- chosen here to be ALMA -- that has both right- and left-hand gain phases fixed to be zero-valued at all times, \dmc independently models the right- and left-hand complex gain quantities for each scan and each station.

\subsection{Geometric modeling}
\label{sec:geometric}

Geometric modeling differs from imaging methods because the reconstructed source structure is restricted to the space defined by a family of simple geometric models. 
Geometric modeling has been used to infer total intensity source structure parameters for \m87 and \sgra, generally showing excellent consistency with the results of imaging \citep{PaperVI, SgrAEHTCIV}. 
The restricted parameter space in geometric modeling is generally easier to constrain than that of many-parameter image reconstructions and their associated regularizers. Hence, geometric modeling provides a useful complement to imaging, especially to analyze observations with sparse $uv$-sampling. However, geometric modeling results will not be accurate if the true source structure is not represented in the model space, so care must be taken when modeling sources where the source structure is unknown and unconstrained from other methods.

In this work, we fit a full-Stokes $m$-ring model  \citepalias{Roelofs_2023} to EHT \m87 data from 2017. The total intensity $m$-ring model \citep{Johnson_2020, SgrAEHTCIV} consists of a ring with diameter $d$, with azimuthal brightness variations that are decomposed in Fourier modes. The image is given by
\begin{align}
\label{eq::Ring_Image}
\mathcal{I}(\rho,\varphi) &= \frac{F}{\pi d}\delta\left(\rho-\frac{d}{2}\right) \sum_{k=-m_{\mathcal{I}}}^{m_{\mathcal{I}}} \beta_k e^{ik\varphi}.
\end{align}
Here, $\rho$ and $\varphi$ are polar sky coordinates centered on the ring, $\beta_k$ are the complex Fourier coefficients, $F$ is the total flux density (we set $\beta_0\equiv1$), and the total intensity $m$-ring order $m_{\mathcal{I}}$ is the maximum nonzero Fourier coefficient. By setting a higher $m_{\mathcal{I}}$, the total number of model parameters increases and an increasingly complex azimuthal structure can be modeled. Because the total intensity is real, $\beta_k=-\beta_k^*$. The $m$-ring is given a finite width by convolving $\mathcal{I}(\rho,\varphi)$ with a circular Gaussian with FWHM $\alpha$. A useful property of this model is that both the image and its corresponding visibility function are straightforward to calculate analytically \citepalias[see, e.g.,][]{Roelofs_2023}.  

The $m$-ring model can be straightforwardly generalized to include all Stokes parameters. In linear polarization, the azimuthal variations are set by $\left\{ \beta_{\mathcal{P},k} \right\}$. Because the linear polarization image is complex, $\beta_{\mathcal{P},k}$ and $\beta_{\mathcal{P},k}^*$ are not related by conjugation symmetry. In circular polarization, the azimuthal variations are set by $\left\{ \beta_{\mathcal{V},k} \right\}$. Because the circular polarization image is real, $\beta_{\mathcal{V},k}=-\beta_{\mathcal{V},k}^*$.

For our $m$-ring fits to synthetic and real EHT data, we use the geometric modeling module in \ehtim \citep{Chael_2016, Chael_2018_Imaging}. In this work, we fit an ($m_{\mathcal{I}}=3, m_{\mathcal{P}}=3, m_{\mathcal{V}}=1$) $m$-ring to the parallel-hand closure amplitudes and closure phases. 
Posterior exploration is done with the ${\tt dynesty}$ sampler \citep{Speagle2020}. We first fit the total intensity and linear polarization structure, after which we fix the linear polarization parameters and fit the total intensity and circular polarization structure simultaneously. Because the closure products do not constrain the net circular polarization $\beta_{\mathcal{V},0} \equiv v_{\rm net}$, we fix this parameter to the ALMA-only estimate of $-0.3\,\%$ \citep{Goddi2021} for the EHT \m87 data, and to the ground-truth values for the synthetic datasets. Our fitting procedure and assumptions are motivated in more detail in \citetalias{Roelofs_2023}, where additional tests are presented, including varying the circular polarization $m$-ring order $m_{\mathcal{V}}$, varying the assumed net circular polarization fraction ($m_{\rm c} \equiv \beta_{\mathcal{V},0}$), and varying the fitted data products (e.g., using closure quantities versus using parallel-hand complex visibility ratios).

%=================================================================================================================================================================
\section{Synthetic Data Tests}
\label{sec:synthdata}
%=================================================================================================================================================================
\begin{figure*}[t]
\centering
\includegraphics[width=\textwidth]{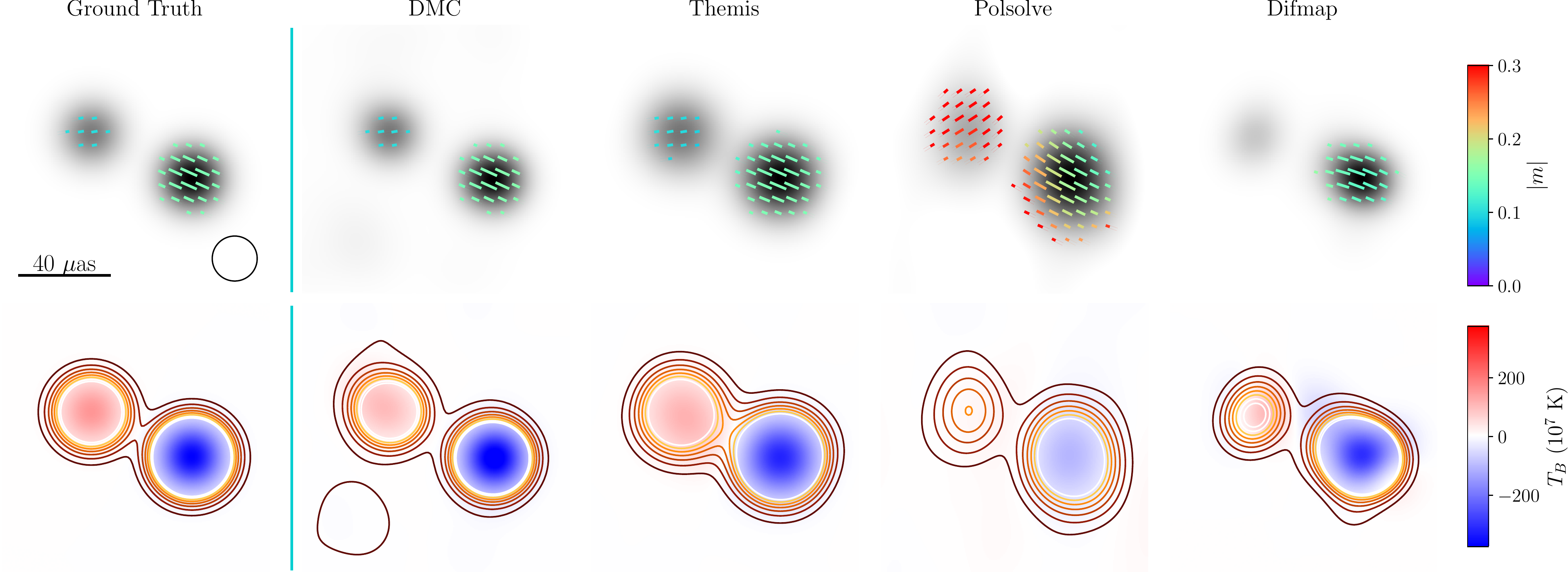}
\caption{Imaging results from the convention test data set, consisting of a highly polarized ($\mathcal{V}_{\rm pk}/\mathcal{I} \approx 10$\%) double source. Images are plotted in the same manner as in \autoref{fig:synthetictests}.} 
\label{fig:conventiontest}
\end{figure*}

\begin{figure*}[t]
\centering
\includegraphics[width=\textwidth]{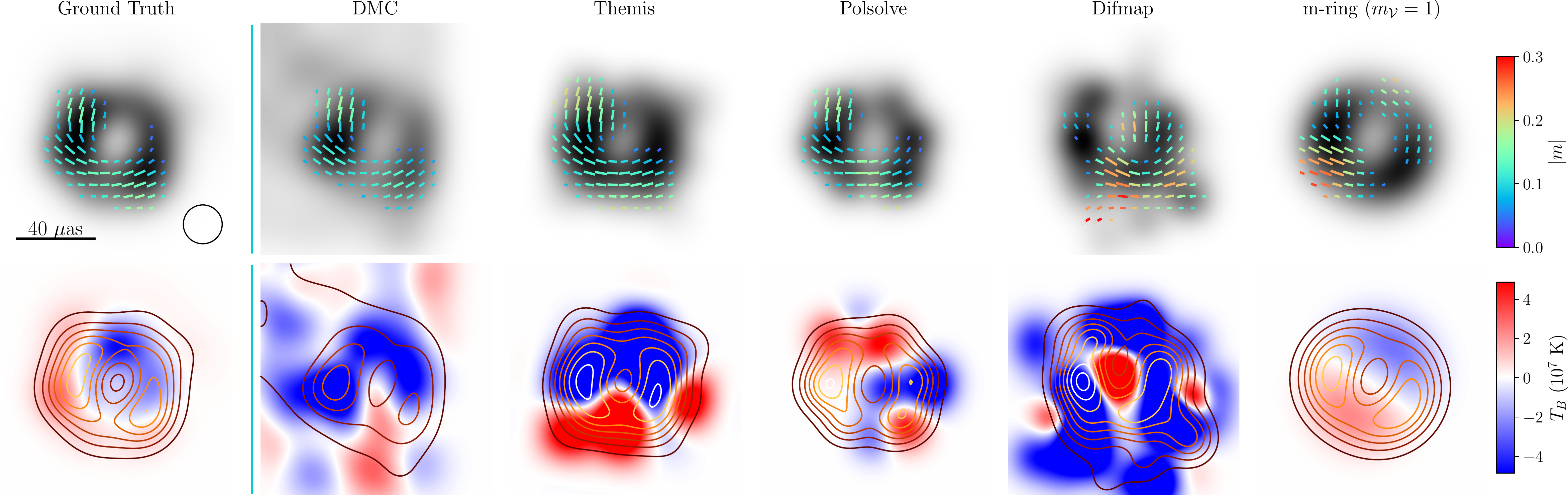} \\
{\color{gray}\rule{\textwidth}{2pt}} \\
\includegraphics[width=\textwidth]{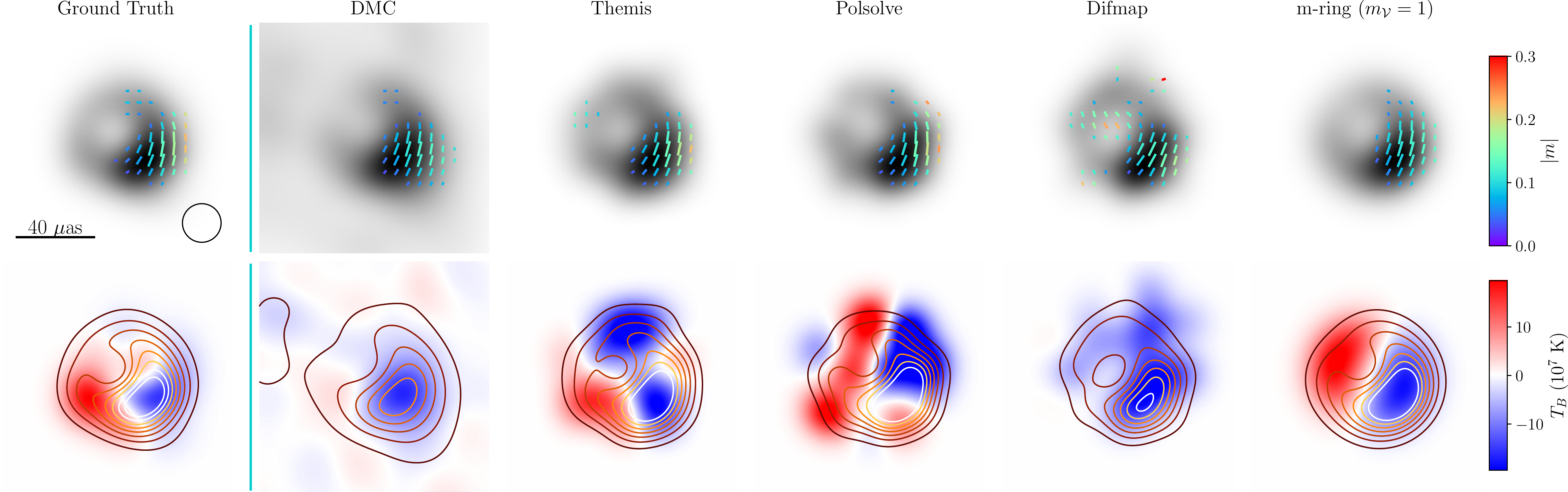} \\
{\color{gray}\rule{\textwidth}{2pt}} \\
\includegraphics[width=\textwidth]{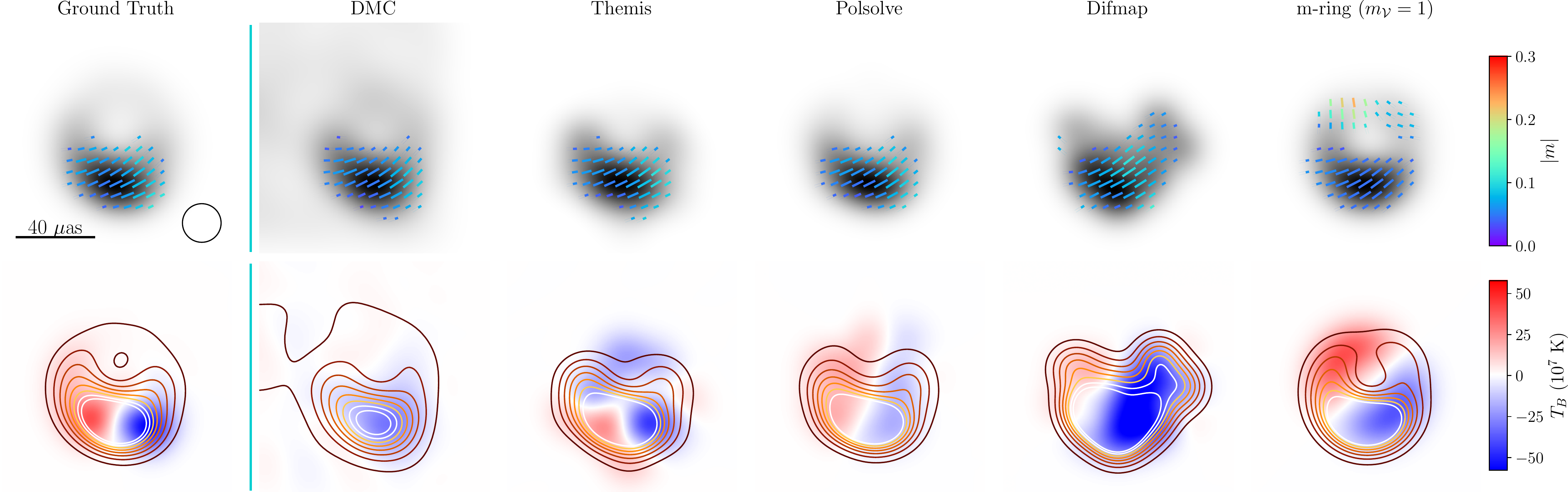} 
\caption{Imaging results from the three synthetic data sets in \autoref{tab:synmodel_summary}. The first row shows the total intensity in grayscale and the fractional linear polarization in the colored ticks. The left column is the ground truth simulation image. From left to right, the next columns show the posterior average images from \dmc and \themis, the final CLEAN images from \polsolve and \difmap, and the posterior average of the $m-$ring model fits. All images are blurred with same $20\mu$as FWHM circular Gaussian beam. The second row shows the total intensity image with 8 contour levels on a linear scale
and the Stokes $\mathcal{V}$ structure in a a diverging colormap. Rows one and two show the results for model 01 (low resolved circular polarization), rows three and four show the results for model 02 (moderate resolved circular polarization), and rows five and six show the results for model 03 (high resolved circular polarization). Note that the colorbar for $\mathcal{V}$ has different maximum values in rows two, four, and six as the GRMHD simulation images become more polarized.
} 
\label{fig:synthetictests}
\end{figure*}

In this Section, we test the ability of our image reconstruction and geometric modeling methods to recover resolved circular polarization from an idealized, high-circular-polarization ``convention test'' as well as representative images from GRMHD simulations of \m87. 

\subsection{Synthetic data generation}
\label{sec:synmodels}

As in \citetalias{PaperVII}, we generated synthetic observations on the \m87 2017 April 11 low-band $(u,v)$ coverage using the \texttt{eht-imaging} software package. After sampling the image Fourier transforms on EHT baselines, the synthetic data were corrupted with time-variable, station-based systematic phase and gain offsets, time-stable polarimetric leakage terms, and thermal noise. 

Unlike in the \citetalias{PaperVII} synthetic data, in this paper we do not assume the complex gains in the right and left circular polarizations are equal: $G_R \neq G_L$. Instead, we introduce time-variable offsets in both the amplitude and phase of the ratio $G_{R/L}$. The magnitude of these offsets and their characteristic timescales were motivated by the a-priori limits described in \autoref{sec:polarimetric_gains}. For most sites, we sample the R/L amplitude gain offset from a normal distribution centered at unity with a standard deviation of $20\,\%$, and the R-L phase offset from a zero-centered normal distribution with a standard deviation of $10\,\deg$. The exceptions are ALMA (where $G_R=G_L$) and the SMA, where a priori limits on the phase offset motivate a use of a $40\deg$ standard deviation. The phase and amplitude gain offsets are correlated in time (as are the overall amplitude and phase gain terms), and we set the correlation timescales for the offsets to be larger than the correlation timescales for the overall amplitude (2 hr vs 1 hr) and phase gains (24 hr vs 0.5 hr). 

\subsection{Convention test results}
Before imaging the GRMHD models presented in \autoref{sec:synresults}, we first tested our methods on an optimistic ``convention test''; a strongly circularly polarized ($v \approx 10\,\%$) double Gaussian source with 1.2\,Jy total flux density. 
In this test, instead of a GRMHD snapshot, a synthetic source image was constructed from two Gaussians with FWHM$\approx20\mu$as separated by a distance $d\approx50\mu$as. The total flux density of the double source was 1.2 Jy, and the fractional circular polarization of each Gaussian component was high, $V_{\rm pk}/I\approx 10$\%. 
We show the results of this initial test in \autoref{fig:conventiontest}.

All participating methods (the $m-$ring modeling group did not participate in this test) were able to recover the two-component structure as well as the approximate degree of circular polarization in each component,
even with systematic gain offsets $G_{R/L}$ present in the synthetic data at similar levels to the GRMHD data sets discussed in the following section. This test demonstrates that our analysis methods would have been able to recover consistent, accurate resolved Stokes $\mathcal{V}$ images from 2017 EHT observations of \m87 if the intrinsic circular polarization brightness was $\approx5-10$ times higher than observed in 2017.

\subsection{GRMHD results}
\label{sec:synresults}

\begin{figure*}[t]
\centering
\includegraphics[width=.9\textwidth]{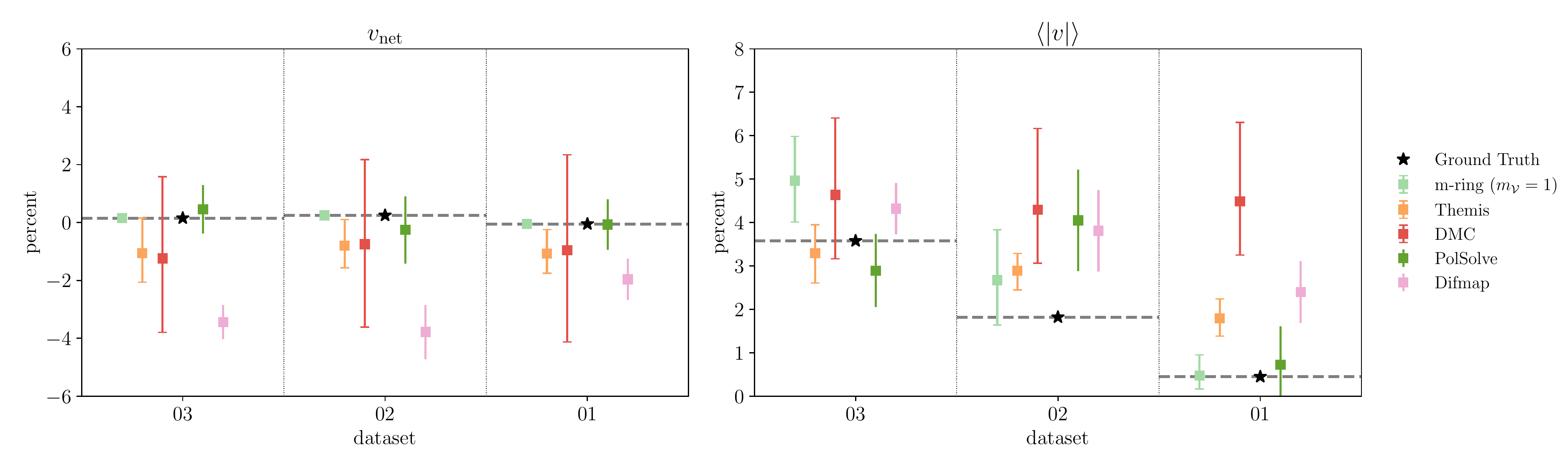} 
\caption{Image-integrated statistics from the results of the synthetic data tests presented in \autoref{fig:synthetictests}. The left panel shows the net circular polarization fraction $v_{\rm net}$ computed from each method for the three synthetic datasets and the right panel shows the average resolved circular polarization fraction $\langle|v|\rangle$ computed after blurring each image with a 20\,$\mu$as kernel. The results from the posterior exploration methods are presented with the median value and 2\,$\sigma$ error bars (note that $m$-ring modeling strictly enforces that $v_{\rm net}$ is equal to the ground truth).   
The CLEAN based methods produce only one image and so only one value of each metric is reported, but 2\,$\sigma$ error bars for \difmap and \polsolve are shown from standard error propagation using the off-source residuals in $\mathcal{V}$ and $\mathcal{I}$.
The ground truth value of each statistic  is indicated with the black star and horizontal dashed line.
}
\label{fig:stats_synthetic}
\end{figure*}

\begin{figure*}[t]
\centering
\includegraphics[width=\textwidth]{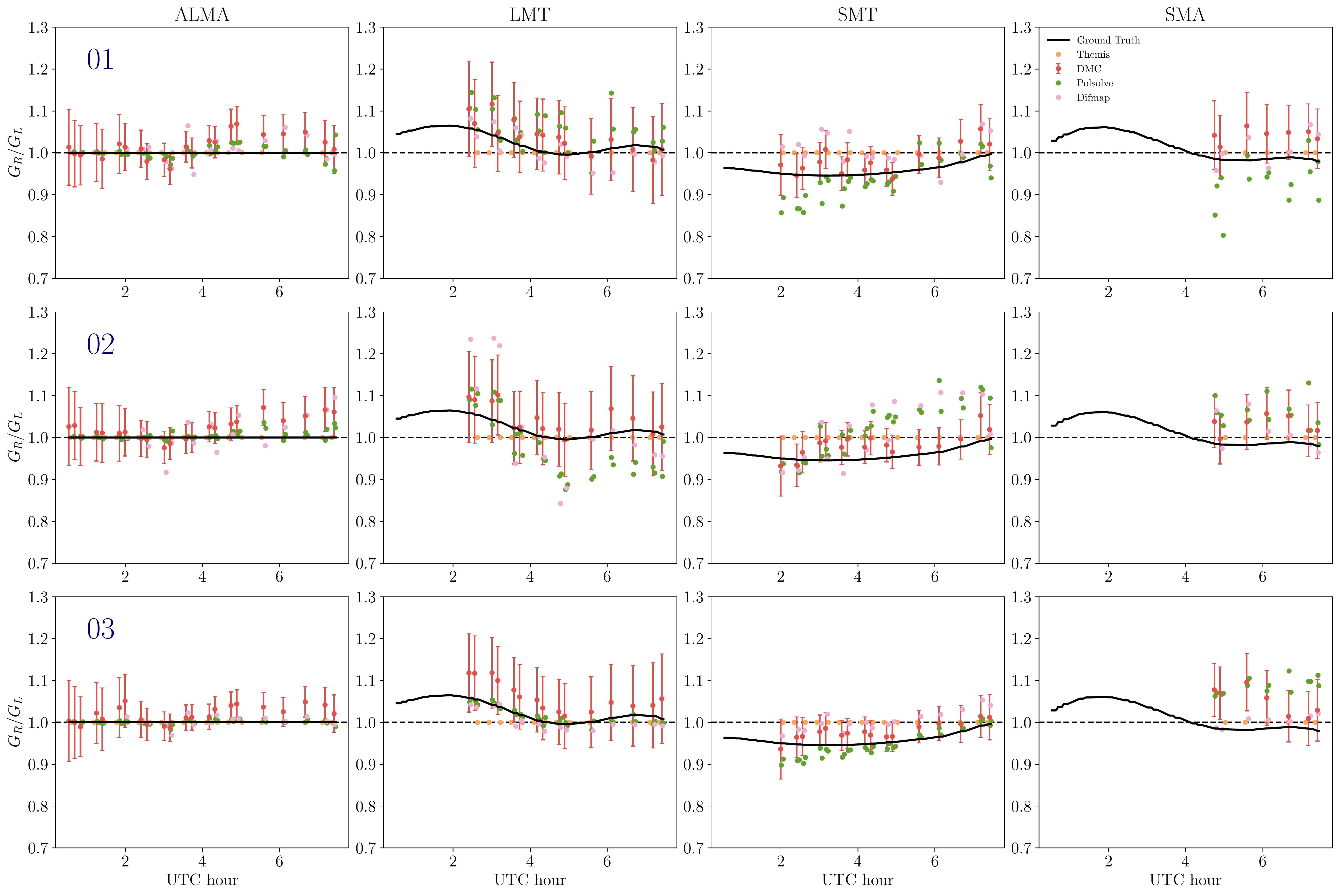} 
\caption{Derived gain ratios $|G_R|/|G_L|$ from different imaging methods applied to synthetic dataset 01 (top), 02 (middle), and 03 (bottom). From left to right - gain ratios are reported for ALMA, the LMT, the SMT, and the SMA sites. The actual applied gain ratios in the synthetic data set (motivated by the limits discussed in \autoref{sec:polarimetric_gains}) are displayed in black. \dmc gains are reported with $2\sigma$ error bars.
}
\label{fig:gains_synthetic}
\end{figure*}

To test our imaging methods on more realistic source models for \m87, we selected three GRMHD images from \citetalias{PaperVIII} that pass all the observational constraints from \citetalias{PaperVII}. We sorted these images by the value of $\langle|v|\rangle$ after blurring to $20\,\mu$as resolution.
We then chose three snapshot images from across this distribution to represent weak, average, and near-maximal $\langle|v|\rangle$ from the distribution of passing \citetalias{PaperVIII} images. All of these images have low image integrated circular polarization $|v_{\rm net}|<0.3\,\%$; images that have stronger polarization at $20\,\mu$as scales have significant regions of both positive and negative circular polarization that cancel in unresolved observations. 

Our three models are: the ``weak polarization'' case (model 01) with an average circular polarization fraction at $20\,\mu$as resolution of $\langle |v| \rangle$ $\approx 0.5\,\%$; the ``moderate polarization'' case (model 02) with $\langle |v| \rangle$ $\approx 2\,\%$; and the ``strong polarization'' case (model 03) with $\langle |v| \rangle$ $ \approx 4\,\%$. In \autoref{tab:synmodel_summary}, we summarize these three models, including the magnetic field configuration and spin of their underlying GRMHD simulation, the net 230 GHz circular polarization fraction $v_{\rm net}$, and the average resolved fraction $\langle|v|\rangle$.

\begin{table}[h]
    \centering
    \begin{tabular}{@{}l|lc|cc}
        \toprule 
        Model & $B-$Field & $a*$ & $v_{\rm net}$ (\%) & $\langle|v|\rangle$ (\%) \\
        \midrule
        01 & MAD & -0.5 & -0.05 & 0.5 \\
        02 & MAD & 0.5 & +0.25 & 1.75 \\
        03 & SANE & 0 & +0.15 & 4.1 \\
        \bottomrule
    \end{tabular}
    \caption{Summary of synthetic data models.}
    \label{tab:synmodel_summary}
\end{table}

\autoref{fig:synthetictests} shows the GRMHD ground truth models and reconstructions from our three synthetic datasets (\autoref{sec:synthdata}), for each of the five imaging or modeling methods described in \autoref{sec:methods}. Because all three GRMHD synthetic data models were chosen from the set of passing \citetalias{PaperVIII} models, all feature relatively similar total linear polarization fractions and EVPA structure, similar to those observed in \m87 by the EHT in \citetalias{PaperVII}. 

For all models, the total intensity and linear polarization structure (top rows) are generally recovered well by all methods. Because the synthetic data included both large overall amplitude gains $G$ on certain stations, the total flux density is not completely constrained by these datasets (as is the case for the real EHT 2017 \m87 data).
The Stokes $\mathcal{V}$ reconstructions (bottom rows) show significant variation between the methods, and most methods have difficulty reconstructing the ground-truth features in circular polarization. 

All methods struggle the most with model 01, the weakly polarized case $(\langle|v|\rangle\approx0.5\,\%)$. In this model, no method correctly recovers the spatial distribution of the sign of circular polarization, and due to the presence of residual relative gains $G_{R/L}$ in the data some methods significantly overproduce the total Stokes $\mathcal{V}$ brightness in the image. In contrast, for the ``high-polarization'' model 03,  $(\langle|v|\rangle\approx4\,\%)$, all methods are able to recover the resolved distribution of $\mathcal{V}$ somewhat more accurately. \themis, \polsolve, and $m$-ring modeling all recover the correct dipolar Stokes $\mathcal{V}$ structure of the image; \dmc and \difmap produce mostly one sign of $\mathcal{V}$ corresponding to the more strongly polarized negative component in the image. The results of model Model 02 are more mixed, with some methods (\dmc, \themis, and \difmap) accurately recovering the sign and approximate magnitude of $\mathcal{V}$ at the total intensity peak, while $m$-ring modeling and \polsolve see an overall dipole in the same orientation as the ground truth. No image reconstruction accurately captures the weak circular polarization in the northern half of the image. 

In all three synthetic models, $m$-ring modeling does the best job of consistently recovering the level of $\mathcal{V}$ in the image and the orientation of the dipolar structure across the ring. 
The limits on structural complexity baked into the $m_{\mathcal{V}}=1$ model likely helps the method lock on to the overall orientation of the source in these synthetic data tests, which all feature an underlying dipolar structure in $\mathcal{V}.$  While many GRMHD images show this dipolar structure at EHT resolution, some show more complicated patterns (see \autoref{sec:theo}). More details and other model choices applied to $m$-ring modeling of these synthetic data sources are presented in \citetalias{Roelofs_2023}). 

\autoref{fig:stats_synthetic} shows the recovery of the image-integrated and resolved circular polarization fractions for all datasets and methods. The low image-integrated circular polarization fraction is recovered decently well by all methods on the three sources. An exception is \difmap, which recovers a relatively large negative integrated circular polarization fraction. 
The $m$-ring modeling method is constrained by construction to produce the ground-truth $v_{\rm net}$. The recovered resolved circular polarization  $\langle|v|\rangle$ shows larger differences between methods and larger deviations from the ground truth. For the high circular polarization case 03, both \dmc and \themis have the ground truth value of $\langle|v|\rangle$ within their $2\sigma$ error bars, while the \polsolve and \difmap single-image results are off by $\approx 1\,\%$. In the moderate and low polarization models 02 and 01, all methods except $m$-ring modeling produce results biased significantly upward from the ground truth. This could indicate that residual uncertainty in the $G_{R/L}$ gains is entering into over-producing the resolved circular polarization in these methods with weak intrinsic $\mathcal{V}$ and low $\mathcal{V}$ S/N in the data, even after self-calibration during imaging. Furthermore, $\langle|v|\rangle$ is biased upwards by uncertainty in the underlying image reconstructions (see \autoref{app:CirPolFracBiases}).
Interestingly, $m$-ring modeling performs the best on datasets 01 and 02, accurately recovering $\langle|v|\rangle$ in both cases, while overproducing $\langle|v|\rangle$ in the ``easier'' case 03. This difference in recovery accuracy could be driven by the limited freedom of the $m$-ring model in the source structures it can recover; the ground-truth structure for model 03 is less dipolar and ring-like than for models 01 and 02, with both total intensity and polarized flux concentrated in the South.

\autoref{fig:gains_synthetic} shows the recovered amplitude gain ratios $|G_{R/L}|$ for selected stations on the three synthetic datasets. Note that \themis does not fit for these gain ratios, and $m$-ring modeling does not fit for any gains as it directly fits to closure quantities. The trends of $|G_{R/L}|$ as a function of time are generally recovered by all methods, although systematic offsets up to $\sim~10\,\%$ do occur. \dmc is the only method that returns uncertainty estimates on these gain ratios; the \dmc results are nearly all within $2\sigma$ of the ground truth. The inability of all methods to accurately recover these gain ratios even in the model with the most circular polarization (03) indicates that residual uncertainty in the solution for $G_{R/L}$ may significantly affect our $\mathcal{V}$ imaging results for \m87. 

%=================================================================================================================================================================

\begin{figure*}[t!]
\centering
\includegraphics[width=\textwidth]{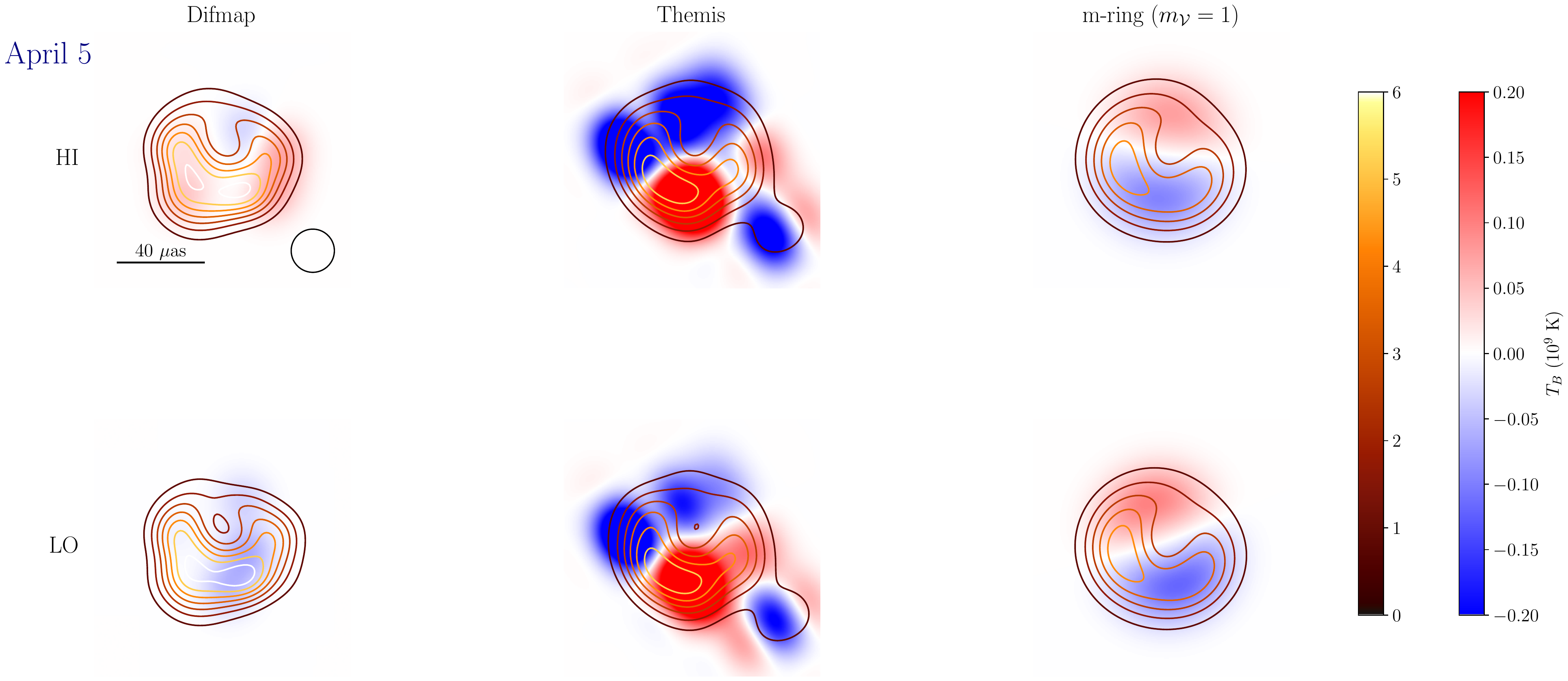} \\
\includegraphics[width=\textwidth]{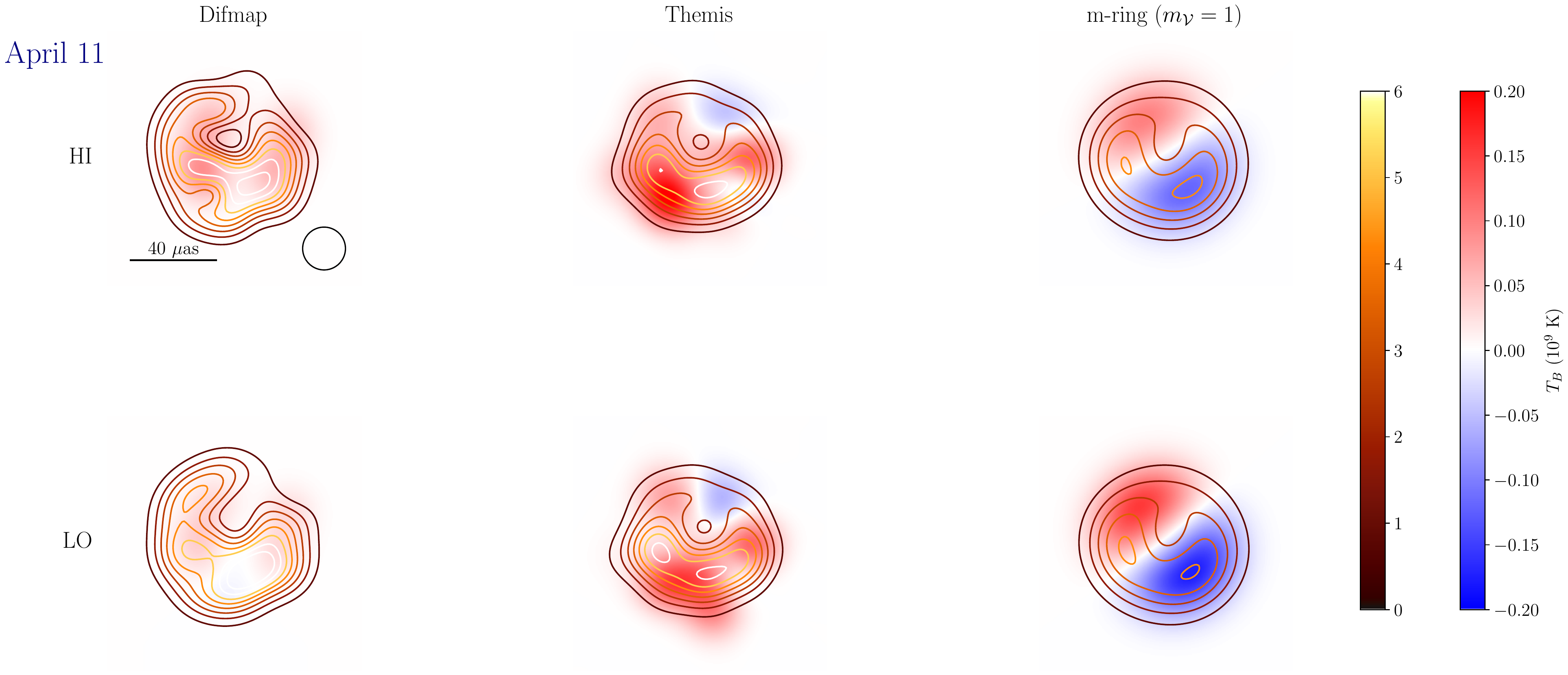} 
\caption{
\m87 Stokes $\mathcal{V}$ imaging results using data reduced with the CASA-based \texttt{rPICARD} pipeline \citep{Janssen_2019} and pre-calibrated with the assumption $\mathcal{V}=0$. The top two rows show results from three methods (\difmap, \themis, and $m$-ring geometric modeling) on April 5 observations, both low and high band. The bottom row shows corresponding results from April 11 observations. For \themis and $m$-ring modeling we show the posterior-average image. Images are plotted in the same manner as in \autoref{fig:images_56}.}
\label{fig:images_casa}
\end{figure*}

\section{M87* Results From CASA-calibrated Data}
\label{append:casa}

\autoref{fig:images_casa} presents Stokes $\mathcal{V}$  imaging results using data reduced with the CASA \texttt{rPICARD} pipeline \citep{Janssen_2019}. The CASA-reduced data calibrated the amplitude gain ratios $G_{R/L}$ assuming $\mathcal{V}=0$, while the fiducial \texttt{HOPS}-reduced data used to produce the results in \autoref{fig:images_56} and \autoref{fig:images_1011} used a multi-day and multi-source fit to calibrate $G_{R/L}$. By producing images with these data, we test if our main conclusions regarding the level of the resolved $\mathcal{V}$ and (in)consistency of the resolved structure in different reconstructions depend strongly on the choices and assumptions made in data reduction, particularly in the way $G_{R/L}$ are calibrated before imaging. 

Three of the imaging pipelines used in this paper (\difmap, \themis, and $m$-ring model fitting) produced Stokes $\mathcal{V}$ images using the CASA data for both high and low band on April 5 and April 11. The results presented in \autoref{fig:images_casa} show that the resulting Stokes $\mathcal{V}$ images from the CASA-calibrated are still inconsistent across reconstruction methods and observing days. Interestingly, the reconstructions using CASA data show somewhat more consistency across the high and low band images than in the reconstructions using the fiducial \texttt{EHT-HOPS} data (\autoref{fig:images_56}, \autoref{fig:images_1011}). This may be a result of the fact that the \texttt{rPICARD} pipeline combines information across the bands for the signal stabilization \citep{2022Janssen} of the data, while the bands are treated entirely separately in the \texttt{EHT-HOPS} reduction. 
\citetalias{Roelofs_2023} presents an analysis of the CASA data $m$-ring reconstructions compared to the \texttt{EHT-HOPS} data reconstructions in more detail and without any $G_{R/L}$ amplitude gain calibration. 

%=================================================================================================================================================================

\section{Biases in Resolved Circular Polarization Fraction}
\label{app:CirPolFracBiases}

For an image composed of $N_\text{res}$ pixels or resolution elements, each of which has angular area $\Delta A$, then \autoref{eq:vavg} can be expressed as a sum

\begin{equation}
\langle | v | \rangle = \frac{\Delta A}{F} \sum_{i=1}^{N_{\text{res}}} \left| \frac{\mathcal{V}_i }{\mathcal{I}_i } \right| \mathcal{I}_i ,
\end{equation}

\noindent where $F$ is the total integrated Stokes $\mathcal{I}$ flux density in the image, and $i$ is a subscript that indicates a particular pixel or resolution element.  If $\mathcal{I}$ is everywhere positive in the image, then $\left| \mathcal{V}_i / \mathcal{I}_i \right| \mathcal{I}_i = |\mathcal{V}_i|$ and we can simplify the expression for $\langle | v | \rangle$ to be

\begin{equation}
\langle | v | \rangle = \frac{\Delta A}{F} \sum_{i=1}^{N_{\text{res}}} \left| \mathcal{V}_i \right| .
\end{equation}

If $\mathcal{V}_i$ is purely noise-like and distributed according to a normal distribution with mean zero and variance $\sigma_i^2$, then the probability distribution function of $|\mathcal{V}_i|$ is given by

\begin{equation}
|\mathcal{V}_i| \sim \begin{cases}
\sqrt{\frac{2}{\pi \sigma_i^2}} \exp\left( -\frac{\mathcal{V}_i^2}{2 \sigma_i^2} \right) , & |\mathcal{V}_i| \geq 0 \\
0 , & |\mathcal{V}_i| < 0
\end{cases} .
\end{equation}

\noindent The mean of this distribution is $\mu_i = \sigma_i \sqrt{2/\pi}$, which is nonzero.  The mean of $\langle | v | \rangle$ will then be given by

\begin{equation}
E\left[ \langle | v | \rangle \right] = \frac{\Delta A}{F} \sqrt{\frac{2}{\pi}} \sum_{i=1}^{N_{\text{res}}} \sigma_i .
\end{equation}

\noindent If each pixel or resolution element has a similar noise variance $\sigma_i^2$, then we can see that $E\left[ \langle | v | \rangle \right]$ will be proportional to $N_{\text{res}}$. Precisely, the value of the resolved circular polarization fraction depends on the number of pixels or resolution elements contained in the image, which means that it will be sensitive to things such as the size of a blurring kernel or the field of view of the image.
%=================================================================================================================================================================
\section{Additional GRMHD Library Generation Details}
\label{app:lib_details}

The GRMHD models in this paper were computed using the code {\sc HARM} \citep{2003ApJ...589..444G,Noble2006} and were first presented in \citetalias{PaperV}.  Simulations were initialized with an idealized Fishbone-Moncrief torus of gas in hydrostatic equilibrium \citep{1976ApJ...207..962F}, within which the magneto-rotational instability \citep{1991ApJ...376..214B, 1998RvMP...70....1B} naturally develops, driving accretion.  Our GRMHD simulations are mainly characterized by two parameters.  The first is the black hole spin $a_*$ ranging from 
\mbox{$ -1 < a_* < 1$}
with positive/negative values representing alignment/anti-alignment between the engines angular momenta.  We only consider models where the BH angular momentum and the disk angular momentum on large scales is either perfectly aligned or anti-aligned, where anti-alignment is denoted with a negative sign.  The second is the absolute magnetic flux in its dimensionless form $\phi \equiv \Phi_\text{BH} (\dot{M}r_g^2c)^{-1/2}$, where $\Phi_\text{BH}$ is the magnetic flux' magnitude traversing one hemisphere of the event horizon \citep[see, e.g.][]{porth_2019} and the mass accretion rate $\dot{M}$ set over the event horizon.  The relative strength of the magnetic flux near the horizon is determined by the dimensionless quantity $\phi$ and affects the dynamics of the accretion flow solutions. This led to a division of the solutions into two categories: the Magnetically Arrested Disk (MAD) state which highly affects the accretion dynamics $\phi \geq 50$ \citep[e.g.][]{tchekhovskoy_2011,mckinney_2012,narayan_2022}, and the Standard and Normal Evolution (SANE) state $\phi \approx 1-5$ (in Gaussian units). SANE simulations were produced on a $288 \times 128 \times 128$ grid, a fluid adiabatic index $\gamma=4/3$, and a boundary of $r_\text{out}=50\,r_g$.  MAD simulations adopt a $384 \times 192 \times 192$ resolution, $\gamma=13/9$, and a domain of $r_\text{out}=10^3\,r_g$.  

To produce images via general relativistic radiative transfer (GRRT), we use the code {\sc ipole} \citep{Moscibrodzka&Gammie2018}.  First, null geodesics are calculated backwards from the camera through the source.  Then, radiative transfer is calculated forwards accounting for polarized synchrotron emission and self-absorption, as well as Faraday rotation and conversion.  We use radiative transfer coefficients appropriate for a thermal relativistic plasma \citep{2011ApJ...737...21L}, with the updated fitting functions from \citet{Marszewski+2021}.  The fast-light approximation is made during ray-tracing. 

In ideal GRMHD, the fluid density, magnetic field strength, and internal energy can be rescaled during post-processing, and we do so to achieve an average flux density of $F_\nu \approx 0.5\,$Jy for our images at $\nu=230\,$GHz \citep[e.g.,][]{PaperV}.  In addition to the fluid scaling and BH parameters, there are three other free parameters we explore at this stage.  Two of them, $R_\mathrm{low}$ and $R_\mathrm{high}$ are associated with the ion-to-electron temperature ratio, where the electron temperature is defined as 
\begin{align}
\label{eq::T_e}
T_e = \frac{2 m_e u_\text{gas}}{3 \rho k_B \left(2 + R\right)}.
\end{align}
In order for the description in Equation \eqref{eq::T_e} to be valid, the ions are assumed to be non-relativistic while electrons be relativistic.
Adopting a prescription in \citet{Moscibrodzka:2016}, the ion-to-electron temperature ratio $R$ is given by
\begin{align}
\label{eq::Temp}
R = \frac{T_i}{T_e} = R_\text{high}\frac{\beta^2}{1+\beta^2} + R_\text{low} \frac{1}{1+\beta^2}
\end{align}
where the components describe the local plasma $\beta=p_\text{gas}/p_\text{mag}$ with $p_\text{gas}=(\gamma -1)u_\text{gas}$, $p_\text{mag} = B^2/8\pi$, and further $u_\text{gas}$ being the total internal energy density. To regulate the ratio in Equation \eqref{eq::Temp}, our models explore $R_\text{low}=1, 10$ and $R_\text{high}=1, 10, 20, 40, 80, 160$ as in \citetalias{PaperVIII}.

Such a parameterization to treat the model as thermal flow is motivated by the fact that mean free path is much larger than the Debye-Length and Larmor radius of the jet’s plasma.
This assumption together with Equation \eqref{eq::Temp} mimic collisionless plasma properties and allows to associate the electron heating with its magnetic properties \citepalias[see][]{PaperVIII}.

%=================================================================================================================================================================
\section{Field Reversal and Linear Polarization}
\label{app:field_reversal}
As we discuss in \ref{sec:lib}, reversing the polarity of the magnetic field has significant effects on circular polarization, both on resolved and unresolved scales.  Our previous studies including linear polarization of this source only included models wherein the poloidal magnetic field and the angular momentum of the disk are aligned.  Here, we briefly explore the effect of flipping the magnetic field on linear polarization metrics.  Overall, we find the effect is minor and does not have a strong impact on linear polarization constraints.

\begin{figure*}
    \centering
    \includegraphics[width=0.9\textwidth]{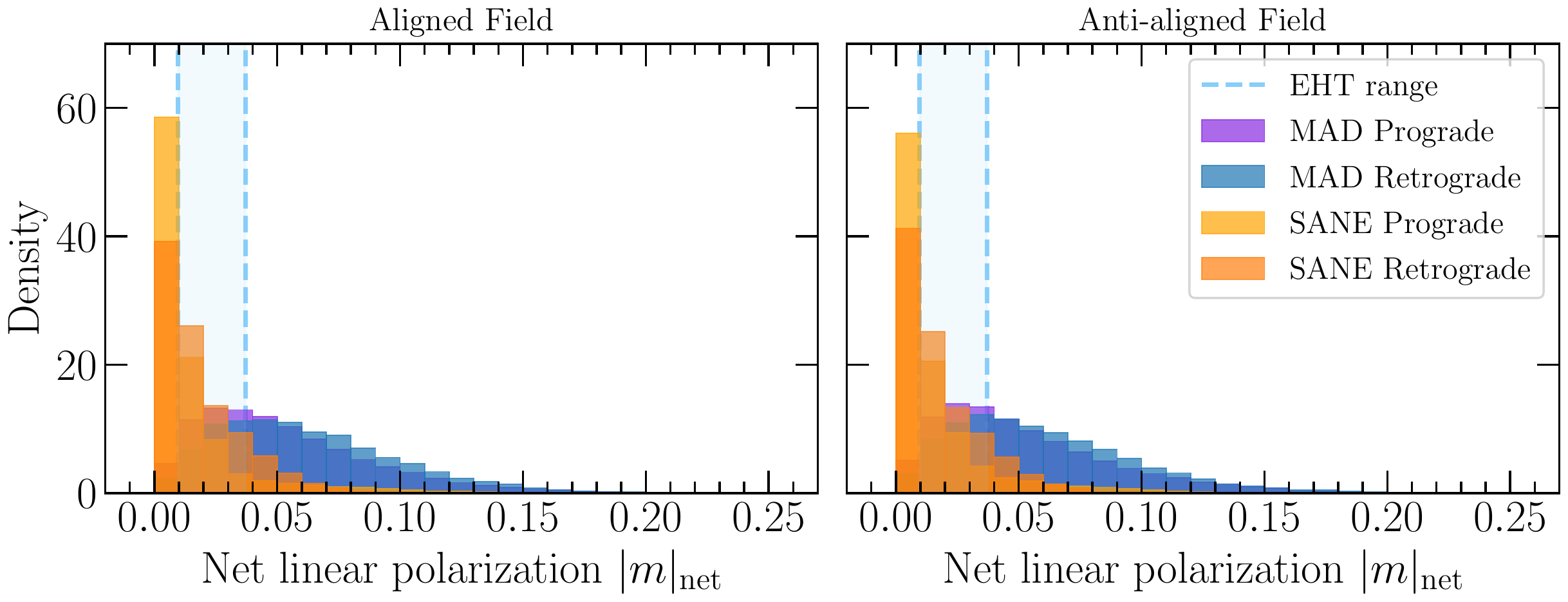} 
    \caption{Distribution of image-integrated net linear polarization fraction with an aligned (left panel) and anti-aligned (right panel) magnetic field respecting all images in the \m87 library. Allowed inferred ranges of EHT image reconstructions are limited by the dashed lines.}
    \label{fig:mnet}
\end{figure*}

\begin{figure*}
    \centering
    \includegraphics[width=0.9\textwidth]{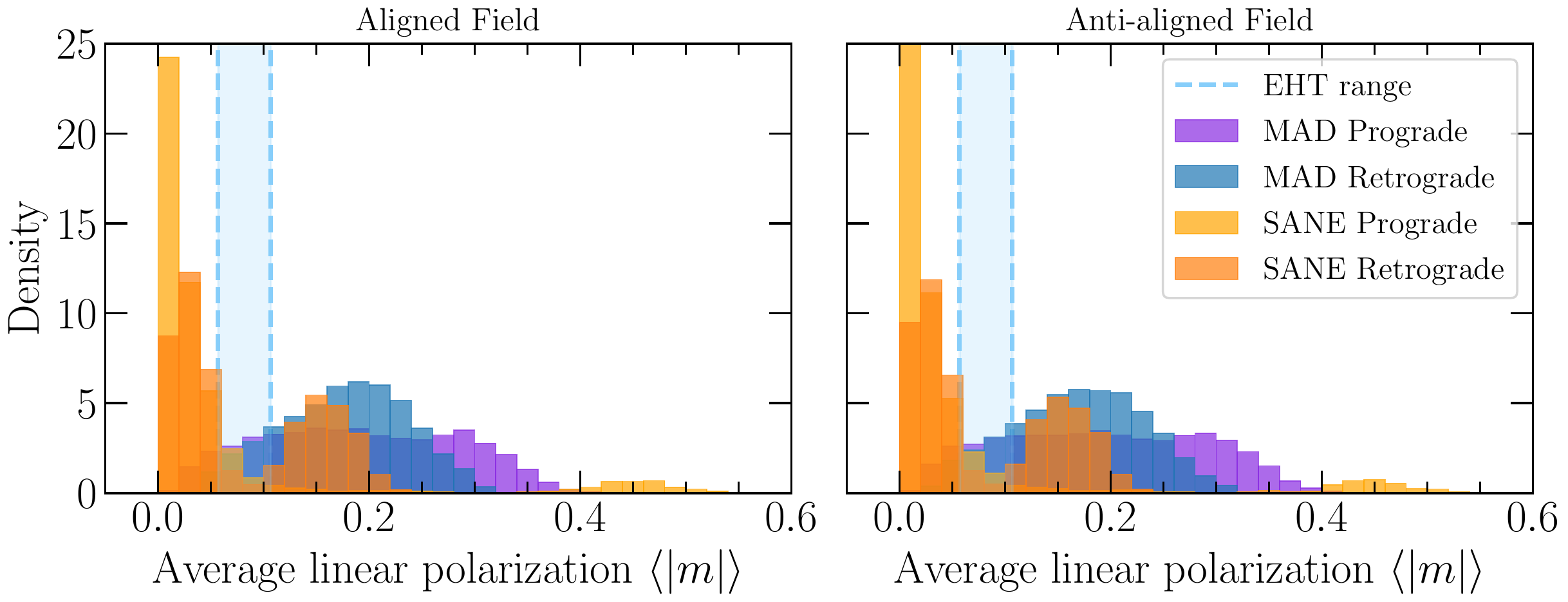} 
    \caption{Distribution of image-averaged fractional linear polarization $\langle |m| \rangle$ with an aligned (left panel) and anti-aligned (right panel) magnetic field respecting all images in the \m87 library blurred with a 20$\,\mu$as FWHM Gaussian beam. Allowed inferred ranges of EHT image reconstructions are limited by the dashed lines.}
    \label{fig:mavg}
\end{figure*}

\begin{figure*}
    \centering
    \includegraphics[width=0.9\textwidth]{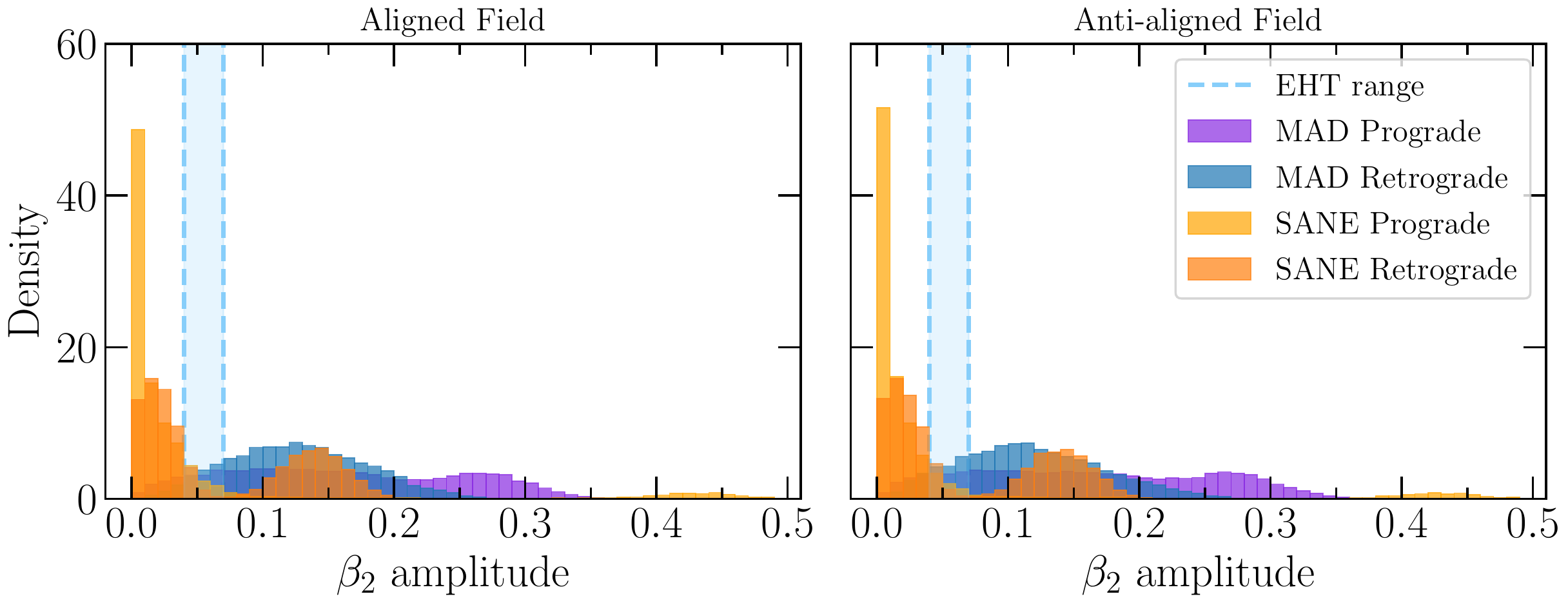} 
    \caption{Distribution of $\beta_2$ amplitude for an aligned (left panel) and anti-aligned (right panel) magnetic field representing values taken from all images in the EHT \m87 library blurred with a $20\,\mu$as beam. Allowed inferred ranges of EHT image reconstructions are limited by the dashed lines.}
    \label{fig:beta2amp}
\end{figure*}

\begin{figure*}
    \centering
    \includegraphics[width=0.9\textwidth]{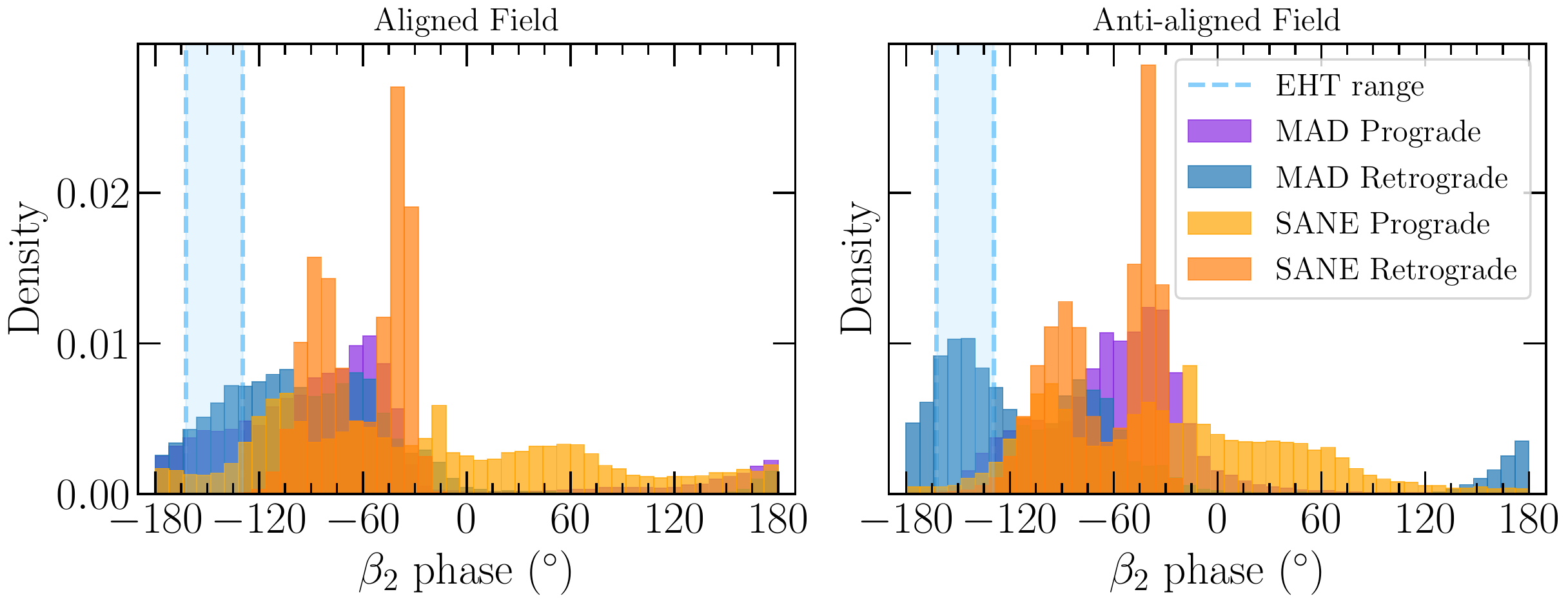} 
    \caption{Distribution of $\beta_2$ phase for an aligned (left panel) and anti-aligned (right panel) magnetic field representing values taken from all images in the EHT \m87 library blurred with a $20\,\mu$as beam. Allowed inferred ranges of EHT image reconstructions are limited by the dashed lines.}
    \label{fig:beta2ang}
\end{figure*}

In Figures \ref{fig:mnet}-\ref{fig:beta2ang}, we plot histograms of observable quantities considered in \citetalias{PaperVIII} for the aligned case (as in previous work) on the left, and for the anti-aligned case on the right.  As expected, most of the linear polarization metrics are insensitive to this flip.  The quantity which shows the most significant, albeit minor, effect is $\angle \beta_2$.  As explored in more detail in \citet{Emami+2022}, on average, Faraday rotation imprints a
small shift in the EVPA pattern corresponding to the line-of-sight magnetic field direction.  For example, since we are viewing \m87 nearly face-on, a poloidal field pointed towards the observer will on average Faraday rotate EVPA ticks in the counter-clockwise direction, and $\beta_2$ is also rotated counter-clockwise.  Similarly, a poloidal field pointed away from the observer will on average rotate ticks $\beta_2$ clockwise.  More complicated behavior can even arise if differences in the Faraday depth among different emitting regions cause summation or cancellation depending on the degree of Faraday rotation.

Ultimately, although flipping the polarity of the magnetic field has an effect on the distributions of $\angle \beta_2$, it has little effect on our linear polarization constraints.  
SANE models do not appear to change significantly in this metric. The MAD models appear to shift slightly in $\angle \beta_2$ when the field polarity is reversed. Prograde models have a slight increase, whereas retrograde models have a slight decrease in $\angle \beta_2$. This does not affect the simultaneous snapshot scoring model results, but causes an increased preference for retrograde MAD models in the joint scoring method. 
With multi-frequency data, a sign flip in the magnetic field polarity should naturally lead to a sign flip in the rotation measure \citep[e.g.][]{Contopoulos+2022}, but we do not consider rotation measure in this work.

%=================================================================================================================================================================

\section{More Details on GRMHD Scoring Results}
\label{app:stokesvscore}

\subsection{Simultaneous Snapshot Scoring}
\label{sec:simultaneous}

For simultaneous scoring, the upper limit (3.7\,\%) for $\langle |v| \rangle$ reported in this paper does not provide any additional information to constrain the GRMHD image parameter space.  That is, all models that pass the other polarimetric constraints already naturally pass this constraint. This is perhaps not surprising, since the upper limit on $v_\mathrm{net}$ 
reported by \citet{Goddi2021} and already considered in \citetalias{PaperVIII}
is more stringent than that on $\langle |v| \rangle$ obtained in this work.
Only 261/184796 snapshots ($0.14\,\%$) fail the $\langle |v| \rangle$ constraint but pass the $v_\mathrm{net}$ constraint. Producing a low $v_{\rm net}$ but a high $\langle|v|\rangle$ requires regions of high resolved circular polarization fraction of both positive and negative sign to nearly cancel across an image. Ultimately when combined with the linear polarization constraints, $\langle |v| \rangle$ adds no discerning power. Consequently, model predictions for \m87 remain consistent with \citetalias{PaperVIII}. 

The minor differences in the simultaneous scoring histogram (left panel of \autoref{fig:theory_model_scoring}) compared to \citetalias{PaperVIII} can be attributed not to the additional constraint on $\langle |v| \rangle$, but rather to the additional parameter probed in this paper, the magnetic field polarity, as well as slight differences in time sampling of the images and updates to radiative transfer coefficients. 

\begin{figure}[ht]
     \centering
     \includegraphics[width=0.45\textwidth]{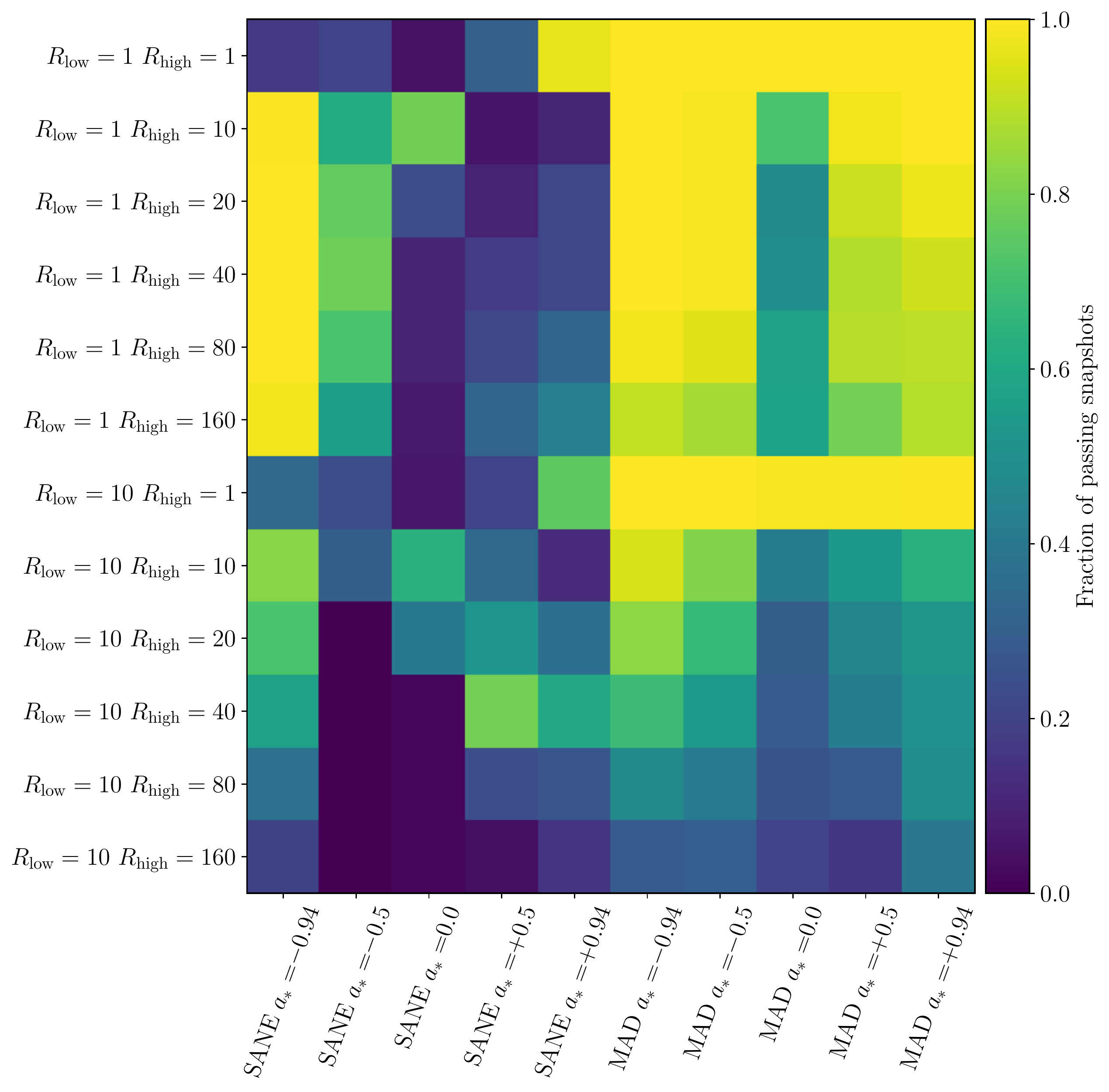}
     \caption{Passing fraction of the $v_\mathrm{net}$ and $\langle |v| \rangle$ constraints for all the models in the simulation library. Aligned and anti-aligned models are scored combined since the individual passing fractions are similar.}
     \label{fig:theory_vconstraint_map}
\end{figure}

To better understand how Stokes $\mathcal{V}$ alone affects our model scoring, \autoref{fig:theory_vconstraint_map} plots the fraction of snapshots of a given model that pass Stokes $\mathcal{V}$ constraints without considering the others. We find that Stokes $\mathcal{V}$ constraints are more likely to rule out SANE models than MADs, since SANE models are more likely to overproduce $v_\mathrm{net}$. The $\langle |v| \rangle$ constraint alone does not rule out many models.  Passing fractions appear similar for both field configurations, and thus they have been combined in the figure.

\subsection{Joint Scoring}
\label{sec:joint}

As shown in Fig. \ref{fig:theory_model_scoring}, the joint scoring method favors MAD $R_\mathrm{low}=10$ models that are not highly spinning. The inclusion of reversed-field images greatly increases the relative likelihood for MAD spin $-0.5$ models, particularly for $\angle \beta_{2\mathrm{,LP}}$. The joint scoring results appear qualitatively different from that given in \citetalias{PaperVIII}, which is likely due to differences in the image library sets rather than just the $\langle |v| \rangle$ constraint.

The inclusion of the $\langle |v| \rangle$ constraint removes small likelihoods for MAD spin $0$,$-0.5$, $R_\mathrm{low}=1$ models but otherwise does not qualitatively change the likelihoods. Note that the midpoint for the $\langle |v| \rangle$ allowed range is $1.85\,\%$, or half of the conservative upper limit. The results from the joint scoring method are more sensitive to this choice than is the case in simultaneous scoring. Using an aggressive upper limit on $\langle|v|\rangle$ of 2\,\% from \themis (\autoref{tab:method_upperlimits}), we find that qualitatively the joint scoring results remain largely the same, but some $R_\mathrm{low}=1$ models now are included as those models have lower $\langle |v| \rangle$. A lower limit on $\langle |v| \rangle$ at 1\,\% tends to increase the likelihood of the MAD spin $+0.5$, $R_\mathrm{low}=$ 10 models.

Comparisons between the image library sets between this paper and \citetalias{PaperVIII} can yield insight into the differences in the joint scoring method. For example, the relatively lower likelihoods for the MAD spin $+0.5$ models in Fig. \ref{fig:theory_model_scoring} can in part be attributed to only scoring images at $17^\circ$ or $163^\circ$ inclination, while \citetalias{PaperVIII} included models at inclinations $\pm 5^\circ$ from $17^\circ$ or $163^\circ$. Models from \citetalias{PaperVIII} scored using only $17^\circ$ or $163^\circ$ inclination cause the relative likelihood for MAD spin $+0.5$ models to decrease. The time sampling in the library sets can also influence the model scoring. In this paper, we sample uniformly for each model at 5M, while in \citetalias{PaperVIII} the number of snapshots per model is kept fixed to 200, with the sampling cadence varying. Using the sampling in \citetalias{PaperVIII} causes a slight increase to the likelihoods for the prograde MAD models. Moreover, the library set in \citetalias{PaperVIII} excludes emission from within 5\% of the event horizon, while the current image library consists of emission from further inside. While this only affects models that are very optically thin (MAD models), it can change the EVPA structure and thus the $\beta_{2\mathrm{,LP}}$ modes. The $\beta_{2\mathrm{,LP}}$ modes appear to be the most discrepant between the two library sets, and most linear and circular polarization constraints otherwise remain consistent.

%=================================================================================================================================================================
\section{Stokes $\mathcal{V}$ Origin in Simulations.}
\label{app:stokesvorigin}
\begin{figure*}
    \centering
    {\includegraphics[width=0.99\linewidth]{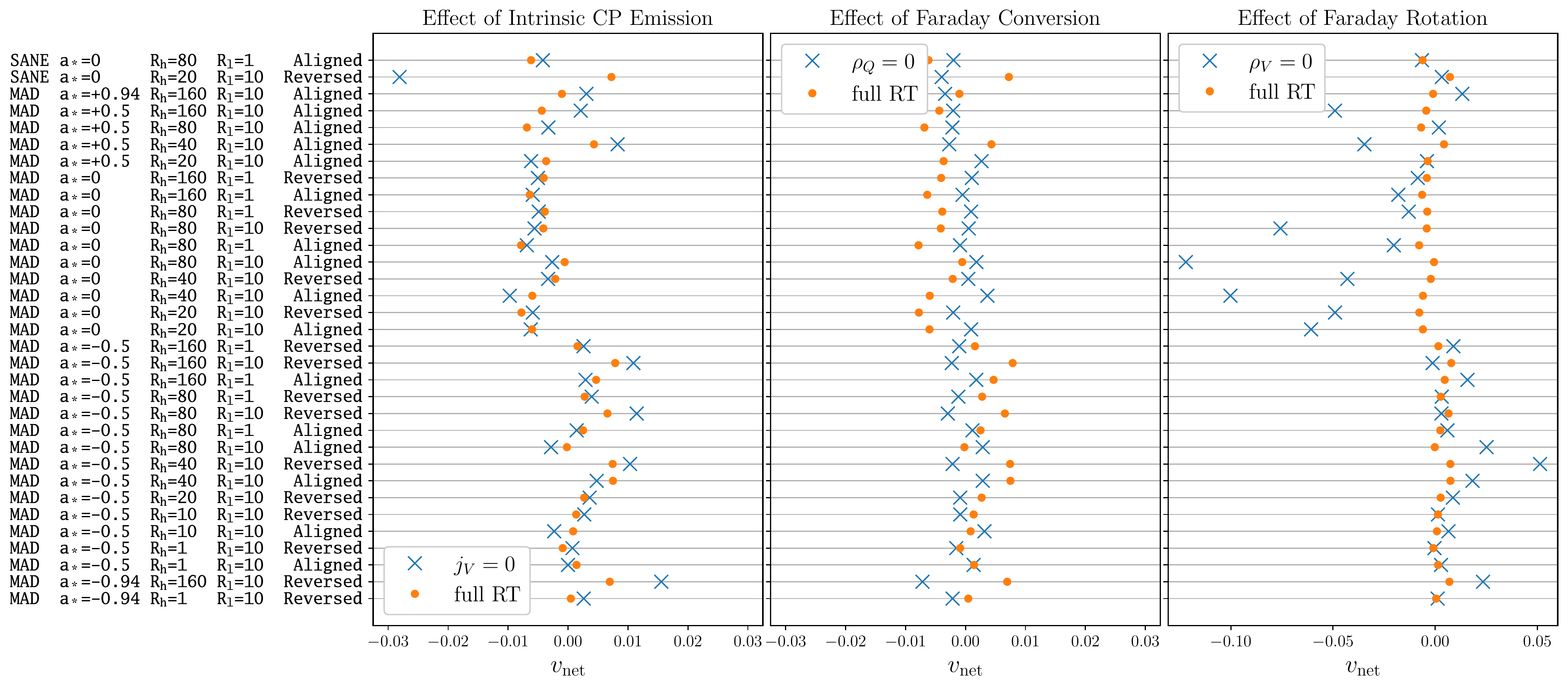}}
    {\includegraphics[width=0.99\linewidth]{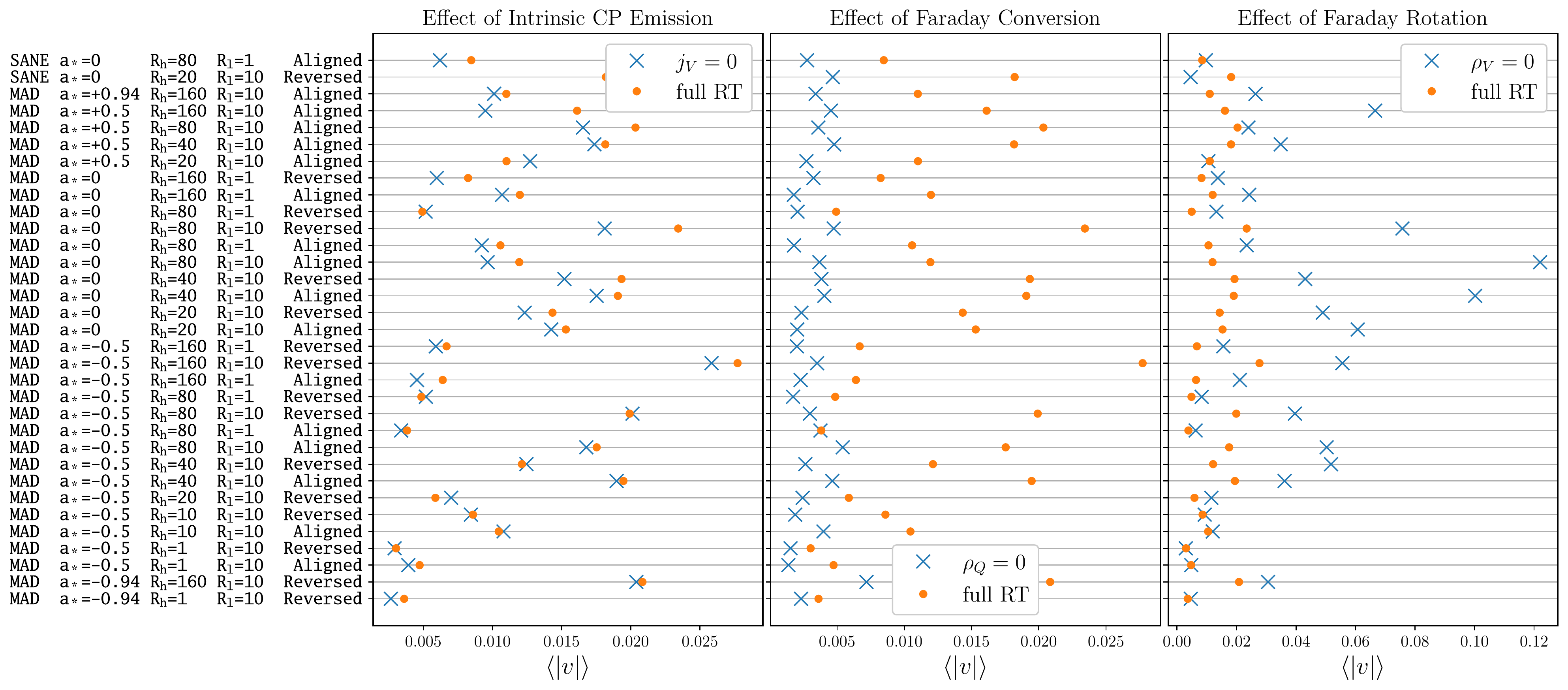}}
    \caption{ Exploration of the effects of various CP production mechanisms on $v_\mathrm{net}$ (top panels) and $\langle |v| \rangle$ (bottom panels). Each column corresponds to a comparison between the full radiative transfer (RT) simulation and the same calculation with one coefficient turned off: $j_\mathcal{V}$=0 (intrinsic emission) for column 1, $\rho_\mathcal{Q}$=0 (Faraday conversion) for column 2, $\rho_\mathcal{V}$=0 (Faraday rotation) for column 3. Each tick on the y axis corresponds to a model for which at least one snapshot passes all the polarimetric constraints. A single snapshot from each of these models has been plotted at each tick.}
    \label{fig:theory_coefficient}
\end{figure*}

Here, we present in more detail the test summarized in \autoref{sec:vorigins} to determine the physical origin of Stokes $\mathcal{V}$ emission in physical models. We investigate the full set of 288 snapshots that simultaneously pass all polarimetric constraints. In {\sc ipole}, our GRRT code, it is straightforward to probe the effects of each physical effect by setting the appropriate radiative transfer coefficient to 0.  Here, the most relevant coefficients are intrinsic circularly polarized emission ($j_\mathcal{V}$), Faraday conversion ($\rho_\mathcal{Q}$), and Faraday rotation ($\rho_\mathcal{V}$).  For details of these coefficients, we refer the reader to \citet{Moscibrodzka&Gammie2018}.  

In this experiment, we re-image all the passing snapshots but each time, turn off one of these coefficients.  
Selecting one representative snapshot per passing model, we compare both $v_\mathrm{net}$ and $\langle |v| \rangle$ against the full radiative transfer solution and plot the results in \autoref{fig:theory_coefficient}.
Since significant cancellation occurs in the image plane for Stokes $\mathcal{V}$, large differences to the image do not necessarily correspond to large changes in $v_\mathrm{net}$. Therefore, $\langle |v| \rangle$ is somewhat more informative, and robust to cancellations, although cancellations within the Gaussian kernel of $20\,\mu\mathrm{as}$ still occur. 

From \autoref{fig:theory_coefficient} we note 3 main results:
\begin{enumerate}
    \item Turning off intrinsic Stokes $\mathcal{V}$ emission ($j_\mathcal{V}$) does not appear to greatly influence either $v_\mathrm{net}$ or $\langle |v| \rangle$ (first column of \autoref{fig:theory_coefficient}). While this is not true for every snapshot in every model, for most of the passing models, we find that intrinsic emission is not the dominant origin of Stokes $\mathcal{V}$.
    \item On the other hand, turning off Faraday conversion ($\rho_\mathcal{Q}=0$) greatly suppresses $\langle|v|\rangle$ (second row, second column in \autoref{fig:theory_coefficient}), suggesting that Faraday conversion is the dominant production mechanism for Stokes $\mathcal{V}$ in passing models.
    \item Interestingly, turning off Faraday rotation increases $\langle|v|\rangle$ for some of the passing models. This implies that scrambling via Faraday rotation is important not only for reducing the linear polarization fraction, but also for indirectly reducing the circular polarization fraction.  In these models, Faraday rotation partly scrambles the linear polarization that is converted into circular.
\end{enumerate}

In summary, Faraday effects are critical for Stokes $\mathcal{V}$ generation in our models, while intrinsic emission plays a sub-dominant, though non-negligible, role.

%=================================================================================================================================================================
\section{Alternative Plasma Models}
\label{app:alternative_plasmas}

Throughout this work, we have modeled a pure ion-electron plasma with thermal electron distributions.  Here, we explore our sensitivity to the details of the plasma by considering the existence of electron-positron pairs as well as non-thermal electron distribution functions.  In brief, our Stokes $\mathcal{V}$ images are indeed sensitive to these details, motivating future studies in these areas. 

\subsection{Ionic vs. pair plasmas}
\label{sec:pairs}

\begin{figure*}[ht!]
    \centering
    \includegraphics[width=\textwidth]{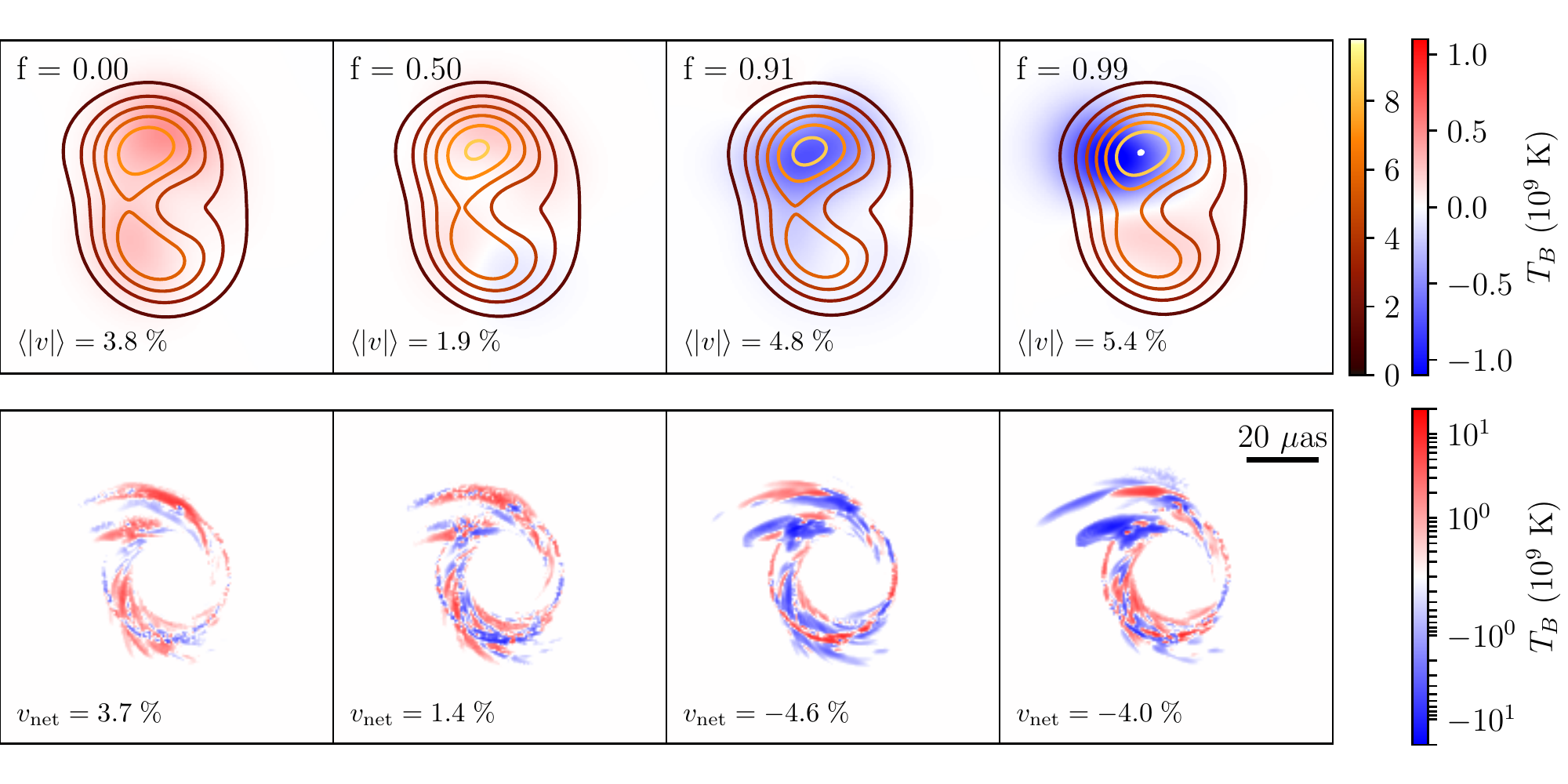}
    \caption{Test of plasma content, where we ray-trace a single snapshot of the MAD $a=+0.5$ $R_\mathrm{high}=80$ $R_\mathrm{low}=10$ aligned field model with an increasing positron-to-electron ratio, denoted as $f$ in the top left of each panel.  The Stokes $\mathcal{V}$ structure clearly evolves as $f$ increases, but we do not observe a clear discriminant of plasma content.}
    \label{fig:positron_test}
\end{figure*}

\begin{figure*}[ht!]
    \centering
    \includegraphics[width=\textwidth]{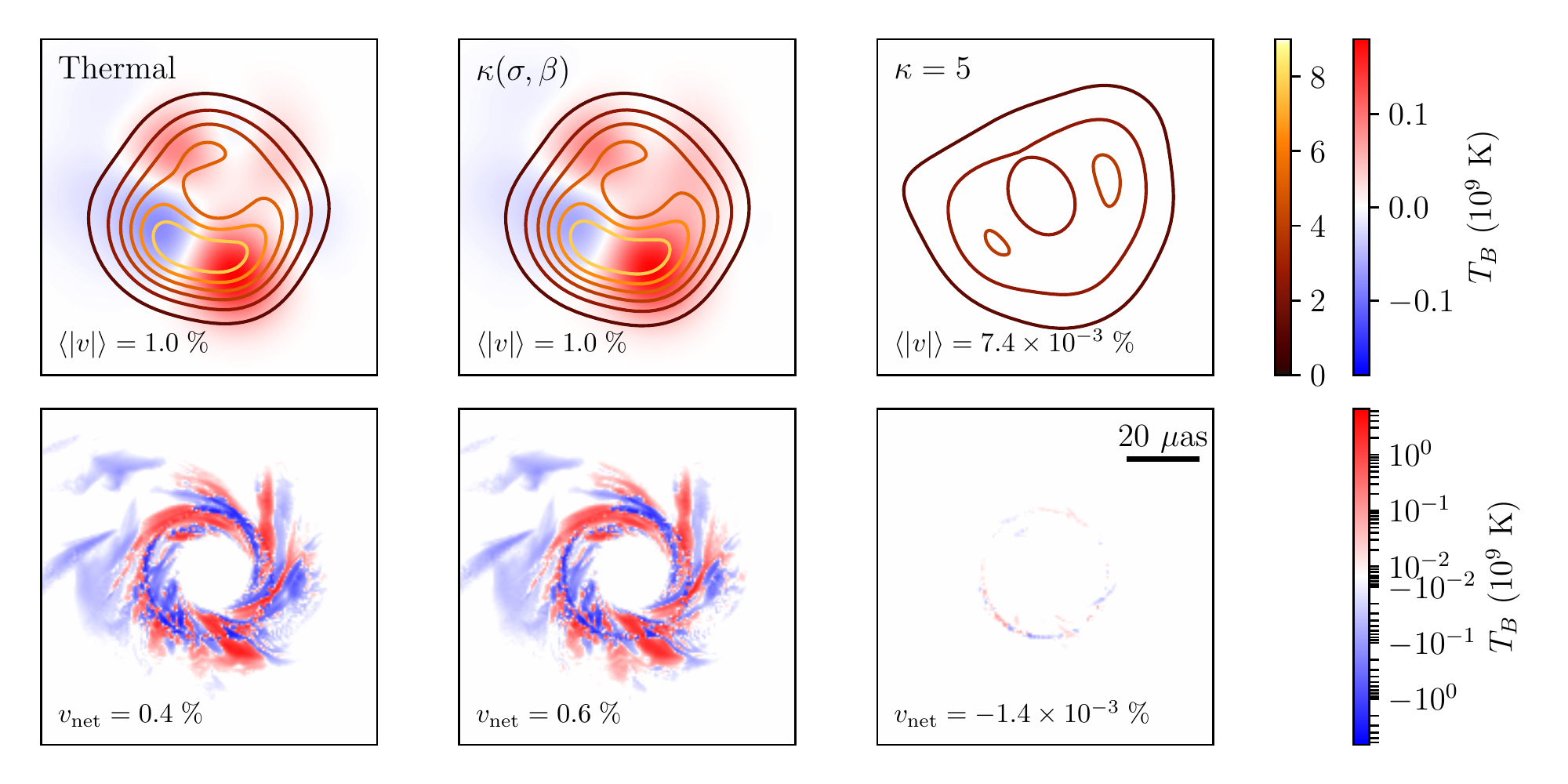}
    \caption{A single GRMHD snapshot (MAD $a=+0.5$ $R_\mathrm{high}=80$ $R_\mathrm{low}=10$ aligned field) ray traced with three different electron distribution functions.  The physically motivated variable kappa model produces both a Stokes $\mathcal{I}$ and Stokes $\mathcal{V}$ image very similar to thermal model.  However, a model with fixed $\kappa=5$ produces a much more diffuse Stokes $\mathcal{I}$  and extremely little Stokes $\mathcal{V}$.}
    \label{fig:nonthermal_test}
\end{figure*}

Unlike an ionic plasma (assumed throughout this paper), a pure pair plasma produces no circular polarization via synchrotron emission and has no Faraday rotation, but can still perform Faraday conversion \citep{Jones_ODell_1977,Jones_1988,Wardle+1998}.  For ionic plasma models, all of the radiative transfer coefficients are believed to be important for producing the Stokes $\mathcal{V}$ image. On EHT scales, the few recent studies on this topic find that increasing the pair fraction can alter both resolved Stokes $\mathcal{V}$ images at 230 GHz as well as the evolution as a function of frequency \citep{Anantua+2020,Emami+2021,Emami+2023}. 

In \autoref{fig:positron_test}, we test a single snapshot of the MAD $a=+0.5$ $R_\mathrm{high}=80$ $R_\mathrm{low}=10$ aligned field model, performing polarized ray-tracing with an increasing positron-to-electron number density ratio, denoted as $f$ in each panel.  To obtain a given value of $f$, we inject electron-positron pairs to each cell, representing a simplistic scenario in which a large fraction of pairs are produced at number densities directly proportional to the number densities of pre-existing electrons.  Since this would drastically increase the number density of emitting leptons in each successive panel, we also find a new value of 
the density normalization
for each value of $f$, such that the total flux is always 0.5 Jy as in the original snapshot.

For a given value of $f$, the radiative transfer coefficients are modified via
\citep[e.g.,][]{Macdonald&Marscher2018,Emami+2021}

\begin{align}
    j_{I,Q,U} &\to (1+f) j_{I,Q,U}, \nonumber \\
    j_{V} &\to (1-f) j_{V}, \nonumber \\
    \alpha_{I,Q,U} &\to (1+f) \alpha_{I,Q,U}, \nonumber \\
    \alpha_{V} &\to (1-f) \alpha_{V}, \nonumber \\
    \rho_{Q,U} &\to (1+f)\rho_{Q,U}, \nonumber \\
    \rho_{V} &\to (1-f)\rho_{V},
\end{align}

\noindent capturing the effects described above.  \autoref{fig:positron_test} illustrates that the circularly polarized morphology of this snapshot clearly and strongly evolves with $f$, consistent with previous works.  However, given the inherent diversity of Stokes $\mathcal{V}$ structures among the library, there is no known signature in Stokes $\mathcal{V}$ that clearly indicates the presence of pairs.  Interestingly, we find that $\langle |v| \rangle$ increases monotonically in the perfect-resolution images, which can be attributed to drastically falling Faraday rotation depth (to be discussed in Anantua et al. in prep.).  However, due to cancellations within the EHT beam, $\langle |v| \rangle$ does {\it not} necessarily increase monotonically at EHT resolution.  In summary, we are sensitive to the content of the emitting plasma, but we are unaware of a clear signature distinguishing a pair plasma from an ionic plasma on event horizon scales where both intrinsic emission and Faraday conversion are important. While not explored here, images originating from GRMHD are also surprisingly sensitive to the atomic composition of a given ionic plasma \citep{Wong&Gammie2022}.  Identifying clear signatures to distinguish different models of plasma content on event horizon scales remains a topic of ongoing research.

%=================================================================================================================================================================
\subsection{Non-thermal electrons}
\label{sec:nonthermal}

Throughout this work, we consider only models with thermal electron distribution functions (eDF). However, it is believed that non-thermal particle acceleration can occur due to magnetic reconnection, MHD turbulence and collective plasma modes.  At present, a consensus model for the presence of non-thermal electrons has not be found, but here we explore two physically motivated implementations.  For a single snapshot, we compare their thermal eDF images to images produced assuming ``kappa'' models, which feature a thermal core and a high-energy non-thermal tail \citep{Vasyliunas1968}.  Such distributions have been observed in the solar wind \citep{Decker&Krimigis2003,Pierrard&Lazar2010} and occur naturally in particle-in-cell (PIC) simulations of particle acceleration in magnetized plasmas \citep{Kunz+2016}.  We try both a constant $\kappa=5$ model, where $\kappa-1=4$ is the power law index of the high-energy tail, and variable kappa model where $\kappa = \kappa(\sigma,\beta)$ as fit by the PIC simulations of \citet{Ball+2016}.  Due to limitations of the fitting functions used, any time the prescription would assign $\kappa > 7$, the code instead uses a radiative transfer coefficient appropriate for a thermal eDF.

In \autoref{fig:nonthermal_test}, we spot-check our sensitivity to our assumption of thermal electrons by ray-tracing a GRMHD snapshot with three different assumptions for the electron distribution function.  A 
new plasma density scale
is found for each image to match the total flux of the image with thermal electrons, 0.7\,Jy.  This snapshot corresponds to the MAD $a=+0.5$ $R_\mathrm{high}=80$ $R_\mathrm{low}=10$ aligned field model.  In the top row, we plot images blurred to EHT resolution, while in the bottom row we plot perfect-resolution images of Stokes $\mathcal{V}$ in symmetric logarithmic scale.  Compared to the leftmost image assuming thermal electrons, we find very little changes in the variable kappa model shown in the central panel.  This is broadly consistent with our findings in \citet{SgrAEHTCV} for \sgra in total intensity.  Thus, at least for this snapshot, exchanging a thermal distribution for a physically motivated non-thermal electron distribution has very little effect on Stokes $\mathcal{V}$. However, we report dramatic differences when switching to a constant kappa model with a value of $\kappa=5$.  As found in many other studies, non-thermal electrons make the image noticeably larger and more diffuse \citep[e.g.,][]{Ozel+2000,Mao+2017,Fromm+2022,Ricarte+2023}.  Intriguingly, although 
the plasma density 
is decreased only by a factor of 4 compared to the thermal model, the Stokes $\mathcal{V}$ signal almost entirely vanishes. This may be due to the fact that Faraday conversion is caused by the coldest relativistic electrons, which occur at a smaller fraction in $\kappa$ models by definition.

\begin{figure*}[ht!]
    \centering
    \includegraphics[width=\textwidth]{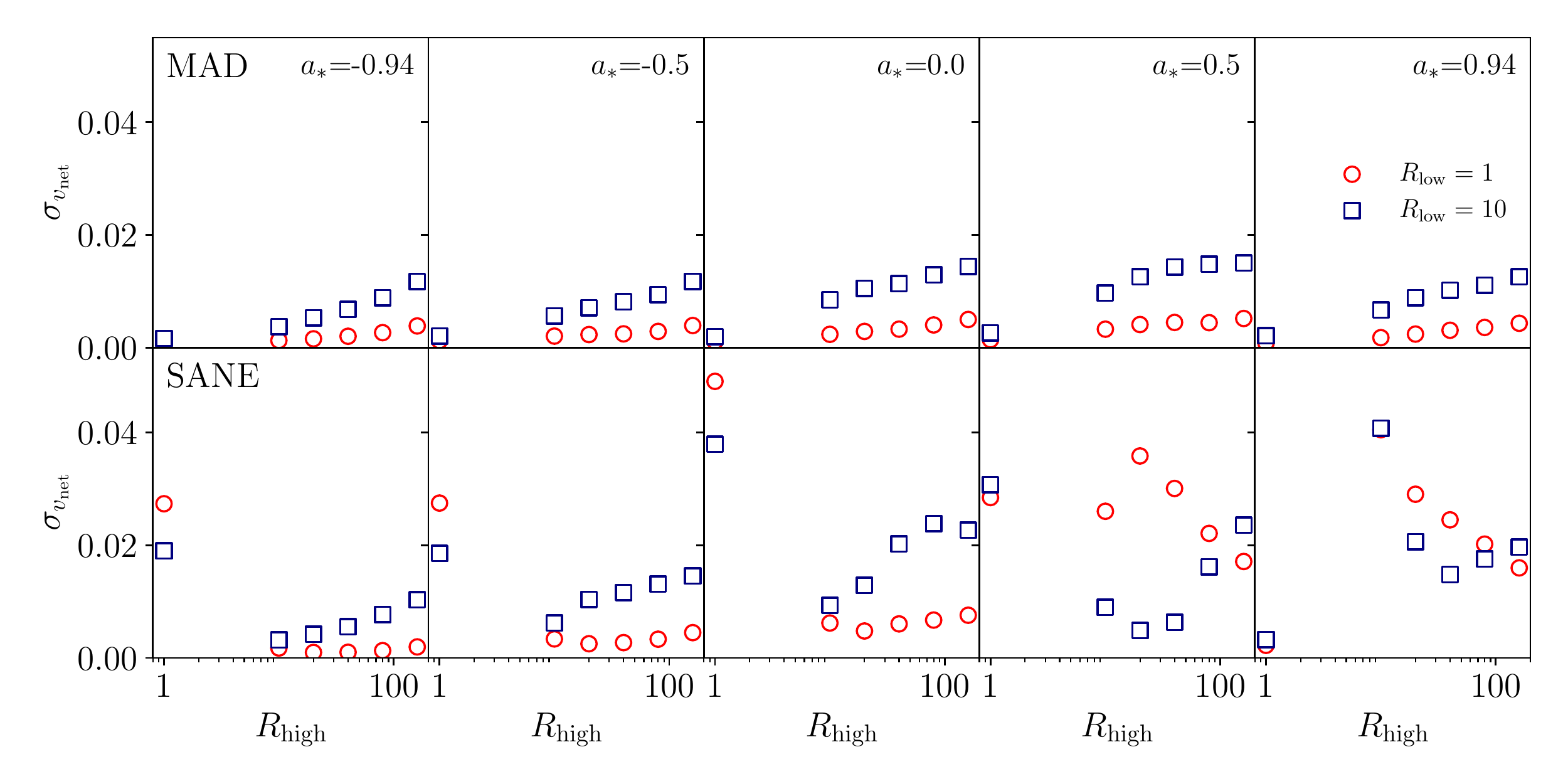}
    \caption{Standard deviations of the distributions of $v_\mathrm{net}$ in GRMHD models over time.  The standard deviations of models with different magnetic field polarities have been averaged together.  Different magnetic field states are plotted in different rows, different spins are plotted in different columns, $R_\mathrm{high}$ is plotted on the $x$-axis of each panel, and the color and shape of the markers encodes $R_\mathrm{low}$.  
    %Characterizing the variability of $v_\mathrm{net}$ with future observations would help distinguish models.  
    In particular, models with $R_\mathrm{low}=10$ are more variable than those with $R_\mathrm{low}=1$, and a similar but weaker trend occurs with $R_\mathrm{high}$.}
    \label{fig:variability_vnet}
\end{figure*}

%=================================================================================================================================================================
\section{Stokes $\mathcal{V}$ Variability in Passing Models}
\label{sec:vvar}

In \autoref{fig:variability_vnet}, we consider each model's distribution of $v_\mathrm{net}$ over time and plot its standard deviation, which we denote as $\sigma_{v_\mathrm{net}}$.  We average together the standard deviations for each magnetic field polarity, since their distributions can be disjoint.  We find that models with larger $R_\mathrm{low}$ are more variable in $v_\mathrm{net}$, with a similar but weaker trend in $R_\mathrm{high}$.  Although not shown, we find that the variability of $\langle |v| \rangle$ is qualitatively similar.  Some models have $\sigma_{v_\mathrm{net}}$ in excess of our upper limit of $0.008$, suggesting that future observations during more favorable conditions may result in higher S/N detections of Stokes $\mathcal{V}$ on EHT baselines and hence more robust image reconstructions of the source.

\clearpage
%=================================================================================================================================================================
%=================================================================================================================================================================

\bibliography{M87_POL.bib}

\begin{thebibliography}{}
\expandafter\ifx\csname natexlab\endcsname\relax\def\natexlab#1{#1}\fi
\providecommand{\url}[1]{\href{#1}{#1}}
\providecommand{\dodoi}[1]{doi:~\href{http://doi.org/#1}{\nolinkurl{#1}}}
\providecommand{\doeprint}[1]{\href{http://ascl.net/#1}{\nolinkurl{http://ascl.net/#1}}}
\providecommand{\doarXiv}[1]{\href{https://arxiv.org/abs/#1}{\nolinkurl{https://arxiv.org/abs/#1}}}

\bibitem[{{Anantua} {et~al.}(2020){Anantua}, {Emami}, {Loeb}, \&
  {Chael}}]{Anantua+2020}
{Anantua}, R., {Emami}, R., {Loeb}, A., \& {Chael}, A. 2020, \apj, 896, 30,
  \dodoi{10.3847/1538-4357/ab9103}

\bibitem[{{Balbus} \& {Hawley}(1991)}]{1991ApJ...376..214B}
{Balbus}, S.~A., \& {Hawley}, J.~F. 1991, \apj, 376, 214,
  \dodoi{10.1086/170270}

\bibitem[{{Balbus} \& {Hawley}(1998)}]{1998RvMP...70....1B}
---. 1998, Reviews of Modern Physics, 70, 1, \dodoi{10.1103/RevModPhys.70.1}

\bibitem[{{Ball} {et~al.}(2016){Ball}, {{\"O}zel}, {Psaltis}, \&
  {Chan}}]{Ball+2016}
{Ball}, D., {{\"O}zel}, F., {Psaltis}, D., \& {Chan}, C.-k. 2016, \apj, 826,
  77, \dodoi{10.3847/0004-637X/826/1/77}

\bibitem[{{Bisnovatyi-Kogan} \&
  {Ruzmaikin}(1974)}]{Bisnovatyi-Kogan&Ruzmaikin1974}
{Bisnovatyi-Kogan}, G.~S., \& {Ruzmaikin}, A.~A. 1974, \apss, 28, 45,
  \dodoi{10.1007/BF00642237}

\bibitem[{{Blackburn} {et~al.}(2020){Blackburn}, {Pesce}, {Johnson}, {Wielgus},
  {Chael}, {Christian}, \& {Doeleman}}]{Blackburn_Closure}
{Blackburn}, L., {Pesce}, D.~W., {Johnson}, M.~D., {et~al.} 2020, \apj, 894,
  31, \dodoi{10.3847/1538-4357/ab8469}

\bibitem[{{Blackburn} {et~al.}(2019){Blackburn}, {Chan}, {Crew}, {Fish},
  {Issaoun}, {Johnson}, {Wielgus}, {Akiyama}, {Barrett}, {Bouman}, {Cappallo},
  {Chael}, {Janssen}, {Lonsdale}, \& {Doeleman}}]{Blackburn_2019}
{Blackburn}, L., {Chan}, C.-k., {Crew}, G.~B., {et~al.} 2019, \apj, 882, 23,
  \dodoi{10.3847/1538-4357/ab328d}

\bibitem[{{Bower} {et~al.}(1999){Bower}, {Falcke}, \& {Backer}}]{Bower+1999}
{Bower}, G.~C., {Falcke}, H., \& {Backer}, D.~C. 1999, \apjl, 523, L29,
  \dodoi{10.1086/312246}

\bibitem[{{Bower} {et~al.}(2002){Bower}, {Falcke}, \& {Mellon}}]{Bower2002}
{Bower}, G.~C., {Falcke}, H., \& {Mellon}, R.~R. 2002, \apjl, 578, L103,
  \dodoi{10.1086/344607}

\bibitem[{{Bower} {et~al.}(2018){Bower}, {Broderick}, {Dexter}, {Doeleman},
  {Falcke}, {Fish}, {Johnson}, {Marrone}, {Moran}, {Moscibrodzka}, {Peck},
  {Plambeck}, \& {Rao}}]{Bower+2018}
{Bower}, G.~C., {Broderick}, A., {Dexter}, J., {et~al.} 2018, \apj, 868, 101,
  \dodoi{10.3847/1538-4357/aae983}

\bibitem[{{Broderick} \& {Pesce}(2020)}]{BroderickPesce_2020}
{Broderick}, A.~E., \& {Pesce}, D.~W. 2020, \apj, 904, 126,
  \dodoi{10.3847/1538-4357/abbd9d}

\bibitem[{{Broderick} {et~al.}(2020{\natexlab{a}}){Broderick}, {Pesce},
  {Tiede}, {Pu}, \& {Gold}}]{Themaging:2020}
{Broderick}, A.~E., {Pesce}, D.~W., {Tiede}, P., {Pu}, H.-Y., \& {Gold}, R.
  2020{\natexlab{a}}, \apj, 898, 9, \dodoi{10.3847/1538-4357/ab9c1f}

\bibitem[{{Broderick} {et~al.}(2020{\natexlab{b}}){Broderick}, {Gold},
  {Karami}, {Preciado-L{\'o}pez}, {Tiede}, {Pu}, {Akiyama}, {Alberdi}, {Alef},
  {Asada}, {Azulay}, {Baczko}, {Balokovi{\'c}}, {Barrett}, {Bintley},
  {Blackburn}, {Boland}, {Bouman}, {Bower}, {Bremer}, {Brinkerink},
  {Brissenden}, {Britzen}, {Broguiere}, {Bronzwaer}, {Byun}, {Carlstrom},
  {Chael}, {Chatterjee}, {Chatterjee}, {Chen}, {Chen}, {Cho}, {Conway},
  {Cordes}, {Crew}, {Cui}, {Davelaar}, {De Laurentis}, {Deane}, {Dempsey},
  {Desvignes}, {Doeleman}, {Eatough}, {Falcke}, {Fish}, {Fomalont},
  {Fraga-Encinas}, {Friberg}, {Fromm}, {Galison}, {Gammie}, {Garc{\'\i}a},
  {Gentaz}, {Georgiev}, {Goddi}, {G{\'o}mez}, {Gu}, {Gurwell}, {Hada}, {Hecht},
  {Hesper}, {Ho}, {Ho}, {Honma}, {Huang}, {Huang}, {Hughes}, {Inoue},
  {Issaoun}, {James}, {Janssen}, {Jeter}, {Jiang}, {Jim{\'e}nez-Rosales},
  {Johnson}, {Jorstad}, {Jung}, {Karuppusamy}, {Kawashima}, {Keating},
  {Kettenis}, {Kim}, {Kim}, {Kino}, {Koay}, {Koch}, {Koyama}, {Kramer},
  {Kramer}, {Krichbaum}, {Kuo}, {Lee}, {Li}, {Li}, {Lindqvist}, {Lico}, {Liu},
  {Liuzzo}, {Lo}, {Lobanov}, {Loinard}, {Lonsdale}, {Lu}, {MacDonald}, {Mao},
  {Marscher}, {Mart{\'\i}-Vidal}, {Matsushita}, {Matthews}, {Menten}, {Mizuno},
  {Mizuno}, {Moran}, {Moriyama}, {Moscibrodzka}, {M{\"u}ller}, {Nagai},
  {Nagar}, {Nakamura}, {Narayan}, {Narayanan}, {Natarajan}, {Neri}, {Ni},
  {Noutsos}, {Okino}, {Olivares}, {Ortiz-Le{\'o}n}, {Oyama}, {Palumbo}, {Park},
  {Pen}, {Pesce}, {Pi{\'e}tu}, {Plambeck}, {PopStefanija}, {Porth}, {Prather},
  {Ramakrishnan}, {Rao}, {Rawlings}, {Raymond}, {Rezzolla}, {Ripperda},
  {Roelofs}, {Rogers}, {Ros}, {Rose}, {Rottmann}, {Ruszczyk}, {Ryan}, {Rygl},
  {S{\'a}nchez}, {S{\'a}nchez-Arguelles}, {Sasada}, {Savolainen}, {Schloerb},
  {Schuster}, {Shao}, {Shen}, {Small}, {Sohn}, {SooHoo}, {Tazaki}, {Tilanus},
  {Titus}, {Toma}, {Torne}, {Traianou}, {Trippe}, {Tsuda}, {van Bemmel}, {van
  Langevelde}, {van Rossum}, {Wagner}, {Wardle}, {Weintroub}, {Wex}, {Wharton},
  {Wielgus}, {Wong}, {Wu}, {Yoon}, {Young}, {Young}, {Younsi}, {Yuan}, {Yuan},
  {Zensus}, {Zhao}, {Zhao}, {Zhu}, \& {Event Horizon Telescope
  Collaboration}}]{Broderick_2019}
{Broderick}, A.~E., {Gold}, R., {Karami}, M., {et~al.} 2020{\natexlab{b}},
  \apj, 897, 139, \dodoi{10.3847/1538-4357/ab91a4}

\bibitem[{{Chael} {et~al.}(2018){Chael}, {Johnson}, {Bouman}, {Blackburn},
  {Akiyama}, \& {Narayan}}]{Chael_2018_Imaging}
{Chael}, A.~A., {Johnson}, M.~D., {Bouman}, K.~L., {et~al.} 2018, \apj, 857,
  23, \dodoi{10.3847/1538-4357/aab6a8}

\bibitem[{{Chael} {et~al.}(2016){Chael}, {Johnson}, {Narayan}, {Doeleman},
  {Wardle}, \& {Bouman}}]{Chael_2016}
{Chael}, A.~A., {Johnson}, M.~D., {Narayan}, R., {et~al.} 2016, \apj, 829, 11,
  \dodoi{10.3847/0004-637X/829/1/11}

\bibitem[{{Chatterjee} {et~al.}(2020){Chatterjee}, {Younsi}, {Liska},
  {Tchekhovskoy}, {Markoff}, {Yoon}, {van Eijnatten}, {Hesp}, {Ingram}, \& {van
  der Klis}}]{Chatterjee+2020}
{Chatterjee}, K., {Younsi}, Z., {Liska}, M., {et~al.} 2020, \mnras, 499, 362,
  \dodoi{10.1093/mnras/staa2718}

\bibitem[{{Contopoulos} \& {Kazanas}(1998)}]{Contopoulos+1998}
{Contopoulos}, I., \& {Kazanas}, D. 1998, \apj, 508, 859,
  \dodoi{10.1086/306426}

\bibitem[{{Contopoulos} {et~al.}(2022){Contopoulos}, {Myserlis}, {Kazanas}, \&
  {Nathanail}}]{Contopoulos+2022}
{Contopoulos}, I., {Myserlis}, I., {Kazanas}, D., \& {Nathanail}, A. 2022,
  Galaxies, 10, 80, \dodoi{10.3390/galaxies10040080}

\bibitem[{{Decker} \& {Krimigis}(2003)}]{Decker&Krimigis2003}
{Decker}, R.~B., \& {Krimigis}, S.~M. 2003, Advances in Space Research, 32,
  597, \dodoi{10.1016/S0273-1177(03)00356-9}

\bibitem[{{Dexter}(2016)}]{Dexter_2016}
{Dexter}, J. 2016, \mnras, 462, 115, \dodoi{10.1093/mnras/stw1526}

\bibitem[{{Dihingia} {et~al.}(2023){Dihingia}, {Mizuno}, {Fromm}, \&
  {Rezzolla}}]{Dihingia+2023}
{Dihingia}, I.~K., {Mizuno}, Y., {Fromm}, C.~M., \& {Rezzolla}, L. 2023,
  \mnras, 518, 405, \dodoi{10.1093/mnras/stac3165}

\bibitem[{{Emami} {et~al.}(2021){Emami}, {Anantua}, {Chael}, \&
  {Loeb}}]{Emami+2021}
{Emami}, R., {Anantua}, R., {Chael}, A.~A., \& {Loeb}, A. 2021, \apj, 923, 272,
  \dodoi{10.3847/1538-4357/ac2950}

\bibitem[{{Emami} {et~al.}(2023{\natexlab{a}}){Emami}, {Ricarte}, {Wong},
  {Palumbo}, {Chang}, {Doeleman}, {Broderick}, {Narayan}, {Wielgus},
  {Blackburn}, {Prather}, {Chael}, {Anantua}, {Chatterjee}, {Marti-Vidal},
  {G{\'o}mez}, {Akiyama}, {Liska}, {Hernquist}, {Tremblay}, {Vogelsberger},
  {Alcock}, {Smith}, {Steiner}, {Tiede}, \& {Roelofs}}]{Emami+2022}
{Emami}, R., {Ricarte}, A., {Wong}, G.~N., {et~al.} 2023{\natexlab{a}}, \apj,
  950, 38, \dodoi{10.3847/1538-4357/acc8cd}

\bibitem[{{Emami} {et~al.}(2023{\natexlab{b}}){Emami}, {Anantua}, {Ricarte},
  {Doeleman}, {Broderick}, {Wong}, {Blackburn}, {Wielgus}, {Narayan},
  {Tremblay}, {Alcock}, {Hernquist}, {Smith}, {Liska}, {Natarajan},
  {Vogelsberger}, {Curd}, \& {Kramer}}]{Emami+2023}
{Emami}, R., {Anantua}, R., {Ricarte}, A., {et~al.} 2023{\natexlab{b}},
  Galaxies, 11, 11, \dodoi{10.3390/galaxies11010011}

\bibitem[{{Event Horizon Telescope Collaboration}(2023)}]{M87poldata}
{Event Horizon Telescope Collaboration}. 2023, EHT M87 Polarized Data,  CyVerse
  Data Commons, \dodoi{10.25739/q46m-m857}

\bibitem[{{Event Horizon Telescope Collaboration}
  {et~al.}(2019{\natexlab{a}})}]{PaperI}
{Event Horizon Telescope Collaboration}, {et~al.} 2019{\natexlab{a}}, \apjl,
  875, L1, \dodoi{10.3847/2041-8213/ab0ec7}

\bibitem[{{Event Horizon Telescope Collaboration}
  {et~al.}(2019{\natexlab{b}})}]{PaperII}
---. 2019{\natexlab{b}}, \apjl, 875, L2, \dodoi{10.3847/2041-8213/ab0c96}

\bibitem[{{Event Horizon Telescope Collaboration}
  {et~al.}(2019{\natexlab{c}})}]{PaperIII}
---. 2019{\natexlab{c}}, \apjl, 875, L3, \dodoi{10.3847/2041-8213/ab0c57}

\bibitem[{{Event Horizon Telescope Collaboration}
  {et~al.}(2019{\natexlab{d}})}]{PaperIV}
---. 2019{\natexlab{d}}, \apjl, 875, L4, \dodoi{10.3847/2041-8213/ab0e85}

\bibitem[{{Event Horizon Telescope Collaboration}
  {et~al.}(2019{\natexlab{e}})}]{PaperV}
---. 2019{\natexlab{e}}, \apjl, 875, L5, \dodoi{10.3847/2041-8213/ab0f43}

\bibitem[{{Event Horizon Telescope Collaboration}
  {et~al.}(2019{\natexlab{f}})}]{PaperVI}
---. 2019{\natexlab{f}}, \apjl, 875, L6, \dodoi{10.3847/2041-8213/ab1141}

\bibitem[{{Event Horizon Telescope Collaboration}
  {et~al.}(2021{\natexlab{a}})}]{PaperVII}
---. 2021{\natexlab{a}}, \apjl, 910, L12, \dodoi{10.3847/2041-8213/abe71d}

\bibitem[{{Event Horizon Telescope Collaboration}
  {et~al.}(2021{\natexlab{b}})}]{PaperVIII}
---. 2021{\natexlab{b}}, \apjl, 910, L13, \dodoi{10.3847/2041-8213/abe4de}

\bibitem[{{Event Horizon Telescope Collaboration}
  {et~al.}(2022{\natexlab{a}})}]{SgrAEHTCI}
---. 2022{\natexlab{a}}, \apjl, 930, L12, \dodoi{10.3847/2041-8213/ac6674}

\bibitem[{{Event Horizon Telescope Collaboration}
  {et~al.}(2022{\natexlab{b}})}]{SgrAEHTCIII}
---. 2022{\natexlab{b}}, \apjl, 930, L14, \dodoi{10.3847/2041-8213/ac6429}

\bibitem[{{Event Horizon Telescope Collaboration}
  {et~al.}(2022{\natexlab{c}})}]{SgrAEHTCIV}
---. 2022{\natexlab{c}}, \apjl, 930, L15, \dodoi{10.3847/2041-8213/ac6736}

\bibitem[{{Event Horizon Telescope Collaboration}
  {et~al.}(2022{\natexlab{d}})}]{SgrAEHTCV}
---. 2022{\natexlab{d}}, \apjl, 930, L16, \dodoi{10.3847/2041-8213/ac6672}

\bibitem[{{Fishbone} \& {Moncrief}(1976)}]{1976ApJ...207..962F}
{Fishbone}, L.~G., \& {Moncrief}, V. 1976, \apj, 207, 962,
  \dodoi{10.1086/154565}

\bibitem[{{Fromm} {et~al.}(2022){Fromm}, {Cruz-Osorio}, {Mizuno}, {Nathanail},
  {Younsi}, {Porth}, {Olivares}, {Davelaar}, {Falcke}, {Kramer}, \&
  {Rezzolla}}]{Fromm+2022}
{Fromm}, C.~M., {Cruz-Osorio}, A., {Mizuno}, Y., {et~al.} 2022, \aap, 660,
  A107, \dodoi{10.1051/0004-6361/202142295}

\bibitem[{{Gabuzda} {et~al.}(2008){Gabuzda}, {Vitrishchak}, {Mahmud}, \&
  {O'Sullivan}}]{Gabuzda+2008}
{Gabuzda}, D.~C., {Vitrishchak}, V.~M., {Mahmud}, M., \& {O'Sullivan}, S.~P.
  2008, \mnras, 384, 1003, \dodoi{10.1111/j.1365-2966.2007.12773.x}

\bibitem[{{Gammie} {et~al.}(2003){Gammie}, {McKinney}, \&
  {T{\'o}th}}]{2003ApJ...589..444G}
{Gammie}, C.~F., {McKinney}, J.~C., \& {T{\'o}th}, G. 2003, \apj, 589, 444,
  \dodoi{10.1086/374594}

\bibitem[{{Goddi} {et~al.}(2019){Goddi}, {Mart{\'\i}-Vidal}, {Messias}, {Crew},
  {Herrero-Illana}, {Impellizzeri}, {Rottmann}, {Wagner}, {Fomalont},
  {Matthews}, {Petry}, {Phillips}, {Tilanus}, {Villard}, {Blackburn},
  {Janssen}, \& {Wielgus}}]{Goddi_2019}
{Goddi}, C., {Mart{\'\i}-Vidal}, I., {Messias}, H., {et~al.} 2019, \pasp, 131,
  075003, \dodoi{10.1088/1538-3873/ab136a}

\bibitem[{{Goddi} {et~al.}(2021){Goddi}, {Mart{\'\i}-Vidal}, {Messias},
  {Bower}, {Broderick}, {Dexter}, {Marrone}, {Moscibrodzka}, {Nagai}, {Algaba},
  {Asada}, {Crew}, {G{\'o}mez}, {Impellizzeri}, {Janssen}, {Kadler},
  {Krichbaum}, {Lico}, {Matthews}, {Nathanail}, {Ricarte}, {Ros}, {Younsi}, \&
  {Akiyama}}]{Goddi2021}
---. 2021, \apjl, 910, L14, \dodoi{10.3847/2041-8213/abee6a}

\bibitem[{{Greisen }(2003)}]{Greisen_2003}
{Greisen }, E.~W. 2003, in Astrophysics and Space Science Library, Vol. 285,
  Information Handling in Astronomy - Historical Vistas, ed. A.~{Heck}, 109,
  \dodoi{10.1007/0-306-48080-8_7}

\bibitem[{{Hamaker} {et~al.}(1996){Hamaker}, {Bregman}, \&
  {Sault}}]{Hamaker_1996}
{Hamaker}, J.~P., {Bregman}, J.~D., \& {Sault}, R.~J. 1996, \aaps, 117, 137

\bibitem[{Harris {et~al.}(2020)Harris, Millman, van~der Walt, Gommers,
  Virtanen, Cournapeau, Wieser, Taylor, Berg, Smith, Kern, Picus, Hoyer, van
  Kerkwijk, Brett, Haldane, del R{\'{i}}o, Wiebe, Peterson,
  G{\'{e}}rard-Marchant, Sheppard, Reddy, Weckesser, Abbasi, Gohlke, \&
  Oliphant}]{numpy_2020}
Harris, C.~R., Millman, K.~J., van~der Walt, S.~J., {et~al.} 2020, Nature, 585,
  357, \dodoi{10.1038/s41586-020-2649-2}

\bibitem[{{H{\"o}gbom}(1974)}]{Hogbom_1974}
{H{\"o}gbom}, J.~A. 1974, \aaps, 15, 417

\bibitem[{{Homan} \& {Lister}(2006)}]{Homan2006}
{Homan}, D.~C., \& {Lister}, M.~L. 2006, \aj, 131, 1262, \dodoi{10.1086/500256}

\bibitem[{{Homan} \& {Wardle}(1999)}]{homan_wardle_1999}
{Homan}, D.~C., \& {Wardle}, J.~F.~C. 1999, \aj, 118, 1942,
  \dodoi{10.1086/301108}

\bibitem[{Hunter(2007)}]{Hunter_2007}
Hunter, J.~D. 2007, Computing In Science \& Engineering, 9, 90,
  \dodoi{10.1109/MCSE.2007.55}

\bibitem[{{Igumenshchev} {et~al.}(2003){Igumenshchev}, {Narayan}, \&
  {Abramowicz}}]{Igumenshchev+2003}
{Igumenshchev}, I.~V., {Narayan}, R., \& {Abramowicz}, M.~A. 2003, \apj, 592,
  1042, \dodoi{10.1086/375769}

\bibitem[{{Issaoun} {et~al.}(2022){Issaoun}, {Wielgus}, {Jorstad}, {Krichbaum},
  {Blackburn}, {Janssen}, {Chan}, {Pesce}, {G{\'o}mez}, {Akiyama},
  {Mo{\'s}cibrodzka}, {Mart{\'\i}-Vidal}, {Chael}, {Lico}, {Liu},
  {Ramakrishnan}, {Lisakov}, {Fuentes}, {Zhao}, {Moriyama}, {Broderick},
  {Tiede}, {MacDonald}, {Mizuno}, {Traianou}, {Loinard}, {Davelaar}, {Gurwell},
  \& {Lu}}]{Issaoun2022}
{Issaoun}, S., {Wielgus}, M., {Jorstad}, S., {et~al.} 2022, \apj, 934, 145,
  \dodoi{10.3847/1538-4357/ac7a40}

\bibitem[{{Janssen} {et~al.}(2022){Janssen}, {Radcliffe}, \&
  {Wagner}}]{2022Janssen}
{Janssen}, M., {Radcliffe}, J.~F., \& {Wagner}, J. 2022, Universe, 8, 527,
  \dodoi{10.3390/universe8100527}

\bibitem[{{Janssen} {et~al.}(2019){Janssen}, {Goddi}, {van Bemmel}, {Kettenis},
  {Small}, {Liuzzo}, {Rygl}, {Mart{\'\i}-Vidal}, {Blackburn}, {Wielgus}, \&
  {Falcke}}]{Janssen_2019}
{Janssen}, M., {Goddi}, C., {van Bemmel}, I.~M., {et~al.} 2019, \aap, 626, A75,
  \dodoi{10.1051/0004-6361/201935181}

\bibitem[{{Janssen} {et~al.}(2021){Janssen}, {Falcke}, {Kadler}, {Ros},
  {Wielgus}, {Akiyama}, {Balokovi{\'c}}, {Blackburn}, {Bouman}, {Chael},
  {Chan}, {Chatterjee}, {Davelaar}, {Edwards}, {Fromm}, {G{\'o}mez}, {Goddi},
  {Issaoun}, {Johnson}, {Kim}, {Koay}, {Krichbaum}, {Liu}, {Liuzzo}, {Markoff},
  {Markowitz}, {Marrone}, {Mizuno}, {M{\"u}ller}, {Ni}, {Pesce},
  {Ramakrishnan}, {Roelofs}, {Rygl}, {van Bemmel}, \& {Event Horizon Telescope
  Collaboration}}]{Janssen2021}
{Janssen}, M., {Falcke}, H., {Kadler}, M., {et~al.} 2021, Nature Astronomy, 5,
  1017, \dodoi{10.1038/s41550-021-01417-w}

\bibitem[{{Jiang} {et~al.}(2023){Jiang}, {Mizuno}, {Fromm}, \&
  {Nathanail}}]{Jiang+2023}
{Jiang}, H.-X., {Mizuno}, Y., {Fromm}, C.~M., \& {Nathanail}, A. 2023, \mnras,
  \dodoi{10.1093/mnras/stad1106}

\bibitem[{{Johnson} {et~al.}(2020){Johnson}, {Lupsasca}, {Strominger}, {Wong},
  {Hadar}, {Kapec}, {Narayan}, {Chael}, {Gammie}, {Galison}, {Palumbo},
  {Doeleman}, {Blackburn}, {Wielgus}, {Pesce}, {Farah}, \&
  {Moran}}]{Johnson_2020}
{Johnson}, M.~D., {Lupsasca}, A., {Strominger}, A., {et~al.} 2020, Science
  Advances, 6, eaaz1310, \dodoi{10.1126/sciadv.aaz1310}

\bibitem[{{Jones}(1988)}]{Jones_1988}
{Jones}, T.~W. 1988, \apj, 332, 678, \dodoi{10.1086/166685}

\bibitem[{{Jones} \& {O'Dell}(1977)}]{Jones_ODell_1977}
{Jones}, T.~W., \& {O'Dell}, S.~L. 1977, \apj, 214, 522, \dodoi{10.1086/155278}

\bibitem[{{Jorstad} {et~al.}(2023){Jorstad}, {Wielgus}, {Lico}, {Issaoun},
  {Broderick}, {Pesce}, {Liu}, {Zhao}, {Krichbaum}, {Blackburn}, {Chan},
  {Janssen}, {Ramakrishnan}, {Akiyama}, {Alberdi}, {Algaba}, {Bouman}, {Cho},
  {Fuentes}, {G{\'o}mez}, {Gurwell}, {Johnson}, {Kim}, {Lu},
  {Mart{\'\i}-Vidal}, {Moscibrodzka}, {P{\"o}tzl}, {Traianou}, {van Bemmel},
  {Alef}, {Anantua}, {Asada}, {Azulay}, {Bach}, {Baczko}, {Ball},
  {Balokovi{\'c}}, {Barrett}, {Baub{\"o}ck}, {Benson}, {Bintley}, {Blundell},
  {Bower}, {Boyce}, {Bremer}, {Brinkerink}, {Brissenden}, {Britzen},
  {Broguiere}, {Bronzwaer}, {Bustamante}, {Byun}, {Carlstrom}, {Ceccobello},
  {Chael}, {Chatterjee}, {Chatterjee}, {Chen}, {Chen}, {Cheng}, {Christian},
  {Conroy}, {Conway}, {Cordes}, {Crawford}, {Crew}, {Cruz-Osorio}, {Cui},
  {Davelaar}, {De Laurentis}, {Deane}, {Dempsey}, {Desvignes}, {Dexter},
  {Dhruv}, {Doeleman}, {Dougal}, {Dzib}, {Eatough}, {Emami}, {Falcke}, {Farah},
  {Fish}, {Fomalont}, {Ford}, {Fraga-Encinas}, {Freeman}, {Friberg}, {Fromm},
  {Galison}, {Gammie}, {Garc{\'\i}a}, {Gentaz}, {Georgiev}, {Goddi}, {Gold},
  {G{\'o}mez-Ruiz}, {Gu}, {Hada}, {Haggard}, {Haworth}, {Hecht}, {Hesper},
  {Heumann}, {Ho}, {Ho}, {Honma}, {Huang}, {Huang}, {Hughes}, {Ikeda},
  {Impellizzeri}, {Inoue}, {James}, {Jannuzi}, {Jeter}, {Jiang},
  {Jim{\'e}nez-Rosales}, {Joshi}, {Jung}, {Karami}, {Karuppusamy}, {Kawashima},
  {Keating}, {Kettenis}, {Kim}, {Kim}, {Kim}, {Kino}, {Koay}, {Kocherlakota},
  {Kofuji}, {Koyama}, {Kramer}, {Kramer}, {Kuo}, {La Bella}, {Lauer}, {Lee},
  {Lee}, {Leung}, {Levis}, {Li}, {Lindahl}, {Lindqvist}, {Lisakov}, {Liu},
  {Liuzzo}, {Lo}, {Lobanov}, {Loinard}, {Lonsdale}, {MacDonald}, {Mao},
  {Marchili}, {Markoff}, {Marrone}, {Marscher}, {Matsushita}, {Matthews},
  {Medeiros}, {Menten}, {Michalik}, {Mizuno}, {Mizuno}, {Moran}, {Moriyama},
  {M{\"u}ller}, {Mus}, {Musoke}, {Myserlis}, {Nadolski}, {Nagai}, {Nagar},
  {Nakamura}, {Narayan}, {Narayanan}, {Natarajan}, {Nathanail}, {Fuentes},
  {Neilsen}, {Neri}, {Ni}, {Noutsos}, {Nowak}, {Oh}, {Okino}, {Olivares},
  {Ortiz-Le{\'o}n}, {Oyama}, {{\"O}zel}, {Palumbo}, {Paraschos}, {Park},
  {Parsons}, {Patel}, {Pen}, {Pi{\'e}tu}, {Plambeck}, {PopStefanija}, {Porth},
  {Prather}, {Preciado-L{\'o}pez}, {Psaltis}, {Pu}, {Rao}, {Rawlings},
  {Raymond}, {Rezzolla}, {Ricarte}, {Ripperda}, {Roelofs}, {Rogers}, {Ros},
  {Romero-Ca{\~n}izales}, {Roshanineshat}, {Rottmann}, {Roy}, {Ruiz},
  {Ruszczyk}, {Rygl}, {S{\'a}nchez}, {S{\'a}nchez-Arg{\"u}elles},
  {S{\'a}nchez-Portal}, {Sasada}, {Satapathy}, {Savolainen}, {Schloerb},
  {Schonfeld}, {Schuster}, {Shao}, {Shen}, {Small}, {Sohn}, {SooHoo},
  {Souccar}, {Sun}, {Tazaki}, {Tetarenko}, {Tiede}, {Tilanus}, {Titus},
  {Torne}, {Trent}, {Trippe}, {Turk}, {van Langevelde}, {van Rossum}, {Vos},
  {Wagner}, {Ward-Thompson}, {Wardle}, {Weintroub}, {Wex}, {Wharton}, {Wiik},
  {Witzel}, {Wondrak}, {Wong}, {Wu}, {Yamaguchi}, {Yoon}, {Young}, {Young},
  {Younsi}, {Yuan}, {Yuan}, {Zensus}, {Zhang}, \& {Zhao}}]{Jorstad2023}
{Jorstad}, S., {Wielgus}, M., {Lico}, R., {et~al.} 2023, \apj, 943, 170,
  \dodoi{10.3847/1538-4357/acaea8}

\bibitem[{{Kettenis} {et~al.}(2006){Kettenis}, {van Langevelde}, {Reynolds}, \&
  {Cotton}}]{Kettenis_2006}
{Kettenis}, M., {van Langevelde}, H.~J., {Reynolds}, C., \& {Cotton}, B. 2006,
  in Astronomical Society of the Pacific Conference Series, Vol. 351,
  Astronomical Data Analysis Software and Systems XV, ed. C.~{Gabriel},
  C.~{Arviset}, D.~{Ponz}, \& S.~{Enrique}, 497

\bibitem[{{Kim} {et~al.}(2020){Kim}, {Krichbaum}, {Broderick}, {Wielgus},
  {Blackburn}, {G{\'o}mez}, {Johnson}, {Bouman}, {Chael}, {Akiyama}, {Jorstad},
  {Marscher}, {Issaoun}, {Janssen}, {Chan}, {Savolainen}, {Pesce}, {{\"O}zel},
  {Alberdi}, {Alef}, {Asada}, {Azulay}, {Baczko}, {Ball}, {Balokovi{\'c}},
  {Barrett}, {Bintley}, {Boland }, {Bower}, {Bremer}, {Brinkerink},
  {Brissenden}, {Britzen}, {Broguiere}, {Bronzwaer}, {Byun}, {Carlstrom},
  {Chatterjee}, {Chatterjee}, {Chen}, {Chen}, {Cho}, {Christian}, {Conway},
  {Cordes}, {Crew}, {Cui}, {Davelaar}, {De Laurentis}, {Deane}, {Dempsey},
  {Desvignes}, {Dexter}, {Doeleman}, {Eatough}, {Falcke}, {Fish}, {Fomalont},
  {Fraga-Encinas}, {Friberg}, {Fromm}, {Galison}, {Gammie}, {Garc{\'\i}a},
  {Gentaz}, {Georgiev}, {Goddi}, {Gold}, {G{\'o}mez-Ruiz}, {Gu}, {Gurwell},
  {Hada}, {Hecht}, {Hesper}, {Ho}, {Ho}, {Honma}, {Huang}, {Huang}, {Hughes},
  {Ikeda}, {Inoue}, {James}, {Jannuzi}, {Jeter}, {Jiang}, {Jimenez-Rosales},
  {Jung}, {Karami}, {Karuppusamy}, {Kawashima}, {Keating}, {Kettenis}, {Kim},
  {Kim}, {Kino}, {Koay}, {Koch}, {Koyama}, {Kramer}, {Kramer}, {Kuo}, {Lauer},
  {Lee}, {Li}, {Li}, {Lindqvist}, {Lico}, {Liu}, {Liuzzo}, {Lo}, {Lobanov},
  {Loinard}, {Lonsdale}, {Lu}, {MacDonald}, {Mao}, {Markoff}, {Marrone},
  {Mart{\'\i}-Vidal}, {Matsushita}, {Matthews}, {Medeiros}, {Menten}, {Mizuno},
  {Mizuno}, {Moran}, {Moriyama}, {Moscibrodzka}, {Musoke}, {M{\"u}ller},
  {Nagai}, {Nagar}, {Nakamura}, {Narayan}, {Narayanan}, {Natarajan}, {Neri},
  {Ni}, {Noutsos}, {Okino}, {Olivares}, {Ortiz-Le{\'o}n}, {Oyama}, {Palumbo},
  {Park}, {Patel}, {Pen}, {Pi{\'e}tu}, {Plambeck}, {PopStefanija}, {Porth},
  {Prather}, {Preciado-L{\'o}pez}, {Psaltis}, {Pu}, {Ramakrishnan}, {Rao},
  {Rawlings}, {Raymond}, {Rezzolla}, {Ripperda}, {Roelofs}, {Rogers}, {Ros},
  {Rose}, {Roshanineshat}, {Rottmann}, {Roy}, {Ruszczyk}, {Ryan}, {Rygl},
  {S{\'a}nchez}, {S{\'a}nchez-Arguelles}, {Sasada}, {Schloerb}, {Schuster},
  {Shao}, {Shen}, {Small}, {Sohn}, {SooHoo}, {Tazaki}, {Tiede}, {Tilanus},
  {Titus}, {Toma}, {Torne}, {Trent}, {Traianou}, {Trippe}, {Tsuda}, {van
  Bemmel}, {van Langevelde}, {van Rossum}, {Wagner}, {Wardle}, {Ward-Thompson},
  {Weintroub}, {Wex}, {Wharton}, {Wong}, {Wu}, {Yoon}, {Young}, {Young},
  {Younsi}, {Yuan}, {Yuan}, {Zensus}, {Zhao}, {Zhao}, {Zhu}, {Algaba},
  {Allardi}, {Amestica}, {Anczarski}, {Bach}, {Baganoff}, {Beaudoin}, {Benson},
  {Berthold}, {Blanchard}, {Blundell}, {Bustamente}, {Cappallo},
  {Castillo-Dom{\'\i}nguez}, {Chang}, {Chang}, {Chang}, {Chen}, {Chilson},
  {Chuter}, {Rosado}, {Coulson}, {Crowley}, {Derome}, {Dexter}, {Dornbusch},
  {Dudevoir}, {Dzib}, {Eckart}, {Eckert}, {Erickson}, {Everett}, {Faber},
  {Farah}, {Fath}, {Folkers}, {Forbes}, {Freund}, {Gale}, {Gao}, {Geertsema},
  {Graham}, {Greer}, {Grosslein}, {Gueth}, {Haggard}, {Halverson}, {Han},
  {Han}, {Hao}, {Hasegawa}, {Henning}, {Hern{\'a}ndez-G{\'o}mez},
  {Herrero-Illana}, {Heyminck}, {Hirota}, {Hoge}, {Huang}, {Violette
  Impellizzeri}, {Jiang}, {John}, {Kamble}, {Keisler}, {Kimura}, {Kono},
  {Kubo}, {Kuroda}, {Lacasse}, {Laing}, {Leitch}, {Li}, {Lin}, {Liu}, {Liu},
  {Lu}, {Marson}, {Martin-Cocher}, {Massingill}, {Matulonis}, {McColl},
  {McWhirter}, {Messias}, {Meyer-Zhao}, {Michalik}, {Monta{\~n}a},
  {Montgomerie}, {Mora-Klein}, {Muders}, {Nadolski}, {Navarro}, {Neilsen},
  {Nguyen}, {Nishioka}, {Norton}, {Nowak}, {Nystrom}, {Ogawa}, {Oshiro},
  {Oyama}, {Parsons}, {Pe{\~n}alver}, {Phillips}, {Poirier}, {Pradel},
  {Primiani}, {Raffin}, {Rahlin}, {Reiland}, {Risacher}, {Ruiz},
  {S{\'a}ez-Mada{\'\i}n}, {Sassella}, {Schellart}, {Shaw}, {Silva}, {Shiokawa},
  {Smith}, {Snow}, {Souccar}, {Sousa}, {Sridharan}, {Srinivasan}, {Stahm},
  {Stark}, {Story}, {Timmer}, {Vertatschitsch}, {Walther}, {Wei}, {Whitehorn},
  {Whitney}, {Woody}, {Wouterloot}, {Wright}, {Yamaguchi}, {Yu}, {Zeballos},
  {Zhang}, {Ziurys}, \& {Event Horizon Telescope Collaboration}}]{Kim:2020}
{Kim}, J.-Y., {Krichbaum}, T.~P., {Broderick}, A.~E., {et~al.} 2020, \aap, 640,
  A69, \dodoi{10.1051/0004-6361/202037493}

\bibitem[{Kluyver {et~al.}(2016)Kluyver, Ragan-Kelley, P{\'e}rez, Granger,
  Bussonnier, Frederic, Kelley, Hamrick, Grout, Corlay, Ivanov, Avila, Abdalla,
  \& Willing}]{jupyter}
Kluyver, T., Ragan-Kelley, B., P{\'e}rez, F., {et~al.} 2016, in Positioning and
  Power in Academic Publishing: Players, Agents and Agendas, ed. F.~Loizides \&
  B.~Schmidt (IOS Press), 87 -- 90

\bibitem[{{Kunz} {et~al.}(2016){Kunz}, {Stone}, \& {Quataert}}]{Kunz+2016}
{Kunz}, M.~W., {Stone}, J.~M., \& {Quataert}, E. 2016, \prl, 117, 235101,
  \dodoi{10.1103/PhysRevLett.117.235101}

\bibitem[{{Leung} {et~al.}(2011){Leung}, {Gammie}, \&
  {Noble}}]{2011ApJ...737...21L}
{Leung}, P.~K., {Gammie}, C.~F., \& {Noble}, S.~C. 2011, \apj, 737, 21,
  \dodoi{10.1088/0004-637X/737/1/21}

\bibitem[{{MacDonald} \& {Marscher}(2018)}]{Macdonald&Marscher2018}
{MacDonald}, N.~R., \& {Marscher}, A.~P. 2018, \apj, 862, 58,
  \dodoi{10.3847/1538-4357/aacc62}

\bibitem[{{Mao} {et~al.}(2017){Mao}, {Dexter}, \& {Quataert}}]{Mao+2017}
{Mao}, S.~A., {Dexter}, J., \& {Quataert}, E. 2017, \mnras, 466, 4307,
  \dodoi{10.1093/mnras/stw3324}

\bibitem[{{Marszewski} {et~al.}(2021){Marszewski}, {Prather}, {Joshi},
  {Pandya}, \& {Gammie}}]{Marszewski+2021}
{Marszewski}, A., {Prather}, B.~S., {Joshi}, A.~V., {Pandya}, A., \& {Gammie},
  C.~F. 2021, \apj, 921, 17, \dodoi{10.3847/1538-4357/ac1b28}

\bibitem[{{Mart{\'\i}-Vidal} \& {Marcaide}(2008)}]{Marti_2008}
{Mart{\'\i}-Vidal}, I., \& {Marcaide}, J.~M. 2008, \aap, 480, 289,
  \dodoi{10.1051/0004-6361:20078690}

\bibitem[{{Mart{\'\i}-Vidal} {et~al.}(2021){Mart{\'\i}-Vidal}, {Mus},
  {Janssen}, {de Vicente}, \& {Gonz{\'a}lez}}]{polsolve}
{Mart{\'\i}-Vidal}, I., {Mus}, A., {Janssen}, M., {de Vicente}, P., \&
  {Gonz{\'a}lez}, J. 2021, \aap, 646, A52, \dodoi{10.1051/0004-6361/202039527}

\bibitem[{{Mart{\'\i}-Vidal} {et~al.}(2016){Mart{\'\i}-Vidal}, {Roy}, {Conway},
  \& {Zensus}}]{Marti_2016}
{Mart{\'\i}-Vidal}, I., {Roy}, A., {Conway}, J., \& {Zensus}, A.~J. 2016, \aap,
  587, A143, \dodoi{10.1051/0004-6361/201526063}

\bibitem[{{Matthews} {et~al.}(2018){Matthews}, {Crew}, {Doeleman}, {Lacasse},
  {Saez}, {Alef}, {Akiyama}, {Amestica}, {Anderson}, {Barkats}, {Baudry},
  {Brogui{\`e}re}, {Escoffier}, {Fish}, {Greenberg}, {Hecht}, {Hiriart},
  {Hirota}, {Honma}, {Ho}, {Impellizzeri}, {Inoue}, {Kohno}, {Lopez},
  {Mart{\'{\i}}-Vidal}, {Messias}, {Meyer-Zhao}, {Mora-Klein}, {Nagar},
  {Nishioka}, {Oyama}, {Pankratius}, {Perez}, {Phillips}, {Pradel}, {Rottmann},
  {Roy}, {Ruszczyk}, {Shillue}, {Suzuki}, \& {Treacy}}]{Matthews_2018}
{Matthews}, L.~D., {Crew}, G.~B., {Doeleman}, S.~S., {et~al.} 2018, \pasp, 130,
  015002, \dodoi{10.1088/1538-3873/aa9c3d}

\bibitem[{{McKinney} {et~al.}(2012){McKinney}, {Tchekhovskoy}, \&
  {Blandford}}]{mckinney_2012}
{McKinney}, J.~C., {Tchekhovskoy}, A., \& {Blandford}, R.~D. 2012, \mnras, 423,
  3083, \dodoi{10.1111/j.1365-2966.2012.21074.x}

\bibitem[{McKinney(2010)}]{pandas}
McKinney, W. 2010, in Proceedings of the 9th Python in Science Conference, ed.
  S.~van~der Walt \& J.~Millman, 51 -- 56

\bibitem[{{Mizuno} {et~al.}(2021){Mizuno}, {Fromm}, {Younsi}, {Porth},
  {Olivares}, \& {Rezzolla}}]{Mizuno+2021}
{Mizuno}, Y., {Fromm}, C.~M., {Younsi}, Z., {et~al.} 2021, \mnras, 506, 741,
  \dodoi{10.1093/mnras/stab1753}

\bibitem[{{Mo{\'s}cibrodzka} {et~al.}(2016){Mo{\'s}cibrodzka}, {Falcke}, \&
  {Shiokawa}}]{Moscibrodzka:2016}
{Mo{\'s}cibrodzka}, M., {Falcke}, H., \& {Shiokawa}, H. 2016, \aap, 586, A38,
  \dodoi{10.1051/0004-6361/201526630}

\bibitem[{{Mo{\'s}cibrodzka} \& {Gammie}(2018)}]{Moscibrodzka&Gammie2018}
{Mo{\'s}cibrodzka}, M., \& {Gammie}, C.~F. 2018, \mnras, 475, 43,
  \dodoi{10.1093/mnras/stx3162}

\bibitem[{{Mo{\'s}cibrodzka} {et~al.}(2021){Mo{\'s}cibrodzka}, {Janiuk}, \& {De
  Laurentis}}]{Moscibrodzka+2021}
{Mo{\'s}cibrodzka}, M., {Janiuk}, A., \& {De Laurentis}, M. 2021, \mnras, 508,
  4282, \dodoi{10.1093/mnras/stab2790}

\bibitem[{{Myserlis} {et~al.}(2018){Myserlis}, {Angelakis}, {Kraus}, {Liontas},
  {Marchili}, {Aller}, {Aller}, {Karamanavis}, {Fuhrmann}, {Krichbaum}, \&
  {Zensus}}]{Myserlis2018}
{Myserlis}, I., {Angelakis}, E., {Kraus}, A., {et~al.} 2018, \aap, 609, A68,
  \dodoi{10.1051/0004-6361/201630301}

\bibitem[{{Narayan} {et~al.}(2022){Narayan}, {Chael}, {Chatterjee}, {Ricarte},
  \& {Curd}}]{narayan_2022}
{Narayan}, R., {Chael}, A., {Chatterjee}, K., {Ricarte}, A., \& {Curd}, B.
  2022, \mnras, 511, 3795, \dodoi{10.1093/mnras/stac285}

\bibitem[{{Narayan} {et~al.}(2003){Narayan}, {Igumenshchev}, \&
  {Abramowicz}}]{Narayan+2003}
{Narayan}, R., {Igumenshchev}, I.~V., \& {Abramowicz}, M.~A. 2003, \pasj, 55,
  L69, \dodoi{10.1093/pasj/55.6.L69}

\bibitem[{{Narayan} {et~al.}(2012){Narayan}, {S\k{a}dowski}, {Penna}, \&
  {Kulkarni}}]{Narayan+2012}
{Narayan}, R., {S\k{a}dowski}, A., {Penna}, R.~F., \& {Kulkarni}, A.~K. 2012,
  \mnras, 426, 3241, \dodoi{10.1111/j.1365-2966.2012.22002.x}

\bibitem[{{Noble} {et~al.}(2006){Noble}, {Gammie}, {McKinney}, \& {Del
  Zanna}}]{Noble2006}
{Noble}, S.~C., {Gammie}, C.~F., {McKinney}, J.~C., \& {Del Zanna}, L. 2006,
  \apj, 641, 626, \dodoi{10.1086/500349}

\bibitem[{{Noble} {et~al.}(2007){Noble}, {Leung}, {Gammie}, \&
  {Book}}]{Noble2007}
{Noble}, S.~C., {Leung}, P.~K., {Gammie}, C.~F., \& {Book}, L.~G. 2007,
  Classical and Quantum Gravity, 24, S259, \dodoi{10.1088/0264-9381/24/12/S17}

\bibitem[{{{\"O}zel} {et~al.}(2000){{\"O}zel}, {Psaltis}, \&
  {Narayan}}]{Ozel+2000}
{{\"O}zel}, F., {Psaltis}, D., \& {Narayan}, R. 2000, \apj, 541, 234,
  \dodoi{10.1086/309396}

\bibitem[{{Palumbo} {et~al.}(2020){Palumbo}, {Wong}, \& {Prather}}]{PWP}
{Palumbo}, D. C.~M., {Wong}, G.~N., \& {Prather}, B.~S. 2020, \apj, 894, 156,
  \dodoi{10.3847/1538-4357/ab86ac}

\bibitem[{{Pesce}(2021)}]{DMC}
{Pesce}, D.~W. 2021, \aj, 161, 178, \dodoi{10.3847/1538-3881/abe3f8}

\bibitem[{{Pierrard} \& {Lazar}(2010)}]{Pierrard&Lazar2010}
{Pierrard}, V., \& {Lazar}, M. 2010, \solphys, 267, 153,
  \dodoi{10.1007/s11207-010-9640-2}

\bibitem[{Pordes {et~al.}(2007)Pordes, Petravick, Kramer, Olson, Livny, Roy,
  Avery, Blackburn, Wenaus, W{\"u}rthwein, Foster, Gardner, Wilde, Blatecky,
  McGee, \& Quick}]{osg07}
Pordes, R., Petravick, D., Kramer, B., {et~al.} 2007, in 78, Vol.~78, J. Phys.
  Conf. Ser., 012057, \dodoi{10.1088/1742-6596/78/1/012057}

\bibitem[{{Porth} {et~al.}(2019){Porth}, {Chatterjee}, {Narayan}, {Gammie},
  {Mizuno}, {Anninos}, {Baker}, {Bugli}, {Chan}, {Davelaar}, {Del Zanna},
  {Etienne}, {Fragile}, {Kelly}, {Liska}, {Markoff}, {McKinney}, {Mishra},
  {Noble}, {Olivares}, {Prather}, {Rezzolla}, {Ryan}, {Stone}, {Tomei},
  {White}, {Younsi}, {Akiyama}, {Alberdi}, {Alef}, {Asada}, {Azulay}, {Baczko},
  {Ball}, {Balokovi{\'c}}, {Barrett}, {Bintley}, {Blackburn}, {Boland},
  {Bouman}, {Bower}, {Bremer}, {Brinkerink}, {Brissenden}, {Britzen},
  {Broderick}, {Broguiere}, {Bronzwaer}, {Byun}, {Carlstrom}, {Chael},
  {Chatterjee}, {Chen}, {Chen}, {Cho}, {Christian}, {Conway}, {Cordes},
  {Geoffrey}, {Crew}, {Cui}, {De Laurentis}, {Deane}, {Dempsey}, {Desvignes},
  {Doeleman}, {Eatough}, {Falcke}, {Fish}, {Fomalont}, {Fraga-Encinas},
  {Freeman}, {Friberg}, {Fromm}, {G{\'o}mez}, {Galison}, {Garc{\'\i}a},
  {Gentaz}, {Georgiev}, {Goddi}, {Gold}, {Gu}, {Gurwell}, {Hada}, {Hecht},
  {Hesper}, {Ho}, {Ho}, {Honma}, {Huang}, {Huang}, {Hughes}, {Ikeda}, {Inoue},
  {Issaoun}, {James}, {Jannuzi}, {Janssen}, {Jeter}, {Jiang}, {Johnson},
  {Jorstad}, {Jung}, {Karami}, {Karuppusamy}, {Kawashima}, {Keating},
  {Kettenis}, {Kim}, {Kim}, {Kim}, {Kino}, {Koay}, {Patrick}, {Koch}, {Koyama},
  {Kramer}, {Kramer}, {Krichbaum}, {Kuo}, {Lauer}, {Lee}, {Li}, {Li},
  {Lindqvist}, {Liu}, {Liuzzo}, {Lo}, {Lobanov}, {Loinard}, {Lonsdale}, {Lu},
  {MacDonald}, {Mao}, {Marrone}, {Marscher}, {Mart{\'\i}-Vidal}, {Matsushita},
  {Matthews}, {Medeiros}, {Menten}, {Mizuno}, {Moran}, {Moriyama},
  {Moscibrodzka}, {M{\"u}ller}, {Nagai}, {Nagar}, {Nakamura}, {Narayanan},
  {Natarajan}, {Neri}, {Ni}, {Noutsos}, {Okino}, {Oyama}, {{\"O}zel},
  {Palumbo}, {Patel}, {Pen}, {Pesce}, {Pi{\'e}tu}, {Plambeck}, {PopStefanija},
  {Preciado-L{\'o}pez}, {Psaltis}, {Pu}, {Ramakrishnan}, {Rao}, {Rawlings},
  {Raymond}, {Ripperda}, {Roelofs}, {Rogers}, {Ros}, {Rose}, {Roshanineshat},
  {Rottmann}, {Roy}, {Ruszczyk}, {Rygl}, {S{\'a}nchez},
  {S{\'a}nchez-Arguelles}, {Sasada}, {Savolainen}, {Schloerb}, {Schuster},
  {Shao}, {Shen}, {Small}, {Sohn}, {SooHoo}, {Tazaki}, {Tiede}, {Tilanus},
  {Titus}, {Toma}, {Torne}, {Trent}, {Trippe}, {Tsuda}, {van Bemmel}, {van
  Langevelde}, {van Rossum}, {Wagner}, {Wardle}, {Weintroub}, {Wex}, {Wharton},
  {Wielgus}, {Wong}, {Wu}, {Young}, {Young}, {Yuan}, {Yuan}, {Zensus}, {Zhao},
  {Zhao}, {Zhu}, \& {Event Horizon Telescope Collaboration}}]{porth_2019}
{Porth}, O., {Chatterjee}, K., {Narayan}, R., {et~al.} 2019, \apjs, 243, 26,
  \dodoi{10.3847/1538-4365/ab29fd}

\bibitem[{{Qiu} {et~al.}(2023){Qiu}, {Ricarte}, {Narayan}, {Wong}, {Chael}, \&
  {Palumbo}}]{Qiu+2023}
{Qiu}, R., {Ricarte}, A., {Narayan}, R., {et~al.} 2023, \mnras,
  \dodoi{10.1093/mnras/stad466}

\bibitem[{{Ricarte} {et~al.}(2023){Ricarte}, {Gammie}, {Narayan}, \&
  {Prather}}]{Ricarte+2023}
{Ricarte}, A., {Gammie}, C., {Narayan}, R., \& {Prather}, B.~S. 2023, \mnras,
  519, 4203, \dodoi{10.1093/mnras/stac3796}

\bibitem[{{Ricarte} {et~al.}(2021){Ricarte}, {Qiu}, \&
  {Narayan}}]{Ricarte+2021}
{Ricarte}, A., {Qiu}, R., \& {Narayan}, R. 2021, \mnras, 505, 523,
  \dodoi{10.1093/mnras/stab1289}

\bibitem[{{Roelofs} {et~al.}(2023){Roelofs}, {Johnson}, {Chael}, {Janssen},
  {Wielgus}, {Broderick}, {Akiyama}, {Alberdi}, {Alef}, {Algaba}, {Anantua},
  {Asada}, {Azulay}, {Bach}, {Baczko}, {Ball}, {Balokovi{\'c}}, {Barrett},
  {Baub{\"o}ck}, {Benson}, {Bintley}, {Blackburn}, {Blundell}, {Bouman},
  {Bower}, {Boyce}, {Bremer}, {Brinkerink}, {Brissenden}, {Britzen},
  {Broguiere}, {Bronzwaer}, {Bustamante}, {Byun}, {Carlstrom}, {Ceccobello},
  {Chan}, {Chang}, {Chatterjee}, {Chatterjee}, {Chen}, {Chen}, {Cheng}, {Cho},
  {Christian}, {Conroy}, {Conway}, {Cordes}, {Crawford}, {Crew}, {Cruz-Osorio},
  {Cui}, {Dahale}, {Davelaar}, {De Laurentis}, {Deane}, {Dempsey}, {Desvignes},
  {Dexter}, {Dhruv}, {Doeleman}, {Dougal}, {Dzib}, {Eatough}, {Emami},
  {Falcke}, {Farah}, {Fish}, {Fomalont}, {Ford}, {Foschi}, {Fraga-Encinas},
  {Freeman}, {Friberg}, {Fromm}, {Fuentes}, {Galison}, {Gammie}, {Garc{\'\i}a},
  {Gentaz}, {Georgiev}, {Goddi}, {Gold}, {G{\'o}mez-Ruiz}, {G{\'o}mez}, {Gu},
  {Gurwell}, {Hada}, {Haggard}, {Haworth}, {Hecht}, {Hesper}, {Heumann}, {Ho},
  {Ho}, {Honma}, {Huang}, {Huang}, {Hughes}, {Ikeda}, {Impellizzeri}, {Inoue},
  {Issaoun}, {James}, {Jannuzi}, {Jeter}, {Jiang}, {Jim{\'e}nez-Rosales},
  {Jorstad}, {Joshi}, {Jung}, {Karami}, {Karuppusamy}, {Kawashima}, {Keating},
  {Kettenis}, {Kim}, {Kim}, {Kim}, {Kim}, {Kino}, {Koay}, {Kocherlakota},
  {Kofuji}, {Koch}, {Koyama}, {Kramer}, {Kramer}, {Kramer}, {Krichbaum}, {Kuo},
  {La Bella}, {Lauer}, {Lee}, {Lee}, {Leung}, {Levis}, {Li}, {Lico}, {Lindahl},
  {Lindqvist}, {Lisakov}, {Liu}, {Liu}, {Liuzzo}, {Lo}, {Lobanov}, {Loinard},
  {Lonsdale}, {Lowitz}, {Lu}, {MacDonald}, {Mao}, {Marchili}, {Markoff},
  {Marrone}, {Marscher}, {Mart{\'\i}-Vidal}, {Matsushita}, {Matthews},
  {Medeiros}, {Menten}, {Michalik}, {Mizuno}, {Mizuno}, {Moran}, {Moriyama},
  {Moscibrodzka}, {Mulaudzi}, {M{\"u}ller}, {M{\"u}ller}, {Mus}, {Musoke},
  {Myserlis}, {Nadolski}, {Nagai}, {Nagar}, {Nakamura}, {Narayan}, {Narayanan},
  {Natarajan}, {Nathanail}, {Fuentes}, {Neilsen}, {Neri}, {Ni}, {Noutsos},
  {Nowak}, {Oh}, {Okino}, {Olivares}, {Ortiz-Le{\'o}n}, {Oyama}, {{\"O}zel},
  {Palumbo}, {Paraschos}, {Park}, {Parsons}, {Patel}, {Pen}, {Pesce},
  {Pi{\'e}tu}, {Plambeck}, {PopStefanija}, {Porth}, {P{\"o}tzl}, {Prather},
  {Preciado-L{\'o}pez}, {Psaltis}, {Pu}, {Ramakrishnan}, {Rao}, {Rawlings},
  {Raymond}, {Rezzolla}, {Ricarte}, {Ripperda}, {Rogers},
  {Romero-Ca{\~n}izales}, {Ros}, {Roshanineshat}, {Rottmann}, {Roy}, {Ruiz},
  {Ruszczyk}, {Rygl}, {S{\'a}nchez}, {S{\'a}nchez-Arg{\"u}elles},
  {S{\'a}nchez-Portal}, {Sasada}, {Satapathy}, {Savolainen}, {Schloerb},
  {Schonfeld}, {Schuster}, {Shao}, {Shen}, {Small}, {Sohn}, {SooHoo},
  {Sosapanta Salas}, {Souccar}, {Sun}, {Tazaki}, {Tetarenko}, {Tiede},
  {Tilanus}, {Titus}, {Torne}, {Toscano}, {Traianou}, {Trent}, {Trippe},
  {Turk}, {van Bemmel}, {van Langevelde}, {van Rossum}, {Vos}, {Wagner},
  {Ward-Thompson}, {Wardle}, {Washington}, {Weintroub}, {Wharton}, {Wiik},
  {Witzel}, {Wondrak}, {Wong}, {Wu}, {Yadlapalli}, {Yamaguchi}, {Yfantis},
  {Yoon}, {Young}, {Young}, {Younsi}, {Yu}, {Yuan}, {Yuan}, {Zensus}, {Zhang},
  {Zhao}, \& {Zhao}}]{Roelofs_2023}
{Roelofs}, F., {Johnson}, M.~D., {Chael}, A., {et~al.} 2023, \apjl, 957, L21,
  \dodoi{10.3847/2041-8213/acff6f}

\bibitem[{Sfiligoi {et~al.}(2009)Sfiligoi, Bradley, Holzman, Mhashilkar, Padhi,
  \& Wurthwein}]{osg09}
Sfiligoi, I., Bradley, D.~C., Holzman, B., {et~al.} 2009, in 2, Vol.~2, 2009
  WRI World Congress on Computer Science and Information Engineering, 428--432,
  \dodoi{10.1109/CSIE.2009.950}

\bibitem[{{Shepherd}(2011)}]{Shepherd_2011}
{Shepherd}, M. 2011, {Difmap: Synthesis Imaging of Visibility Data},
  Astrophysics Source Code Library.
\newblock \doeprint{1103.001}

\bibitem[{{S{\k{a}}dowski} {et~al.}(2013){S{\k{a}}dowski}, {Narayan}, {Penna},
  \& {Zhu}}]{Sadowski+2013}
{S{\k{a}}dowski}, A., {Narayan}, R., {Penna}, R., \& {Zhu}, Y. 2013, \mnras,
  436, 3856, \dodoi{10.1093/mnras/stt1881}

\bibitem[{{Smirnov}(2011)}]{Smirnov_2011}
{Smirnov}, O.~M. 2011, \aap, 527, A106, \dodoi{10.1051/0004-6361/201016082}

\bibitem[{{Speagle}(2020)}]{Speagle2020}
{Speagle}, J.~S. 2020, \mnras, 493, 3132,
  \dodoi{10.1093/mnras/staa27810.48550/arXiv.1904.02180}

\bibitem[{{Steel} {et~al.}(2019){Steel}, {Wielgus}, {Blackburn}, {Issaoun}, \&
  {Johnson}}]{Steel2019}
{Steel}, S., {Wielgus}, M., {Blackburn}, L., {Issaoun}, S., \& {Johnson}, M.
  2019, EHT Memo Series, 2019-CE-03

\bibitem[{Tange(2011)}]{GNU}
Tange, O. 2011, ;login: The USENIX Magazine, 36, 42.
\newblock \url{http://www.gnu.org/s/parallel}

\bibitem[{{Tchekhovskoy} {et~al.}(2011){Tchekhovskoy}, {Narayan}, \&
  {McKinney}}]{tchekhovskoy_2011}
{Tchekhovskoy}, A., {Narayan}, R., \& {McKinney}, J.~C. 2011, \mnras, 418, L79,
  \dodoi{10.1111/j.1745-3933.2011.01147.x}

\bibitem[{{The Astropy Collaboration} {et~al.}(2013){The Astropy
  Collaboration}, {Robitaille}, {Tollerud}, {Greenfield}, {Droettboom}, {Bray},
  {Aldcroft}, {Davis}, {Ginsburg}, {Price-Whelan}, {Kerzendorf}, {Conley},
  {Crighton}, {Barbary}, {Muna}, {Ferguson}, {Grollier}, {Parikh}, {Nair},
  {Unther}, {Deil}, {Woillez}, {Conseil}, {Kramer}, {Turner}, {Singer}, {Fox},
  {Weaver}, {Zabalza}, {Edwards}, {Azalee Bostroem}, {Burke}, {Casey},
  {Crawford}, {Dencheva}, {Ely}, {Jenness}, {Labrie}, {Lim}, {Pierfederici},
  {Pontzen}, {Ptak}, {Refsdal}, {Servillat}, \& {Streicher}}]{astropy_2013}
{The Astropy Collaboration}, {Robitaille}, T.~P., {Tollerud}, E.~J., {et~al.}
  2013, \aap, 558, A33, \dodoi{10.1051/0004-6361/201322068}

\bibitem[{{The Astropy Collaboration} {et~al.}(2018){The Astropy
  Collaboration}, {Price-Whelan}, {Sip{\H o}cz}, {G{\"u}nther}, {Lim},
  {Crawford}, {Conseil}, {Shupe}, {Craig}, {Dencheva}, {Ginsburg},
  {VanderPlas}, {Bradley}, {P{\'e}rez-Su{\'a}rez}, {de Val-Borro}, {Aldcroft},
  {Cruz}, {Robitaille}, {Tollerud}, {Ardelean}, {Babej}, {Bach}, {Bachetti},
  {Bakanov}, {Bamford}, {Barentsen}, {Barmby}, {Baumbach}, {Berry}, {Biscani},
  {Boquien}, {Bostroem}, {Bouma}, {Brammer}, {Bray}, {Breytenbach},
  {Buddelmeijer}, {Burke}, {Calderone}, {Cano Rodr{\'{\i}}guez}, {Cara},
  {Cardoso}, {Cheedella}, {Copin}, {Corrales}, {Crichton}, {D'Avella}, {Deil},
  {Depagne}, {Dietrich}, {Donath}, {Droettboom}, {Earl}, {Erben}, {Fabbro},
  {Ferreira}, {Finethy}, {Fox}, {Garrison}, {Gibbons}, {Goldstein}, {Gommers},
  {Greco}, {Greenfield}, {Groener}, {Grollier}, {Hagen}, {Hirst}, {Homeier},
  {Horton}, {Hosseinzadeh}, {Hu}, {Hunkeler}, {Ivezi{\'c}}, {Jain}, {Jenness},
  {Kanarek}, {Kendrew}, {Kern}, {Kerzendorf}, {Khvalko}, {King}, {Kirkby},
  {Kulkarni}, {Kumar}, {Lee}, {Lenz}, {Littlefair}, {Ma}, {Macleod},
  {Mastropietro}, {McCully}, {Montagnac}, {Morris}, {Mueller}, {Mumford},
  {Muna}, {Murphy}, {Nelson}, {Nguyen}, {Ninan}, {N{\"o}the}, {Ogaz}, {Oh},
  {Parejko}, {Parley}, {Pascual}, {Patil}, {Patil}, {Plunkett}, {Prochaska},
  {Rastogi}, {Reddy Janga}, {Sabater}, {Sakurikar}, {Seifert}, {Sherbert},
  {Sherwood-Taylor}, {Shih}, {Sick}, {Silbiger}, {Singanamalla}, {Singer},
  {Sladen}, {Sooley}, {Sornarajah}, {Streicher}, {Teuben}, {Thomas},
  {Tremblay}, {Turner}, {Terr{\'o}n}, {van Kerkwijk}, {de la Vega}, {Watkins},
  {Weaver}, {Whitmore}, {Woillez}, {Zabalza}, \& {Astropy
  Contributors}}]{astropy_2018}
{The Astropy Collaboration}, {Price-Whelan}, A.~M., {Sip{\H o}cz}, B.~M.,
  {et~al.} 2018, \aj, 156, 123, \dodoi{10.3847/1538-3881/aabc4f}

\bibitem[{{Thompson} {et~al.}(2017){Thompson}, {Moran}, \& {Swenson}}]{TMS}
{Thompson}, A.~R., {Moran}, J.~M., \& {Swenson}, Jr., G.~W. 2017,
  {Interferometry and Synthesis in Radio Astronomy, 3rd Edition} (Springer
  International Publishing), \dodoi{10.1007/978-3-319-44431-4}

\bibitem[{{Thum} {et~al.}(2018){Thum}, {Agudo}, {Molina}, {Casadio},
  {G{\'o}mez}, {Morris}, {Ramakrishnan}, \& {Sievers}}]{Thum2018}
{Thum}, C., {Agudo}, I., {Molina}, S.~N., {et~al.} 2018, \mnras, 473, 2506,
  \dodoi{10.1093/mnras/stx2436}

\bibitem[{{Tsunetoe} {et~al.}(2022){Tsunetoe}, {Mineshige}, {Kawashima},
  {Ohsuga}, {Akiyama}, \& {Takahashi}}]{Tsunetoe+2022}
{Tsunetoe}, Y., {Mineshige}, S., {Kawashima}, T., {et~al.} 2022, \apj, 931, 25,
  \dodoi{10.3847/1538-4357/ac66dd}

\bibitem[{{Tsunetoe} {et~al.}(2020){Tsunetoe}, {Mineshige}, {Ohsuga},
  {Kawashima}, \& {Akiyama}}]{Tsunetoe+2020}
{Tsunetoe}, Y., {Mineshige}, S., {Ohsuga}, K., {Kawashima}, T., \& {Akiyama},
  K. 2020, \pasj, 72, 32, \dodoi{10.1093/pasj/psaa008}

\bibitem[{{Tsunetoe} {et~al.}(2021){Tsunetoe}, {Mineshige}, {Ohsuga},
  {Kawashima}, \& {Akiyama}}]{Tsunetoe+2021}
---. 2021, \pasj, 73, 912, \dodoi{10.1093/pasj/psab054}

\bibitem[{{Vasyliunas}(1968)}]{Vasyliunas1968}
{Vasyliunas}, V.~M. 1968, \jgr, 73, 2839, \dodoi{10.1029/JA073i009p02839}

\bibitem[{Virtanen {et~al.}(2020)Virtanen, Gommers, Oliphant, Haberland, Reddy,
  Cournapeau, Burovski, Peterson, Weckesser, Bright, {van der Walt}, Brett,
  Wilson, Millman, Mayorov, Nelson, Jones, Kern, Larson, Carey, Polat, Feng,
  Moore, {VanderPlas}, Laxalde, Perktold, Cimrman, Henriksen, Quintero, Harris,
  Archibald, Ribeiro, Pedregosa, {van Mulbregt}, \& {SciPy 1.0
  Contributors}}]{scipy}
Virtanen, P., Gommers, R., Oliphant, T.~E., {et~al.} 2020, Nature Methods, 17,
  261, \dodoi{10.1038/s41592-019-0686-2}

\bibitem[{{Vitrishchak} {et~al.}(2008){Vitrishchak}, {Gabuzda}, {Algaba},
  {Rastorgueva}, {O'Sullivan}, \& {O'Dowd}}]{Vitrishchak_2008}
{Vitrishchak}, V.~M., {Gabuzda}, D.~C., {Algaba}, J.~C., {et~al.} 2008, \mnras,
  391, 124, \dodoi{10.1111/j.1365-2966.2008.13919.x}

\bibitem[{{Wardle} \& {Homan}(2003)}]{Wardle&Homan2003}
{Wardle}, J. F.~C., \& {Homan}, D.~C. 2003, \apss, 288, 143,
  \dodoi{10.1023/B:ASTR.0000005001.80514.0c}

\bibitem[{{Wardle} {et~al.}(1998){Wardle}, {Homan}, {Ojha}, \&
  {Roberts}}]{Wardle+1998}
{Wardle}, J.~F.~C., {Homan}, D.~C., {Ojha}, R., \& {Roberts}, D.~H. 1998, \nat,
  395, 457, \dodoi{10.1038/26675}

\bibitem[{{Wielgus} {et~al.}(2020){Wielgus}, {Akiyama}, {Blackburn}, {Chan},
  {Dexter}, {Doeleman}, {Fish}, {Issaoun}, {Johnson}, {Krichbaum}, {Lu},
  {Pesce}, {Wong}, {Bower}, {Broderick}, {Chael}, {Chatterjee}, {Gammie},
  {Georgiev}, {Hada}, {Loinard}, {Markoff}, {Marrone}, {Plambeck}, {Weintroub},
  {Dexter}, {MacMahon}, \& {Wright}}]{Wielgus2020}
{Wielgus}, M., {Akiyama}, K., {Blackburn}, L., {et~al.} 2020, \apj, 901, 67,
  \dodoi{10.3847/1538-4357/abac0d}

\bibitem[{{Wielgus} {et~al.}(2022{\natexlab{a}}){Wielgus}, {Moscibrodzka},
  {Vos}, {Gelles}, {Mart{\'\i}-Vidal}, {Farah}, {Marchili}, {Goddi}, \&
  {Messias}}]{Wielgus2022}
{Wielgus}, M., {Moscibrodzka}, M., {Vos}, J., {et~al.} 2022{\natexlab{a}},
  \aap, 665, L6, \dodoi{10.1051/0004-6361/202244493}

\bibitem[{{Wielgus} {et~al.}(2022{\natexlab{b}}){Wielgus}, {Marchili},
  {Mart{\'\i}-Vidal}, {Keating}, {Ramakrishnan}, {Tiede}, {Fomalont},
  {Issaoun}, {Neilsen}, {Nowak}, {Blackburn}, {Gammie}, {Goddi}, {Haggard},
  {Lee}, {Moscibrodzka}, {Tetarenko}, {Bower}, {Chan}, {Chatterjee}, {Chesler},
  {Dexter}, {Doeleman}, {Georgiev}, {Gurwell}, {Johnson}, {Marrone}, {Mus},
  {Psaltis}, {Ripperda}, {Witzel}, {Akiyama}, {Alberdi}, {Alef}, {Algaba},
  {Anantua}, {Asada}, {Azulay}, {Bach}, {Baczko}, {Ball}, {Balokovi{\'c}},
  {Barrett}, {Baub{\"o}ck}, {Benson}, {Bintley}, {Blundell}, {Boland},
  {Bouman}, {Boyce}, {Bremer}, {Brinkerink}, {Brissenden}, {Britzen},
  {Broderick}, {Broguiere}, {Bronzwaer}, {Bustamante}, {Byun}, {Carlstrom},
  {Ceccobello}, {Chael}, {Chatterjee}, {Chen}, {Chen}, {Cho}, {Christian},
  {Conroy}, {Conway}, {Cordes}, {Crawford}, {Crew}, {Cruz-Osorio}, {Cui},
  {Davelaar}, {De Laurentis}, {Deane}, {Dempsey}, {Desvignes}, {Dhruv}, {Dzib},
  {Eatough}, {Emami}, {Falcke}, {Farah}, {Fish}, {Ford}, {Fraga-Encinas},
  {Freeman}, {Friberg}, {Fromm}, {Fuentes}, {Galison}, {Garc{\'\i}a}, {Gentaz},
  {Gold}, {G{\'o}mez-Ruiz}, {G{\'o}mez}, {Gu}, {Hada}, {Haworth}, {Hecht},
  {Hesper}, {Ho}, {Ho}, {Honma}, {Huang}, {Huang}, {Hughes}, {Ikeda},
  {Impellizzeri}, {Inoue}, {James}, {Jannuzi}, {Janssen}, {Jeter}, {Jiang},
  {Jim{\'e}nez-Rosales}, {Jorstad}, {Joshi}, {Jung}, {Karami}, {Karuppusamy},
  {Kawashima}, {Kettenis}, {Kim}, {Kim}, {Kim}, {Kim}, {Kino}, {Koay},
  {Kocherlakota}, {Kofuji}, {Koch}, {Koyama}, {Kramer}, {Kramer}, {Krichbaum},
  {Kuo}, {La Bella}, {Lauer}, {Lee}, {Leung}, {Levis}, {Li}, {Lico}, {Lindahl},
  {Lindqvist}, {Lisakov}, {Liu}, {Liu}, {Liuzzo}, {Lo}, {Lobanov}, {Loinard},
  {Lonsdale}, {Lu}, {Mao}, {Markoff}, {Marscher}, {Matsushita}, {Matthews},
  {Medeiros}, {Menten}, {Michalik}, {Mizuno}, {Mizuno}, {Moran}, {Moriyama},
  {M{\"u}ller}, {Musoke}, {Myserlis}, {Nadolski}, {Nagai}, {Nagar}, {Nakamura},
  {Narayan}, {Narayanan}, {Natarajan}, {Nathanail}, {Navarro Fuentes}, {Neri},
  {Ni}, {Noutsos}, {Oh}, {Okino}, {Olivares}, {Ortiz-Le{\'o}n}, {Oyama},
  {{\"O}zel}, {Palumbo}, {Paraschos}, {Park}, {Parsons}, {Patel}, {Pen},
  {Pesce}, {Pi{\'e}tu}, {Plambeck}, {PopStefanija}, {Porth}, {P{\"o}tzl},
  {Prather}, {Preciado-L{\'o}pez}, {Pu}, {Rao}, {Rawlings}, {Raymond},
  {Rezzolla}, {Ricarte}, {Roelofs}, {Rogers}, {Ros}, {Romero-Canizales},
  {Roshanineshat}, {Rottmann}, {Roy}, {Ruiz}, {Ruszczyk}, {Rygl},
  {S{\'a}nchez}, {S{\'a}nchez-Arg{\"u}elles}, {S{\'a}nchez-Portal}, {Sasada},
  {Satapathy}, {Savolainen}, {Schloerb}, {Schuster}, {Shao}, {Shen}, {Small},
  {Won Sohn}, {SooHoo}, {Souccar}, {Sun}, {Tazaki}, {Tilanus}, {Titus},
  {Torne}, {Traianou}, {Trent}, {Trippe}, {van Bemmel}, {van Langevelde}, {van
  Rossum}, {Vos}, {Wagner}, {Ward-Thompson}, {Wardle}, {Weintroub}, {Wex},
  {Wharton}, {Wiik}, {Wondrak}, {Wong}, {Wu}, {Yamaguchi}, {Yoon}, {Young},
  {Young}, {Younsi}, {Yuan}, {Yuan}, {Zensus}, {Zhang}, {Zhao}, \&
  {Zhao}}]{Wielgus2022_light_curves}
{Wielgus}, M., {Marchili}, N., {Mart{\'\i}-Vidal}, I., {et~al.}
  2022{\natexlab{b}}, \apjl, 930, L19, \dodoi{10.3847/2041-8213/ac6428}

\bibitem[{{Wong} \& {Gammie}(2022)}]{Wong&Gammie2022}
{Wong}, G.~N., \& {Gammie}, C.~F. 2022, \apj, 937, 60,
  \dodoi{10.3847/1538-4357/ac854d}

\bibitem[{{Wong} {et~al.}(2022){Wong}, {Prather}, {Dhruv}, {Ryan},
  {Mo{\'s}cibrodzka}, {Chan}, {Joshi}, {Yarza}, {Ricarte}, {Shiokawa},
  {Dolence}, {Noble}, {McKinney}, \& {Gammie}}]{Wong_2022}
{Wong}, G.~N., {Prather}, B.~S., {Dhruv}, V., {et~al.} 2022, \apjs, 259, 64,
  \dodoi{10.3847/1538-4365/ac582e}

\bibitem[{{Yoon} {et~al.}(2020){Yoon}, {Chatterjee}, {Markoff}, {van
  Eijnatten}, {Younsi}, {Liska}, \& {Tchekhovskoy}}]{Yoon+2020}
{Yoon}, D., {Chatterjee}, K., {Markoff}, S.~B., {et~al.} 2020, \mnras, 499,
  3178, \dodoi{10.1093/mnras/staa3031}

\end{thebibliography}

\allauthors

\end{document}